\documentclass[aps,pra,twocolumn,groupaddresses,superscriptaddress,longbibliography]{revtex4-1}
\usepackage{graphics}
\usepackage{amsmath}
\usepackage{graphicx}
\usepackage{amsfonts}
\usepackage{amssymb}
\newtheorem{theorem}{Theorem}
\newtheorem{lemma}[theorem]{Lemma}
\newtheorem{proposition}[theorem]{Proposition}

\DeclareMathOperator*{\argmin}{argmin}
\DeclareMathOperator*{\argmax}{argmax}
\DeclareMathOperator*{\Tr}{Tr}
\newcommand{\ket}[1]{|#1 \rangle}
\newcommand{\bra}[1]{\langle #1|}
\begin{document}

\title{Optimization and learning of quantum programs}
\author{Leonardo Banchi}
\affiliation{Department of Physics and Astronomy, University of
Florence, via G. Sansone 1, I-50019 Sesto Fiorentino (FI), Italy}
\affiliation{ INFN Sezione di Firenze, via G.Sansone 1, I-50019
Sesto Fiorentino (FI), Italy }
\author{Jason Pereira}
\affiliation{Department of Computer Science, University of York, York YO10 5GH, UK}
\author{Seth Lloyd}
\affiliation{Department of Mechanical Engineering, Massachusetts Institute of Technology (MIT), Cambridge MA 02139, USA}
\affiliation{Research Laboratory of Electronics, Massachusetts Institute of Technology (MIT), Cambridge MA 02139, USA}
\author{Stefano Pirandola}
\affiliation{Department of Computer Science, University of York, York YO10 5GH, UK}
\affiliation{Research Laboratory of Electronics, Massachusetts Institute of Technology (MIT), Cambridge MA 02139, USA}

\begin{abstract}
A programmable quantum processor is a fundamental model of quantum computation. In this model, any quantum channel can be approximated by applying a fixed universal quantum operation onto an input state and a quantum
``program''  state, whose role is to condition the operation performed by the processor. It is known that perfect channel simulation is only possible in the limit of infinitely large program states, so that finding the best program state represents an open problem in the presence of realistic finite-dimensional resources. Here we prove that the search for the optimal quantum program is a convex optimization problem. This can be solved either exactly, by minimizing a diamond distance cost function via semi-definite programming, or approximately, by minimizing other cost functions via gradient-based machine learning methods. We apply this general result to a number of different designs for the programmable quantum processor, from the shallow protocol of quantum teleportation, to deeper schemes relying on port-based teleportation and parametric quantum circuits. We benchmark the various designs by investigating their optimal performance in simulating arbitrary unitaries, Pauli and amplitude damping channels.
\end{abstract}
\maketitle

%\title{Quantum learning of channels via a programmable quantum processor}

\section{Introduction}

Today the field of quantum computing~\cite{nielsen2000quantum} is becoming
more and more mature, also thanks to the combined efforts of academic and
industrial researchers. From a theoretical point of view, this endeavour is
supported by increasing interconnections with other rapidly-advancing fields,
such as machine learning~\cite{MLbook}. For instance, we have recently
witnessed the development of new hybrid areas of investigation, such as
quantum-enhanced machine
learning~\cite{QML1,QML2,dunjko2018machine,schuld2015introduction,ciliberto2018quantum}
(e.g., quantum neural
networks, quantum annealing etc.), protocols of quantum-inspired machine
learning (e.g., for recommendation systems~\cite{QIML1} or component analysis
and supervised clustering~\cite{QIML2}) and classical learning methods applied
to quantum computers, as explored here in this manuscript.

In quantum computing, a fundamental model is the programmable quantum gate
array or programmable quantum processor~\cite{nielsen1997programmable}. This
is a quantum processor where a fixed quantum operation is applied to an input
state and a program state. The role of the program state is to condition the
quantum operation in such a way to apply some target quantum gate or channel
to the input state. This is a very flexible scheme but not actually universal:
an arbitrary quantum channel cannot be programmed exactly, unless the program
state is allowed to have an infinite number of qubits. For instance, a
possible design relies on port-based teleportation
(PBT)~\cite{ishizaka2008asymptotic,ishizaka2009quantum,ishizaka2015some},
where an input state is subject to certain local operations and classical
communication (LOCCs) that are programmed by a tensor product of $N$ bipartite
states. For infinite $N$, any quantum channel can be simulated by copies of
its Choi matrix~\cite{pirandola2018fundamental} but, for any finite $N$, this
simulation is not perfect.

Despite this fundamental model of quantum computation is known since 1997, a
quantitative characterization of its actual performance in terms of gate
implementation or channel simulation is still missing. Given a target quantum
gate or channel, it is not yet known what degree of approximation can be
reached and what kind of optimization procedure must be employed to choose the
program state. After more than 20 years, the solution to these open problems
comes from a suitable application of techniques of semidefinite programming
(SDP) and machine learning (ML).

In our work, we quantify the error between an arbitrary target channel and its
programmable simulation in terms of the diamond distance and other suitable
cost functions, including the trace distance and the quantum fidelity. For all
the considered cost functions, we are able to show that the minimization of
the simulation error is a convex optimization problem in the space of the
program states. This already solves an outstanding problem which affects
various models of quantum computers (e.g., variational quantum circuits) where
the optimization over classical parameters is non-convex and therefore not
guaranteed to converge to a global optimum. By contrast, because our problem
is proven to be convex, we can use SDP to minimize the diamond distance and
always find the optimal program state for the simulation of a target channel,
therefore optimizing the programmable quantum processor. Similarly, we may
find suboptimal solutions by minimizing the trace distance or the quantum
fidelity by means of gradient-based ML techniques, such as the projected
subgradient method~\cite{boyd2003subgradient} and the conjugate gradient
method~\cite{jaggi2011convex,jaggi2013revisiting}.
We note indeed that the minimization of the $\ell_1$-norm, mathematically related
to the quantum trace distance, is widely employed
in many ML tasks \cite{duchi2008efficient,liu2013tensor}, so many of those techniques
can be adapted for learning program states.

With these general results in our hands, we first discuss the optimal learning
of arbitrary unitaries with a generic programmable quantum processor. Then, we
consider specific designs of the processor, from a shallow scheme based on the
teleportation protocol, to higher-depth designs based on
PBT~\cite{ishizaka2008asymptotic,ishizaka2009quantum,ishizaka2015some} and
parametric quantum circuits (PQCs)~\cite{lloyd1996universal}, introducing a
suitable convex reformulation of the latter. In the various cases, we
benchmark the processors for the simulation of basic unitary gates (qubit
rotations) and various basic channels, including the amplitude damping channel
which is known to be the most difficult to
simulate~\cite{pirandola2017fundamental,commREVIEW}. For the deeper designs,
we find that the optimal program states do not correspond to the Choi matrices
of the target channels, which is rather counter-intuitive and unexpected.

The paper is structured as follows. In Sec.~\ref{Sec1} we discuss the general
notion of programmable channel simulation, the various cost functions and a
suitable Choi-reduction of the problem. In Sec.~\ref{Sec2} we then show that
the optimization of a generic programmable quantum processor is convex in the
space of the program states. In Sec.~\ref{Sec3} we consider the optimization
of the diamond distance via SDP and the minimization of the other cost
functions via gradient descent. In particular, in Sec.~\ref{Sec4} we provide
the details of the gradient-based ML algorithms to be used, together with a
discussion of smoothing techniques. In Sec.~\ref{Sec5}, we discuss the optimal
learning of arbitrary unitaries. We then move to discuss the various specific
designs based on teleportation (Sec.~\ref{Sec6}), PBT (Sec.~\ref{Sec7}) and
PQC (Sec.~\ref{Sec8}). Sec.~\ref{Sec9} is for conclusions.

\section{Programmable simulation\label{Sec1}}

\subsection{General problem}

\begin{figure}[t]
\vspace{-0.5cm}
\par
\begin{center}
\includegraphics[width=0.35\textwidth]{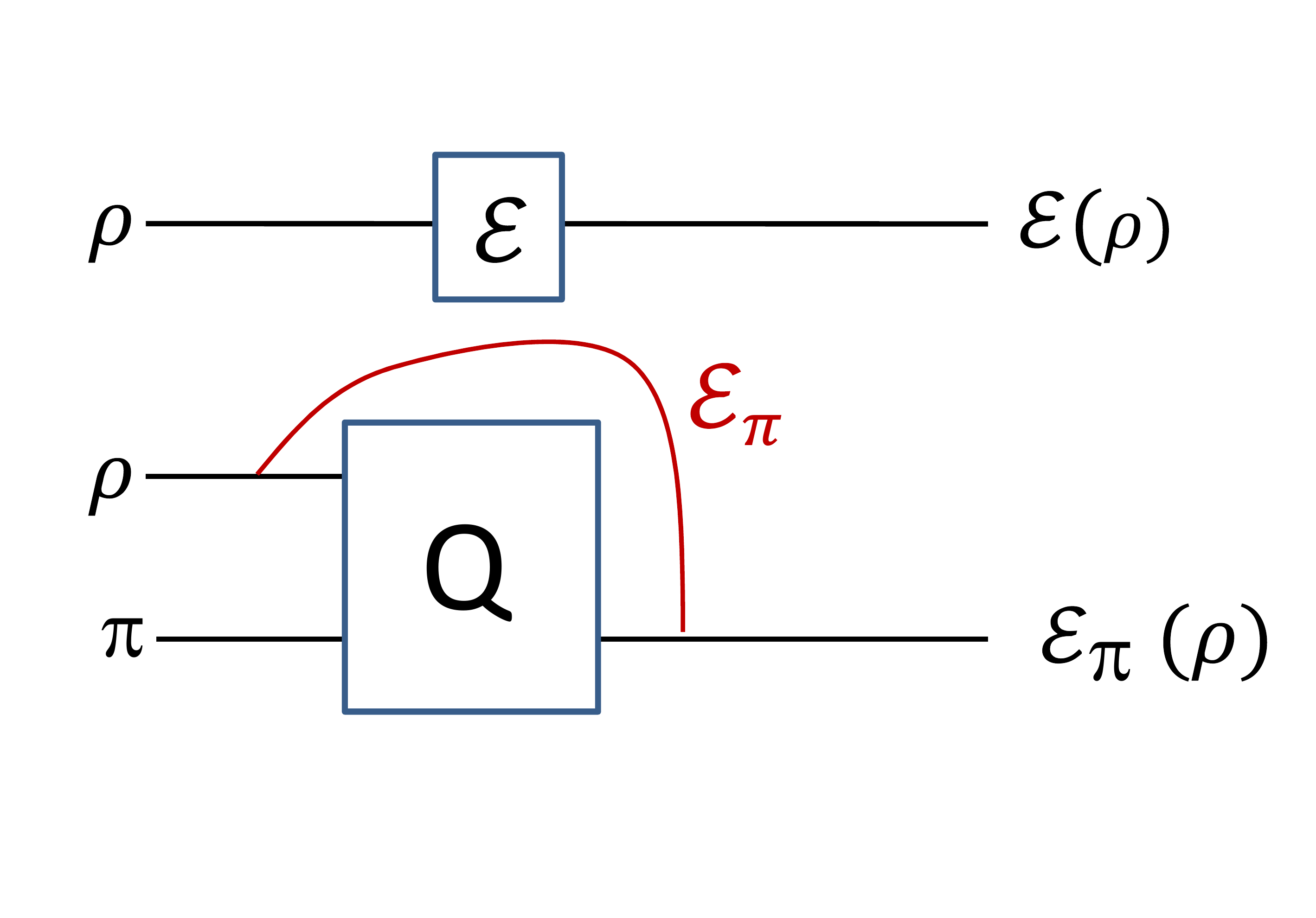}
\end{center}
\par
\vspace{-0.9cm}\caption{Arbitrary quantum channel $\mathcal{E}$ and its
simulation $\mathcal{E}_{\pi}$ via a quantum processor $Q$ applied to a
program state $\pi$.}%
\label{QMLprogram0}%
\end{figure}

Consider an arbitrary but known quantum channel $\mathcal{E}$ from dimension
$d$ to dimension $d^{\prime}$~\cite{watrous2018theory,nielsen2000quantum}. We
want to simulate $\mathcal{E}$ using a programmable quantum
processor~\cite{nielsen1997programmable} that we simply call \textquotedblleft
quantum processor\textquotedblright\ (see Fig.~\ref{QMLprogram0}). This is
represented by a completely positive trace-preserving (CPTP) universal map $Q$
which is assumed to be fixed and applied to the arbitrary input $\rho$ of the
channel together with a program state $\pi$ (which may be varied). In this
way, the quantum processor generates an approximate channel $\mathcal{E}_{\pi
}$ as
\begin{equation}
\mathcal{E}_{\pi}(\rho)=\mathrm{Tr}_{2}\left[  Q(\rho\otimes\pi)\right]  .
\label{channelDEF}%
\end{equation}

Our goal is to find the program state $\pi$ for which the simulation
$\mathcal{E}_{\pi}$ is the closest to $\mathcal{E}$, i.e., so that we minimize
the following cost function
\begin{equation}
C_{\diamond}(\pi):=\left\Vert \mathcal{E}-\mathcal{E}_{\pi}\right\Vert
_{\diamond}\leq2,
\end{equation}
in terms of the diamond
distance~\cite{kitaev2002classical,watrous2004advanced}. In other words,
\begin{equation}
\text{{Find} }\tilde{\pi}\text{ {such that} }C_{\diamond}(\tilde{\pi}%
)=\min_{\pi}C_{\diamond}(\pi). \label{sol1}%
\end{equation}
From theory~\cite{nielsen1997programmable,knill2001scheme} we know that we
cannot achieve $C_{\diamond}=0$ for arbitrary $\mathcal{E}$ unless $\pi$ and
$Q$ have infinite dimensions. As a result, for any finite-dimensional
realistic design of the quantum processor, finding the optimal program state
$\tilde{\pi}$ is an open problem.

Recall that the diamond distance is defined by the following maximization%
\begin{equation}
\left\Vert \mathcal{E}-\mathcal{E}_{\pi}\right\Vert _{\diamond}=\max_{\varphi
}\left\Vert \mathcal{I}\otimes\mathcal{E}(\varphi)-\mathcal{I}\otimes
\mathcal{E}_{\pi}(\varphi)\right\Vert _{1}, \label{diamondDEF}%
\end{equation}
where $\left\Vert O\right\Vert _{1}:=\mathrm{Tr}\sqrt{O^{\dagger}O}$ is the
trace norm~\cite{watrous2018theory}. Because the trace norm is convex over
mixed states, one may reduce the maximization in Eq. (\ref{diamondDEF}) to
bipartite pure states $\varphi=\left\vert \varphi\right\rangle \left\langle
\varphi\right\vert $. In general, we therefore need to consider a min-max
optimization, i.e., find $\tilde{\pi}$ and (pure) $\tilde{\varphi}$ such that%
\begin{align}
&  \left\Vert \mathcal{I}\otimes\mathcal{E}(\tilde{\varphi})-\mathcal{I}%
\otimes\mathcal{E}_{\tilde{\pi}}(\tilde{\varphi})\right\Vert _{1}\nonumber\\
&  =\min_{\pi}\max_{\varphi}\left\Vert \mathcal{I}\otimes\mathcal{E}%
(\varphi)-\mathcal{I}\otimes\mathcal{E}_{\pi}(\varphi)\right\Vert _{1}~.
\label{minmax}%
\end{align}

Also recall that the diamond distance can be computed using SDP~\cite{Watrous}%
. In particular, due to strong duality, it may be computed via a minimization
rather than a maximization, so that the min-max optimization problem in
Eq.~(\ref{minmax}) can be transformed into a more convenient minimization
problem (more details in Sec.~\ref{s:sdpmin}). An alternative solution is to reduce the
general problem into a weaker one which is expressed in terms of the Choi
matrix of the channel (see following section). In this way, we also avoid the
maximization in $\varphi$ but with the downside of using a larger cost
function, the trace distance.

\subsection{Processor map and Choi reduction}

\begin{figure}[t]
\vspace{-0.1cm}
\par
\begin{center}
\includegraphics[width=0.40\textwidth] {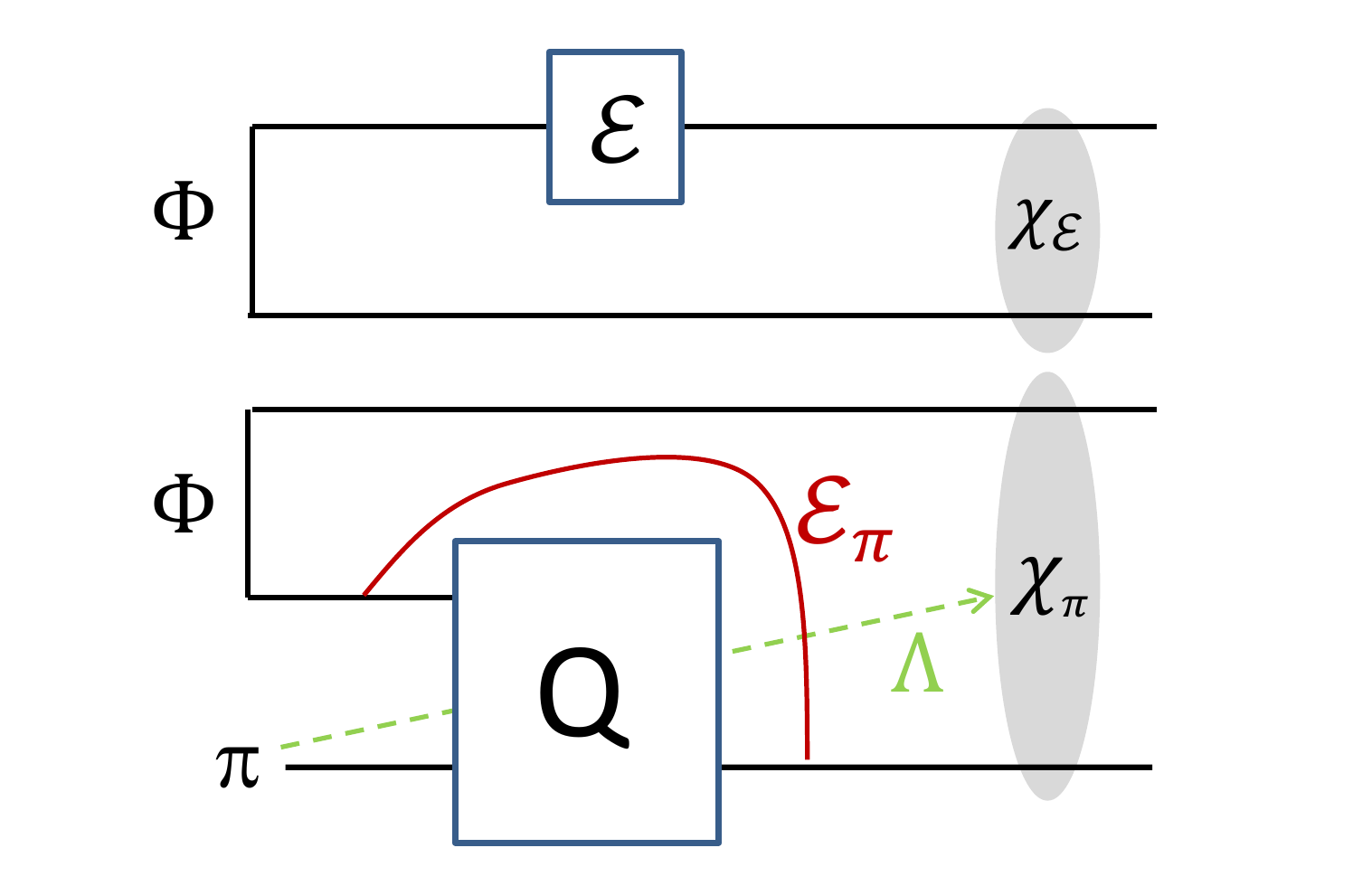}
\end{center}
\par
\vspace{-0.5cm}\caption{Map of the processor and learning of Choi matrices.
Consider an arbitrary (but known) quantum channel $\mathcal{E}$ and its
associated Choi matrix $\chi_{\mathcal{E}}$, generated by propagating part of
a maximally-entangled state $\Phi$. Then, consider a quantum processor $Q$
with program state $\pi$\ which generates the simulated channel $\mathcal{E}%
_{\pi}$\ and, therefore, the corresponding Choi matrix $\chi_{\pi}%
:=\chi_{\mathcal{E}_{\pi}}$ upon propagating part of $\Phi$ as input state.
The map of the processor is the CPTP map $\Lambda$ from the program state
$\pi$ to the output Choi matrix $\chi_{\pi}$. In a simplified version of our
problem, we may optimize the program $\pi$ in such a way to minimize the trace
distance $C_{1}(\pi):=\left\Vert \chi_{\mathcal{E}}-\chi_{\pi}\right\Vert
_{1}$.}%
\label{QMLprogram2}%
\end{figure}

It is known that a quantum channel $\mathcal{E}$ is one-to-one with its Choi
matrix $\chi_{\mathcal{E}}:=\mathcal{I}\otimes\mathcal{E}(\Phi)$, where
$\Phi:=\left\vert \Phi\right\rangle \!\left\langle \Phi\right\vert $ is
$d$-dimensional maximally-entangled state, i.e.,
\begin{equation}
\left\vert \Phi\right\rangle :=d^{-1/2}\sum_{i}\left\vert i,i\right\rangle .
\end{equation}
Using the channel definition of Eq.~(\ref{channelDEF}), we may write
\begin{align}
\chi_{\mathcal{E}_{\pi}}  &  =\mathcal{I}\otimes\mathcal{E}_{\pi}%
(\Phi)\nonumber\\
&  =d^{-1}%
%TCIMACRO{\tsum \nolimits_{ij}}%
%BeginExpansion
{\textstyle\sum\nolimits_{ij}}
%EndExpansion
\left\vert i\right\rangle \!\left\langle j\right\vert \otimes\mathrm{Tr}%
_{2}\left[  Q(\left\vert i\right\rangle \!\left\langle j\right\vert \otimes
\pi)\right]  .
\end{align}
From this expression, it is clear that the Choi matrix $\chi_{\mathcal{E}%
_{\pi}}$ is linear in the program state $\pi$. More precisely, the Choi matrix
$\chi_{\mathcal{E}_{\pi}}$ at the output of the processor $Q$ can be directly
written as a CPTP linear map $\Lambda$ acting on the space of the program
states $\pi$, i.e.,%
\begin{equation}
\chi_{\pi}:=\chi_{\mathcal{E}_{\pi}}=\Lambda(\pi). \label{lambdadef}%
\end{equation}
This map is also depicted in Fig.~\ref{QMLprogram2}.

We may connect the minimization of the diamond distance $C_{\diamond}(\pi)$ to
the minimization of the trace distance%
\begin{equation}
C_{1}(\pi):=\left\Vert \chi_{\mathcal{E}}-\chi_{\pi}\right\Vert _{1}%
,\label{traceD}%
\end{equation}
between the Choi matrices $\chi_{\mathcal{E}}$ and $\chi_{\pi}$. In fact, we
may write the sandwich relation~\cite{watrous2018theory}%
\begin{equation}
C_{1}(\pi)\leq C_{\diamond}(\pi)\leq d~C_{1}(\pi).\label{mainEQ}%
\end{equation}
While the lower bound is immediate from the definition of
Eq.~(\ref{diamondDEF}), the upper bound can be proven using the following
equivalent form of the diamond distance
\begin{equation}
\Vert\mathcal{E}-\mathcal{E}_{\pi}\Vert_{\diamond}=\sup_{\rho_{0},\rho_{1}%
}d\Vert(\sqrt{\rho_{0}}\otimes\openone)(\chi_{\mathcal{E}}-\chi_{\pi}%
)(\sqrt{\rho_{1}}\otimes\openone)\Vert_{1},\label{diamondSDPstate}%
\end{equation}
where the optimization is done over the density matrices $\rho_{0}$ and
$\rho_{1}$~\cite[Theorem 3.1]{Watrous}. In fact, consider the Frobenius norm
$\Vert A\Vert_{2}:=\sqrt{\Tr[A^\dagger A]}$ and the spectral norm
\begin{equation}
\left\Vert A\right\Vert _{\infty}:=\max\{\left\Vert Au\right\Vert
:u\in\mathbb{C}^{d},\left\Vert u\right\Vert \leq1\},
\end{equation}
which satisfy the following properties~\cite{watrous2018theory}
\begin{align}
\Vert ABC\Vert_{1} &  \leq\Vert A\Vert_{\infty}\Vert B\Vert_{1}\Vert
C\Vert_{\infty}~,\label{prop1}\\
\Vert A\otimes\openone\Vert_{\infty} &  =\Vert A\Vert_{\infty}\leq\Vert
A\Vert_{2}.\label{prop2}%
\end{align}
Then, from Eqs.~(\ref{diamondSDPstate}), (\ref{prop1}) and~(\ref{prop2}), one
gets
\begin{align}
\Vert\mathcal{E}-\mathcal{E}_{\pi}\Vert_{\diamond} &  \leq\sup_{\rho_{0}%
,\rho_{1}}d\sqrt{\mathrm{Tr}\rho_{0}\mathrm{Tr}\rho_{1}}\Vert\chi
_{\mathcal{E}}-\chi_{\pi}\Vert_{1}\nonumber\\
&  =d\Vert\chi_{\mathcal{E}}-\chi_{\pi}\Vert_{1}.
\end{align}

Thanks to Eq.~(\ref{mainEQ}), we may avoid the maximization step in the
definition of the diamond distance and simplify the original problem to
approximating the Choi matrix $\chi_{\mathcal{E}}$ of the channel by varying
the program state $\pi$. This is a process of learning Choi matrices as
depicted in Fig.~\ref{QMLprogram2}. Because the simpler cost function
$C_{1}(\pi)$ is an upper bound, its minimization generally provides a
sub-optimal solution for the program state. 

\subsection{Other cost functions}

Besides $C_{\diamond}$\ and $C_{1}$ we can introduce other cost functions.
First of all, using the Fuchs-van de Graaf
inequality~\cite{fuchs1999cryptographic}, we may write
\begin{equation}
C_{1}(\pi)\leq2\sqrt{C_{F}(\pi)},~~C_{F}(\pi)=1-F(\pi)^{2}, \label{Cf}%
\end{equation}
where $F(\pi)$ is Bures' fidelity between the two Choi matrices $\chi
_{\mathcal{E}}$ and $\chi_{\pi}$, i.e.,%
\begin{equation}
F(\pi):=\left\Vert \sqrt{\chi_{\mathcal{E}}}\sqrt{\chi_{\pi}}\right\Vert
_{1}=\mathrm{Tr}\sqrt{\sqrt{\chi_{\mathcal{E}}}\chi_{\pi}\sqrt{\chi
_{\mathcal{E}}}}. \label{fidCC}%
\end{equation}
Another possible upper bound can be written using the quantum Pinsker's
inequality~\cite{pinsker1964information,carlen2012bounds}. In fact, we may
write $C_{1}(\pi)\leq(2\ln\sqrt{2})\sqrt{C_{R}(\pi)}$, where%
\begin{equation}
C_{R}(\pi):=\min\left\{  S(\chi_{\mathcal{E}}||\chi_{\pi}),S(\chi_{\pi}%
||\chi_{\mathcal{E}})\right\}  , \label{REcost}%
\end{equation}
and $S(\rho||\sigma):=\mathrm{Tr}[\rho(\log_{2}\rho-\log_{2}\sigma)]$ is the
quantum relative entropy between $\rho$ and $\sigma$.

Finally we may consider other cost functions in terms of any Shatten p-norm
$C_{p}(\pi):=\Vert\chi_{\mathcal{E}}-\chi_{\pi}\Vert_{p}$, even though this
option provides lower bounds instead of upper bounds for the trace distance.
Recall that, given an operator $O$ and a real number $p\geq1$, we may define
its Schatten p-norm as~\cite{watrous2018theory}%
\begin{equation}
\left\Vert O\right\Vert _{p}=(\mathrm{Tr}|O|^{p})^{1/p},
\end{equation}
where $|O|=\sqrt{O^{\dagger}O}$. For any $1\leq p\leq q\leq\infty$, one has
the monotony $\left\Vert O\right\Vert _{p}\geq\left\Vert O\right\Vert _{q}$,
so that $\left\Vert O\right\Vert _{\infty}\leq\ldots\leq\left\Vert
O\right\Vert _{1}$. An important property is duality. For each pair of
operators $A$ and $B$, and each pair of parameters $p,q\in\lbrack1,\infty]$
such that $p^{-1}+q^{-1}=1$, we may write~\cite{watrous2018theory}
\begin{equation}
\left\Vert A\right\Vert _{p}=\sup_{\left\Vert B\right\Vert _{q}\leq
1}\left\vert \left\langle B,A\right\rangle \right\vert
\equiv \sup_{\left\Vert B\right\Vert _{q}\leq
1}\left\langle B,A\right\rangle
,
\label{dualityNORM}%
\end{equation}
where $\left\langle B,A\right\rangle =\mathrm{Tr}(B^{\dagger}A)$ is the
Hilbert-Schmidt product, and the second inequality follows since we can arbitrarily
change the sign of $B$.

\section{Convexity\label{Sec2}}

In this section, we show that the minimization of the main cost functions
$C_{\diamond}$, $C_{1}$ and $C_{F}$ is a convex optimization problem in the
space of the program states $\pi$. This means that we can find the optimal
program state $\tilde{\pi}$ by minimizing $C_{\diamond}$ or, alternatively,
sub-optimal program states can be found by minimizing either $C_{1}$ or
$C_{F}$. For the sake of generality we prove the result for all the cost
functions discussed in the previous section.

\begin{theorem}
\label{t:convexC1CF} The minimization of the generic cost function
$C=C_{\diamond}$, $C_{1}$, $C_{F}$, $C_{R}$ or $C_{p}$ for any $p>1$ is a
convex optimization problem in the space of program states. In particular, the
global minimum $\tilde{\pi}$ can always be found as a local minimum of
$C_{\diamond}$. Alternatively, this optimal program state can be approximated
by minimizing $C_{1}$ or $C_{F}$.
\end{theorem}

\noindent\textit{Proof.}~~Let us start to show the result for the diamond
distance $C_{\diamond}$. In this case, we can write the following
\begin{align}
&  C_{\diamond}[p\pi+(1-p)\pi^{\prime}]\nonumber\\
&  :=\left\Vert \mathcal{E}-\mathcal{E}_{p\pi+(1-p)\pi^{\prime}}\right\Vert
_{\diamond}\nonumber\\
&  \overset{(1)}{=}\left\Vert (p{+}1{-}p)\mathcal{E}-p\mathcal{E}_{\pi}%
-(1{-}p)\mathcal{E}_{\pi^{\prime}}\right\Vert _{\diamond}\nonumber\\
&  \overset{(2)}{\leq}\left\Vert p\mathcal{E}-p\mathcal{E}_{\pi}\right\Vert
_{\diamond}+\left\Vert (1{-}p)\mathcal{E}-(1{-}p)\mathcal{E}_{\pi^{\prime}%
}\right\Vert _{\diamond}\nonumber\\
&  \overset{(3)}{\leq}p\left\Vert \mathcal{E}-\mathcal{E}_{\pi}\right\Vert
_{\diamond}+(1-p)\left\Vert \mathcal{E}-\mathcal{E}_{\pi^{\prime}}\right\Vert
_{\diamond}\nonumber\\
&  =pC_{\diamond}(\pi)+(1-p)C_{\diamond}(\pi^{\prime}),
\end{align}
where we use $(1)$~the linearity of $\mathcal{E}$, $(2)$~ the triangle
inequality and $(3)$ the property $\Vert xA\Vert_{1}=|x|\Vert A\Vert_{1}$,
valid for any operator $A$ and coefficient $x$.

For any Schatten p-norm $C_{p}$ with $p\geq1$, we may exploit the dual
representation in Eq.~(\ref{dualityNORM}) with $A=\chi_{\mathcal{E}}%
-\Lambda(\pi)$, so that
\begin{equation}
C_{p}(\pi)=\sup_{\left\Vert B\right\Vert _{q}\leq1}\left\vert \mathrm{Tr}%
\{B^{\dagger}[\chi_{\mathcal{E}}-\Lambda(\pi)]\}\right\vert .
\end{equation}
For any convex combination $\bar{\pi}:=p_{0}\pi_{0}+p_{1}\pi_{1}$, with $p_0+p_1=1$, we have
$\Lambda(\bar{\pi})=p_{0}\Lambda(\pi_{0})+p_{1}\Lambda(\pi_{1})$ by linearity,
and we may write%
\begin{align}
&  C_{p}(\bar{\pi})\nonumber\\
&  =\sup_{\left\Vert B\right\Vert _{q}\leq1}\left\vert \mathrm{Tr}%
\{B^{\dagger}[p_{0}\chi_{\mathcal{E}}+p_{1}\chi_{\mathcal{E}}-p_{0}\Lambda
(\pi_{0})-p_{1}\Lambda(\pi_{1})]\}\right\vert \nonumber\\
&  \overset{(1)}{=}\sup_{\left\Vert B\right\Vert _{q}\leq1}\left\vert
\mathrm{Tr}\{p_{0}B^{\dagger}[\chi_{\mathcal{E}}-\Lambda(\pi_{0}%
)]+p_{1}B^{\dagger}[\chi_{\mathcal{E}}-\Lambda(\pi_{1})]\}\right\vert
\nonumber\\
&  \overset{(2)}{=}\sup_{\left\Vert B\right\Vert _{q}\leq1}\left\vert
p_{0}\mathrm{Tr}\{B^{\dagger}[\chi_{\mathcal{E}}-\Lambda(\pi_{0})]\}\right.
+\nonumber\\
&  ~~~~~~~~~~~~~~ \left.  +p_{1}\mathrm{Tr}\{B^{\dagger}[\chi_{\mathcal{E}}-\Lambda(\pi
_{1})]\}\right\vert \nonumber\\
&  \overset{(3)}{\leq}\sup_{\left\Vert B\right\Vert _{q}\leq1}\left\vert
p_{0}\mathrm{Tr}\{B^{\dagger}[\chi_{\mathcal{E}}-\Lambda(\pi_{0}%
)]\}\right\vert + \nonumber\\
& ~~~~~~~~~~~~~~ \ +\left\vert p_{1}\mathrm{Tr}\{B^{\dagger}[\chi_{\mathcal{E}}-\Lambda
(\pi_{1})]\}\right\vert \nonumber\\
&  \overset{(4)}{\leq}p_{0}\sup_{\left\Vert B\right\Vert _{q}\leq1}\left\vert
\mathrm{Tr}\{B^{\dagger}[\chi_{\mathcal{E}}-\Lambda(\pi_{0})]\}\right\vert
+\nonumber\\
&  ~~~~ +p_{1}\sup_{\left\Vert C\right\Vert _{q}\leq1}\mathrm{Tr}\{C^{\dagger}%
[\chi_{\mathcal{E}}-\Lambda(\pi_{1})]\}\nonumber\\
&  =p_{0}C(\pi_{0})+p_{1}C(\pi_{1}),
\end{align}
where we use: (1)~linearity of the operator $B$, (2)~linearity of the trace,
(3)~triangle inequality, (4)~and the inequality $\sup_{B}f(B)+g(B)\leq\sup
_{B}f(B)+\sup_{C}g(C)$ for the optimization of two functions.

To show the convexity of $C_{F}$, defined in Eq.~\eqref{Cf}, we note that the
fidelity function $F(\rho,\sigma)$ satisfies the following concavity relation
\cite{uhlmann1976transition}
\begin{equation}
F\left(  \sum_{k}p_{k}\rho_{k},\sigma\right)  ^{2}\geq\sum_{k}p_{k}F(\rho
_{k},\sigma)^{2}~.
\end{equation}
Due to the linearity of $\chi_{\pi}=\Lambda(\pi)$, the fidelity in
Eq.~\eqref{fidCC} satisfies $F_{\bar{\pi}}^{2}\geq\sum_{k}p_{k}F_{\pi_{k}}%
^{2}$ for $\bar{\pi}:=\sum_{k}p_{k}\pi_{k}$. Accordingly, we get the following
convexity result
\begin{equation}
C_{F}\left(  \sum_{k}p_{k}\pi_{k}\right)  \leq\sum_{k}p_{k}C_{F}(\pi_{k})~.
\end{equation}
For the cost function $C_{R}$, the result comes from the linearity of
$\Lambda(\pi)$ and the joint convexity of the relative entropy. In fact, for
$\bar{\pi}:=p_{0}\pi_{0}+p_{1}\pi_{1}$, we may write
\begin{align}
S[\Lambda(\bar{\pi})||\chi_{\mathcal{E}}] &  =S[p_{0}\Lambda(\pi_{0}%
)+p_{1}\Lambda(\pi_{1})||\chi_{\mathcal{E}}]\nonumber\\
&  =S[p_{0}\Lambda(\pi_{0})+p_{1}\Lambda(\pi_{1})||p_{0}\chi_{\mathcal{E}%
}+p_{1}\chi_{\mathcal{E}}]\nonumber\\
&  \leq p_{0}S[\Lambda(\pi_{0}),\chi_{\mathcal{E}}]+p_{1}S[\Lambda(\pi
_{1}),\chi_{\mathcal{E}}],
\end{align}
with symmetric proof for $S[\chi_{\mathcal{E}}||\Lambda(\bar{\pi})]$. This
implies the convexity of $C_{R}(\pi)$ in Eq.~(\ref{REcost}).~$\blacksquare$

\subsection{Convex classical parametrizations}

The result of the theorem can certainly be extended to any convex
parametrization of program states. For instance, assume that $\pi
=\pi(\boldsymbol{\lambda})$, where $\boldsymbol{\lambda}=\{\lambda_{i}\}$ is a
probability distribution. This means that, for $0\leq p\leq1$ and any two
parametrizations, $\boldsymbol{\lambda}$ and $\boldsymbol{\lambda}^{\prime}$,
we may write%
\begin{equation}
\pi\lbrack p\boldsymbol{\lambda}+(1-p)\boldsymbol{\lambda}^{\prime}%
]=p\pi(\boldsymbol{\lambda})+(1-p)\pi(\boldsymbol{\lambda}^{\prime}).
\label{eqGIA}%
\end{equation}
Then the problem remains convex in $\boldsymbol{\lambda}$ and we may therefore
find the global minimum in these parameters. It is clear that this global
minimum $\boldsymbol{\tilde{\lambda}}$\ identifies a program state
$\pi(\boldsymbol{\tilde{\lambda}})$ which is not generally the optimal state
$\tilde{\pi}$ in the entire program space $\mathcal{S}$, even though the
solution may be a convenient solution for experimental applications.

Note that a possible classical parametrization consists of using classical
program states, of the form
\begin{equation}
\pi(\boldsymbol{\lambda})=\sum_{i}\lambda_{i}\left\vert \varphi_{i}%
\right\rangle \left\langle \varphi_{i}\right\vert ,
\end{equation}
where $\{\left\vert \varphi_{i}\right\rangle \}$ is an orthonormal basis in
the program space. Convex combinations of probability distributions therefore
define a convex set of classical program states
\begin{equation}
\mathcal{S}_{\text{class}}=\{\pi:\pi=\sum_{i}\lambda_{i}\left\vert \varphi
_{i}\right\rangle \left\langle \varphi_{i}\right\vert ,~\left\langle
\varphi_{i}\right\vert \left.  \varphi_{j}\right\rangle =\delta_{ij}\}.
\end{equation}
Optimizing over this specific subspace corresponds to optimizing the
programmable quantum processor over classical programs. It is clear that
global minima in $\mathcal{S}_{\text{class}}$ and $\mathcal{S}$ are expected
to be very different. For instance, $\mathcal{S}_{\text{class}}$ cannot
certainly include Choi matrices which are usually very good quantum programs.

\section{Convex optimization\label{Sec3}}

\subsection{SDP minimization}\label{s:sdpmin}

Once we have Theorem~\ref{t:convexC1CF} in our hands, we can successfully
minimize the various cost functions in the search of the optimal program
state. In other words, for a generic cost function $C$ we want to solve
$\min_{\pi\in\mathcal{S}}C(\pi)$. The solution is exact if we directly use the
diamond-distance cost $C_{\diamond}(\pi)=\Vert\mathcal{E}-\mathcal{E}_{\pi
}\Vert_{\diamond}$ and we minimize it via SDP.

Let us introduce the linear map $\Omega_{\pi}:=\mathcal{E}-\mathcal{E}_{\pi}$
with corresponding Choi matrix
\begin{equation}
\chi_{\Omega_{\pi}}=\chi_{\mathcal{E}}-\chi_{\pi}=\chi_{\mathcal{E}}%
-\Lambda(\pi).
\end{equation}
Thanks to the property of strong duality of the diamond norm, for any program
$\pi$ we can compute the cost function $C_{\diamond}(\pi)=\Vert\Omega_{\pi
}\Vert_{\diamond}$ via the following SDP~\cite{watrous2009semidefinite}
\begin{gather}
\mathrm{Minimize~}\frac{1}{2}\left(  \left\Vert \mathrm{Tr}_{2}M_{0}%
\right\Vert _{\infty}+\left\Vert \mathrm{Tr}_{2}M_{1}\right\Vert _{\infty
}\right)  ,\nonumber\\
\text{\textrm{Subject to}}\mathrm{~}%
\begin{pmatrix}
M_{0} & -d~\chi_{\Omega_{\pi}}\\
-d~\chi_{\Omega_{\pi}}^{\dagger} & M_{1}%
\end{pmatrix}
\geq0,
\end{gather}
where $M_{0}\geq0$ and $M_{1}\geq0$ in $\mathbb{C}^{d\times d^{\prime}}$, and
the spectral norm $\Vert O\Vert_{\infty}$ equals the maximum singular value of
$O$.

Moreover, because $\chi_{\Omega_{\pi}}$ is Hermitian, the above SDP can be
simplified into
\begin{gather}
\mathrm{Minimize~}2\left\Vert \mathrm{Tr}_{2}Z\right\Vert _{\infty
},\nonumber\\
\text{\textrm{Subject to}~}Z\geq0~\text{\textrm{and} }Z\geq d~\chi
_{\Omega_{\pi}}. \label{minZ}%
\end{gather}
Not only this procedure computes $C_{\diamond}(\pi)$ but also provides the
upper bound $C_{\diamond}(\pi)\leq d\left\Vert \mathrm{Tr}_{2}\left\vert
\chi_{\mathcal{E}}-\chi_{\pi}\right\vert \right\Vert _{\infty}$~\cite{Karol}.
In fact, it is sufficient to choose $Z=d~\chi_{\Omega_{\pi}}^{+}$, where
$\chi^{+}=(\chi+|\chi|)/2$ is the positive part of $\chi$. Using
$\mathrm{Tr}_{2}\chi_{\Omega_{\pi}}=0$, we may write $\mathrm{Tr}_{2}Z\leq
d\mathrm{Tr}_{2}\chi_{\Omega_{\pi}}^{+}=\frac{d}{2}\mathrm{Tr}_{2}%
|\chi_{\Omega_{\pi}}|$.

The SDP\ form in Eq.~(\ref{minZ}) is particularly convenient for finding the
optimal program. In fact, suppose now that $\pi$ is not fixed but we want to
optimize on this state too, so as to compute the optimal program state
$\tilde{\pi}$ such that $C_{\diamond}(\tilde{\pi})=\min_{\pi\in\mathcal{S}%
}C_{\diamond}(\pi)$. The problem is therefore mapped into the following unique
minimization%
\begin{gather}
\mathrm{Minimize~}2\left\Vert \mathrm{Tr}_{2}Z\right\Vert _{\infty
},\nonumber\\
\text{\textrm{Subject to}~}Z\geq0,~\pi\geq0,~\mathrm{Tr}(\pi)=1,~{Z}\geq
d~\chi_{\Omega_{\pi}}.
\end{gather}
This algorithm can be used to optimize the performance of any programmable
quantum processor.

\subsection{Gradient descent}

An alternative approach (useful for deeper processors) consists in the
optimization of the larger but easier-to-compute cost function $C=C_{1}$
(trace distance) or $C_{F}$ (infidelity). According to
Theorem~\ref{t:convexC1CF}, the cost function $C:\mathcal{S}\rightarrow
\mathbb{R}$ is convex over the program space $\mathcal{S}$\ and, therefore, we
can solve the optimization $\min_{\pi\in\mathcal{S}}C(\pi)$ by using
gradient-based ML algorithms. This means that we need to compute the
derivatives of $C$ and use gradient descent in order to converge to a local
(global) minimum.

The sub-differential of $C$ at the generic point $\pi\in\mathcal{S}$ is
defined as
\begin{equation}
\partial C(\pi)=\{Z:C(\sigma)-C(\pi)\geq\text{\textrm{Tr}}[Z(\sigma
-\pi)],~\forall\sigma\in\mathcal{S}\} \label{subgradient}%
\end{equation}
where $Z$ is Hermitian~\cite{nesterov2013introductory,coutts2018certifying}.
In the points where $C$ is not only convex but also differentiable, then
\begin{equation}
\partial C(\pi)=\{\nabla C(\pi)\}, \label{fgrad}%
\end{equation}
namely the subgradient contains a single element, the gradient $\nabla C$,
that can be obtained as the Fr\'{e}chet derivative of $C$ (for more details
see Appendix~\ref{a:calculus}). In the points where $C$ is not differentiable,
then the gradient still provides an element of the subgradient to be used in
the gradient-based minimization process.

In order to compute the gradient $\nabla C$, it is convenient to consider the
Kraus decomposition of the processor map $\Lambda$. Let us write%
\begin{equation}
\Lambda(\pi)=\sum_{k}A_{k}\pi A_{k}^{\dagger},
\end{equation}
with Kraus operators $A_{k}$. We then define the dual map $\Lambda^{\ast}$ of
the processor as the one (generally non-trace-preserving) which is given by
the following decomposition%
\begin{equation}
\Lambda^{\ast}(\rho)=\sum_{k}A_{k}^{\dagger}\rho A_{k}.
\end{equation}
With these definitions in hands, we prove the following.

\begin{theorem}
\label{t:gradients}Suppose we use a quantum processor $Q$ with map
$\Lambda(\pi)=\chi_{\pi}$\ in order to approximate the Choi matrix
$\chi_{\mathcal{E}}$ of an arbitrary channel $\mathcal{E}$. Then, the
gradients of the trace distance $C_{1}(\pi)$ and the infidelity $C_{F}(\pi
)$\ are given by the following analytical formulas%
\begin{align}
\nabla C_{1}(\pi) &  =\sum_{k}\mathrm{sign}(\lambda_{k})\Lambda^{\ast}%
(P_{k}),\label{traceDproof}\\
\nabla C_{F}(\pi) &  =-2\sqrt{1-C_{F}(\pi)}\nabla F(\pi),\label{Cfgrad}\\
\nabla F(\pi) &  =\frac{1}{2}\Lambda^{\ast}\left[  \sqrt{\chi_{\mathcal{E}}%
}\left(  \sqrt{\chi_{\mathcal{E}}}\,\Lambda(\pi)\,\sqrt{\chi_{\mathcal{E}}%
}\right)  ^{-\frac{1}{2}}\sqrt{\chi_{\mathcal{E}}}\right]
,\label{fidelitygrad}%
\end{align}
where $\lambda_{k}$ ($P_{k}$) are the eigenvalues (eigenprojectors) of the
Hermitian operator $\chi_{\pi}-\chi_{\mathcal{E}}$. When $C_{1}(\pi)$ or
$C_{F}(\pi)$ are not differentiable at $\pi$, then the above expressions
provide an element of the subgradient $\partial C(\pi)$.
\end{theorem}

\noindent\textbf{Proof}.~~We prove the above theorem assuming that the
functions are differentiable for program $\pi$. For non-differentiable points,
the only difference is that the above analytical expressions are not unique
and provide only one of the possibly infinite elements of the subgradient.
Further details of this mathematical proof are given in
Appendix~\ref{a:calculus}. Following matrix differentiation (see
Appendix~\ref{matrixDIFF}), for any function $f(A)=\Tr[g(A)]$ of a matrix $A$,
we may write
\begin{equation}
d\text{\textrm{Tr}}[g(A)]=\text{\textrm{Tr}}[g^{\prime}(A)dA],\label{nablag}%
\end{equation}
and the gradient is $\nabla f(A)=g^{\prime}(A)$. Both the trace-distance and
fidelity cost functions can be written in this form. To find the explicit
gradient of the fidelity function we first note that, by linearity, we may
write%
\begin{equation}
\Lambda({\pi+\delta\pi})=\Lambda({\pi})+\Lambda(\delta\pi)~,\label{chilinear}%
\end{equation}
and therefore the following expansion
\begin{gather}
\sqrt{\chi_{\mathcal{E}}}\Lambda({\pi+\delta\pi})\sqrt{\chi_{\mathcal{E}}%
}=\nonumber\\
\sqrt{\chi_{\mathcal{E}}}\Lambda({\pi})\sqrt{\chi_{\mathcal{E}}}+\sqrt
{\chi_{\mathcal{E}}}\Lambda({\delta\pi})\sqrt{\chi_{\mathcal{E}}}~.
\end{gather}
From this equation and differential calculations of the fidelity (see
Appendix~\ref{fidelityDIFF} for details), we find
\begin{equation}
dF=\frac{1}{2}\text{\textrm{Tr}}\left[  (\sqrt{\chi_{\mathcal{E}}}\Lambda
({\pi})\sqrt{\chi_{\mathcal{E}}})^{-\frac{1}{2}}\sqrt{\chi_{\mathcal{E}}%
}\Lambda({\delta\pi})\sqrt{\chi_{\mathcal{E}}}\right]  ~,
\end{equation}
where $dF=F(\pi+\delta\pi)-F(\pi)$. Then, using the cyclic property of the
trace, we get
\begin{equation}
dF=\frac{1}{2}\text{\textrm{Tr}}\left[  \Lambda^{\ast}\left[  \sqrt
{\chi_{\mathcal{E}}}(\sqrt{\chi_{\mathcal{E}}}\Lambda(\pi)\sqrt{\chi
_{\mathcal{E}}})^{-\frac{1}{2}}\sqrt{\chi_{\mathcal{E}}}\right]  \delta
\pi\right]  .
\end{equation}
Exploiting this expression in Eq.~\eqref{nablag} we get the gradient $\nabla
F(\pi)$ as in Eq.~(\ref{fidelitygrad}). The other Eq.~\eqref{Cfgrad} simply
follows from applying the definition in Eq.~\eqref{Cf}.

For the trace distance, let us write the eigenvalue decomposition
\begin{equation}
\chi_{\pi}-\chi_{\mathcal{E}}=\sum_{k}\lambda_{k}P_{k}~.\label{tracediag}%
\end{equation}
Then using linearity of Eq.~\eqref{chilinear}, the definition of processor map
of Eq.~\eqref{lambdadef} and differential calculations of the trace distance
(see Appendix~\ref{traceDIFF} for details), we can write
\begin{align}
dC_{1}(\pi) &  =\sum_{k}\mathrm{sign}(\lambda_{k})\text{\textrm{Tr}}%
[P_{k}\Lambda(d\pi)]\nonumber\\
&  =\sum_{k}\mathrm{sign}(\lambda_{k})\text{\textrm{Tr}}[\Lambda^{\ast}%
(P_{k})d\pi]\nonumber\\
&  =\text{\textrm{Tr}}\left\{  \Lambda^{\ast}[\mathrm{sign}(\chi_{\pi}%
-\chi_{\mathcal{E}})]d\pi\right\}  ~.
\end{align}
From the definition of the gradient in Eq.~\eqref{nablag}, we finally get%
\begin{equation}
\nabla C_{1}(\pi)=\Lambda^{\ast}[\mathrm{sign}(\chi_{\pi}-\chi_{\mathcal{E}%
})],
\end{equation}
which leads to the result in Eq.~(\ref{traceDproof}).~$\blacksquare$

The above results in Eqs.~(\ref{Cfgrad}) and~(\ref{traceDproof}) can be used
together with the projected subgradient method~\cite{boyd2003subgradient} or
conjugate gradient algorithm~\cite{jaggi2011convex,jaggi2013revisiting} to
iteratively find the optimal program state in the minimization of $\min
_{\pi\in\mathcal{S}}C(\pi)$ for $C=C_{1}$ or $C_{F}$. In the following section
we present the details of the two mentioned gradient-based ML\ algorithms and
how they can be adapted for the learning of program states.

\section{Gradient-based convex optimization techniques\label{Sec4}}

Gradient-based convex optimization is at the heart of many popular ML
techniques such as, online learning in a high-dimensional feature space
\cite{duchi2008efficient}, missing value estimation problems
\cite{liu2013tensor}, text classification, image ranking, and optical
character recognition \cite{duchi2011adaptive}, to name a few. In all the
above applications, \textquotedblleft learning\textquotedblright\ corresponds
to the following minimization problem $\min_{x\in\mathcal{S}}f(x)$, where
$f(x)$ is a convex function and $\mathcal{S}$ is a convex set. Quantum
learning falls into this category, as the space of program states is convex
due to the linearity of quantum mechanics and cost functions are typically
convex in this space (see Theorem~\ref{t:convexC1CF}). Gradient-based
approaches are among the most applied methods for convex optimization of
non-linear, possibly non-smooth functions~\cite{nesterov2013introductory}.
Here we present two algorithms, the projected subgradient method and the
conjugate gradient method, and show how that can be adapted to our problem.

Projected subgradient methods have the advantage of simplicity and the ability
to optimize non-smooth functions, but can be slower, with a convergence rate
$\mathcal{O}\left(  \epsilon^{-2}\right)  $ for a desired accuracy $\epsilon$.
Conjugate gradient methods~\cite{jaggi2011convex,jaggi2013revisiting} have a
faster convergence rate $\mathcal{O}\left(  \epsilon^{-1}\right)  $, provided
that the cost function is smooth. This convergence rate can be improved even
further to $\mathcal{O}\left(  \epsilon^{-1/2}\right)  $ for strongly convex
functions~\cite{garber2015faster} or using Nesterov's accelerated gradient
method~\cite{nesterov2005smooth}. The technical difficulty in the adaptation
of these methods for learning program states comes because the latter is a
constrained optimization problem, namely at each iteration step the optimal
program must be a proper quantum state, and the cost functions coming from
quantum information theory are, generally, non-smooth.

\subsection{Projected subgradient method}

Given the space $\mathcal{S}$ of program states, let us define the projection
$\mathcal{P}_{\mathcal{S}}$ onto $\mathcal{S}$ as%
\begin{equation}
    \mathcal{P}_{\mathcal{S}}(X)=\argmin_{\pi\in S}\Vert X-\pi\Vert
_{2}~,\label{proj}%
\end{equation}
where argmin is the argument of the minimum, namely the closest state $\pi\in\mathcal S$
to the operator $X$.
Then, a first order algorithm to solve $\min_{\pi\in\mathcal{S}}C(\pi)$ is to
apply the projected subgradient
method~\cite{nesterov2013introductory,boyd2003subgradient}, which iteratively
applies the following steps
\begin{equation}%
\begin{array}
[c]{l}%
1)~\mathrm{Select~an~operator~}g_{i}{~\mathrm{from~}}\partial C(\pi_{i}),\\
2)~\text{\textrm{Update}~}\pi_{i+1}=\mathcal{P}_{\mathcal{S}}\left(  \pi
_{i}-\alpha_{i}g_{i}\right)  ,
\end{array}
\label{projsubgrad}%
\end{equation}
where $i$ is the iteration index and $\alpha_{i}$ a learning rate.

The above algorithm differs from standard gradient methods in two aspects: i)
the update rule is based on the subgradient, which is defined even for
non-smooth functions; ii) the operator $\pi_{i}-\alpha_{i}g_{i}$ is generally
not a quantum state, so the algorithm fixes this issue by projecting that
operator back to the closest quantum state, via Eq.~\eqref{proj}. The
algorithm converges to the optimal solution $\pi_{\ast}$ (approximating the
optimal program $\tilde{\pi}$) as~\cite{boyd2003subgradient}
\begin{equation}
C(\pi_{i})-C(\pi_{\ast})\leq\frac{e_{1}+G\sum_{k=1}^{i}\alpha_{k}^{2}}%
{2\sum_{k=1}^{i}\alpha_{k}}=:\epsilon,
\end{equation}
where $e_{1}=\Vert\pi_{1}-\pi_{\ast}\Vert_{2}^{2}$ is the initial error (in
Frobenius norm) and $G$ is such that $\Vert g\Vert_{2}^{2}\leq G$ for any
$g\in\partial C$. Popular choices for the learning rate that assure
convergence are $\alpha_{k}\propto1/\sqrt{k}$ and $\alpha_{k}=a/(b+k)$ for
some $a,b>0$.

In general, the projection step is the major drawback, which often limits the applicability
of the projected subgradient method to practical
problems. Indeed, projections like Eq.~\eqref{proj} require another full
optimization at each iteration that might be computationally intensive.
Nonetheless, we show in the following theorem that this issue does not occur
in learning quantum states, because the resulting optimization can be solved analytically.

\begin{theorem}
\label{t:proj} Let $X$ be a Hermitian operator in a $d$-dimensional Hilbert
space with spectral decomposition $X=UxU^{\dagger}$, where the eigenvalues
$x_{j}$ are ordered in decreasing order. Then $\mathcal{P}_{\mathcal{S}}(X)$
of Eq.~\eqref{proj} is given by
\begin{equation}
\mathcal{P}_{\mathcal{S}}(X)=U\lambda U^{\dagger},~~\lambda_{i}=\max
\{x_{i}-\theta,0\},
\end{equation}
where $\theta=\frac{1}{s}\sum_{j=1}^{s}\left(  x_{j}-1\right)  $ and
\begin{equation}
s=\max\left\{  k\in\lbrack1,...,d]:x_{k}>\frac{1}{k}\sum_{j=1}^{k}\left(
x_{j}-1\right)  \right\}  .
\end{equation}

\end{theorem}

\noindent\textbf{Proof}.~~Any quantum (program) state can be written in the
diagonal form $\pi=V\lambda V^{\dagger}$ where $V$ is a unitary matrix, and
$\lambda$ is the vector of eigenvalues in decreasing order, with $\lambda
_{j}\geq0$ and $\sum_{j}\lambda_{j}=1$. To find the optimal state, it is
required to find both the optimal unitary $V$ and the optimal eigenvalues
$\lambda$ with the above property, i.e.,
\begin{equation}
\mathcal{P}_{\mathcal{S}}(X)=\argmin_{V,\lambda}\Vert X-V\lambda V^{\dagger
}\Vert_2~.\label{proj2}%
\end{equation}
For any unitarily-invariant norm, the following inequality holds~\cite[Eq.
IV.64]{bhatia2013matrix}
\begin{equation}
\Vert X-\pi\Vert_2\geq\Vert x-\lambda\Vert_2~,
\end{equation}
with equality when $U=V$, where $X=UxU^{\dagger}$ is a spectral decomposition
of $X$ such that the $x_{j}$'s are in decreasing order. This shows that the
optimal unitary in Eq.~\eqref{proj2} is the diagonalization matrix of the
operator $X$. The eigenvalues of any density operator form a probability
simplex. The optimal eigenvalues $\lambda$ are then obtained thanks to
Algorithm 1 from Ref.~\cite{duchi2008efficient}.~$\blacksquare$

In the following section we present an alternative algorithm with faster
convergence rates, but stronger requirements on the function to be optimized.

\subsection{Conjugate gradient method}

The conjugate gradient method~\cite{jaggi2011convex,nesterov2013introductory},
sometimes called Frank-Wolfe algorithm, has been developed to provide better
convergence speed and to avoid the projection step at each iteration. Although
the latter can be explicitly computed for quantum states (thanks to our
Theorem~\ref{t:proj}), having a faster convergence rate is important,
especially with  higher dimensional Hilbert spaces. The downside of this method
is that it necessarily requires a differentiable cost function $C$, with
gradient $\nabla C$.

In its standard form, the conjugate gradient method to approximate the
solution of $\argmin_{\pi\in\mathcal{S}}C(\pi)$ is defined by the following
iterative rule%
\begin{equation}%
\begin{array}
[c]{l}%
1)~\mathrm{Find~}\argmin_{\sigma\in\mathcal{S}}\text{\textrm{Tr}}[\sigma\nabla
C(\pi_{i})],\\
2)~\pi_{i+1}=\pi_{i}+\frac{2}{i+2}(\sigma-\pi_{i})=\frac{i}{i+2}\pi_{i}%
+\frac{2}{i+2}\sigma.
\end{array}
\label{FrankWolfe0}%
\end{equation}
The first step in the above iteration rule is solved by finding the smallest
eigenvector $|\sigma\rangle$ of $\nabla C(\pi_{i})$. Indeed, since $\pi$ is an
operator and $C(\pi)$ a scalar, the gradient $\nabla C$ is an operator with
the same dimension of $\pi$. Therefore, for learning quantum programs
we find the following iterationfollowing
\begin{equation}%
\begin{array}
[c]{l}%
1)~\text{\textrm{Find the smallest eigenvalue}}~\left\vert \sigma
_{i}\right\rangle ~\text{\textrm{of}}~\nabla C(\pi_{i}),\\
2)~\pi_{i+1}=\frac{i}{i+2}\pi_{i}+\frac{2}{i+2}\left\vert \sigma
_{i}\right\rangle \left\langle \sigma_{i}\right\vert .
\end{array}
\label{FrankWolfe}%
\end{equation}
%read this introduction https://www.di.ens.fr/~aspremon/PDF/HPOPT12.pdf
When the gradient of $C$ is Lipschitz continuous with constant $L$, the
conjugate gradient method converges after $\mathcal{O}(L/\epsilon)$
steps~\cite{jaggi2013revisiting,nesterov2005smooth}. The
following iteration with adaptive learning rate $\alpha_{i}$ has even faster
convergence rates, provided that $C$ is strongly
convex~\cite{garber2015faster}:
\begin{equation}%
\begin{array}
[c]{l}%
1)~\text{\textrm{Find the smallest eigenvalue }}\left\vert \sigma
_{i}\right\rangle \text{\textrm{ of~}}\nabla C(\pi_{i}),\\
2)~\text{\textrm{Find }}\alpha_{i}=\argmin_{\alpha\in\lbrack0,1]}\alpha
\langle\tau_{i},\nabla C(\pi_{i})\rangle\\
~\text{\textrm{~~}}+\alpha^{2}\frac{\beta_C}{2}\Vert\tau_{i}\Vert^{2}_C%
,~\text{\textrm{for }}\tau_{i}=|\sigma_{i}\rangle\langle\sigma_{i}|-\pi_{i},\\
3)~\pi_{i+1}=(1-\alpha_{i})\pi_{i}+\alpha_{i}\left\vert \sigma_{i}%
\right\rangle \left\langle \sigma_{i}\right\vert .
\end{array}
\label{FFrankWolfe}%
\end{equation}
where the constant $\beta_C$ and norm $\|\cdot\|_C$ depend on $C$ \cite{garber2015faster}.

In spite of the faster convergence rate, conjugate gradient methods require
smooth cost functions (so that the gradient $\nabla C$ is well defined at
every point). However, cost functions based on trace distance \eqref{traceD}
are not smooth. For instance, the trace distance in one-dimensional spaces reduces
to the absolute value function $|x|$ that is non-analytic at $x=0$.
When some eigenvalues are close to zero, conjugate gradient methods may display
unexpected behaviors, though we have numerically observed that
convergence is always obtained with a careful choice of the learning rate.
Moreover,
in the next section we will show how to formally justify the applicability
of the conjugate gradient method, following
Nesterov's smoothing prescription~\cite{nesterov2005smooth}.

\subsection{Smooth trace distance}

The conjugate gradient method converges to the global optimum after
$\mathcal{O}\left(  \frac{L}{\epsilon}\right)  $ steps, provided that the
gradient of $C$ is $L$-Lipschitz continuous~\cite{nesterov2005smooth}.
However, the constant $L$ can diverge for non-smooth functions like the trace
distance \eqref{traceD} so the convergence of the algorithm cannot be formally
stated, although it may still be observed in numerical simulations, as we will
show. To solidify the convergence proof (see also Appendix~\ref{a:smoothtrace}%
) we introduce a smooth approximation to the trace distance. This is defined
by the following cost function that is differentiable at every point
\begin{equation}
C_{\mu}(\pi)=\Tr\left[  h_{\mu}\left(  \chi_{\pi}-\chi_{\mathcal{E}}\right)
\right]  =\sum_{j}h_{\mu}(\lambda_{j})~,\label{Dmu}%
\end{equation}
where $\lambda_{j}$ are the eigenvalues of $\chi_{\pi}-\chi_{\mathcal{E}}$ and
$h_{\mu}$ is the so-called Huber penalty function
\begin{equation}
h_{\mu}(x):=%
\begin{cases}
\frac{x^{2}}{2\mu} & \mathrm{~if~}|x|<\mu~,\\
|x|-\frac{\mu}{2} & \mathrm{~if~}|x|\leq\mu~.
\end{cases}
\label{HuberPenalty}%
\end{equation}
The previous definition of the trace distance, $C_{1}$ in Eq.~\eqref{traceD},
is recovered for $\mu\rightarrow0$ and, for any non-zero $\mu$, the $C_{\mu}$
bounds $C_{1}$ as follows
\begin{equation}
C_{\mu}(\pi)\leq C_{1}(\pi)\leq C_{\mu}(\pi)+\frac{\mu d}{2},\label{Dmuineq}%
\end{equation}
where $d$ is the dimension of the program state $\pi$. In
Appendix~\ref{a:smoothtrace} we then prove the following result

\begin{theorem}
The smooth cost function $C_{\mu}(\pi)$ is a convex function over program
states and its gradient is given by
\begin{equation}
\nabla C_{\mu}(\pi)=\Lambda^{\ast}[h_{\mu}^{\prime}(\chi_{\pi}-\chi
_{\mathcal{E}})],
\end{equation}
where $h_{\mu}^{\prime}$ is the derivative of $h_{\mu}$. Moreover, the
gradient is $L$-Lipschitz continuous with
\begin{equation}
L=\frac{d}{\mu}~,\label{LipschitzConst}%
\end{equation}
where $d$ is the dimension of the program state.
\end{theorem}

Being Lipschitz continuous, the conjugate gradient algorithm and its
variants~\cite{nesterov2005smooth,garber2015faster} converge up to an accuracy
$\epsilon$ after $\mathcal{O}(L/\epsilon)$ steps. In some applications, it is
desirable to analyze the convergence in trace distance in the limit of large
program states, namely for $d\rightarrow\infty$. The parameter $\mu$ can be
chosen such that the smooth trace distance converges to the trace distance,
namely $C_{\mu}\rightarrow C_{1}$ for $d\rightarrow\infty$. Indeed, given the
inequality \eqref{Dmuineq}, a possibility is to set $\mu=\mathcal{O}%
(d^{-(1+\eta)})$ for some $\eta>0$ so that, from Eq.~\eqref{LipschitzConst},
the convergence to the trace norm is achieved after $\mathcal{O}(d^{2+\eta})$ steps.

\section{Learning of arbitrary unitaries\label{Sec5}}
The simulation of quantum gates or, more generally, unitary transformations
is crucial for quantum computing applications \cite{lloyd1996universal} so
ML techniques have been developed for this purpose
\cite{banchi2016quantum,innocenti2018supervised,mitarai2018quantum,arrazola2018machine}.
Here we consider the more general setting of simulating
an arbitrary finite-dimensional unitary $U$ by means
of a programmable quantum processor with map $\Lambda$. For a unitary $U$ the
Choi matrix is a maximally-entangled pure state $\chi_{\mathcal{E}}=|\chi
_{U}\rangle\langle\chi_{U}|$. Therefore, $\sqrt{\chi_{\mathcal{E}}}%
=\chi_{\mathcal{E}}$ is a one-dimensional projector and
Eq.~\eqref{fidelitygrad} is drastically simplified to
\begin{equation}
\nabla F(\pi)=\frac{\Lambda^{\ast}\left[  |\chi_{U}\rangle\langle\chi
_{U}|\right]  }{2\sqrt{\langle\chi_{U}|\Lambda(\pi)|\chi_{U}\rangle}%
}~.\label{fidelitygradpure}%
\end{equation}
Therefore the gradient \eqref{Cfgrad} of the convex cost function $C_F$,
\begin{equation}
\nabla C_F(\pi)=-{\Lambda^{\ast}\left[  |\chi_{U}\rangle\langle\chi
_{U}|\right]  }~,
\end{equation}
is independent of $\pi$. When we employ the conjugate gradient method,
the state $|\sigma_{k}\rangle$ is the same for each iteration step.
This implies that conjugate gradient is
converging towards one eigenvector of $-\Lambda^{\ast}\left[  |\chi
_{U}\rangle\langle\chi_{U}|\right]  $ with minimum eigenvalue.
In other terms, the fixed point of the
iteration in Eq.~\eqref{FrankWolfe}, namely the optimal program state
$\tilde{\pi}_{F}$ (according to the fidelity cost function) is pure and equal
to the eigenvector of $\Lambda^{\ast}\left[  |\chi_{U}\rangle
\langle\chi_{U}|\right]  $ with maximum eigenvalue.

The above result can be proven as follows. Let $\pi_{1}$ be the initial guess
for the program state. After $k$ iterations of Eq.~\eqref{FrankWolfe}, we find
the following approximation to the optimal program state
\begin{equation}
\pi_{k}=\frac{2}{k+k^{2}}\pi_{1}+\left(  1-\frac{2}{k+k^{2}}\right)
\tilde{\pi}_{F}~,
\end{equation}
where $\frac{2}{k+k^{2}}=\prod_{j=1}^{k-1}\frac{j}{j+2}$. The above equation
shows that $\pi_{k}\rightarrow\tilde{\pi}_{F}$ for $k\rightarrow\infty$, with
error in trace distance
\begin{equation}
\Vert\pi_{k}-\tilde{\pi}_{F}\Vert_{1}=\frac{2}{k+k^{2}}\left\Vert \pi
_{1}-\tilde{\pi}_{F}\right\Vert _{1}=\mathcal{O}(k^{-2})~.\label{itera}%
\end{equation}

For learning arbitrary unitaries, the fidelity cost function provides a
convenient choice where the optimal program can be found analytically.
Moreover, this example shows that the convergence rate $\mathcal{O}%
(\epsilon^{-1})$ of the conjugate method provides a worst case instance
that can be beaten in some applications with some suitable cost functions.
From Eq.~\eqref{itera} we see that $\epsilon=k^{-2}$ for learning arbitrary unitaries
via the minimization of $C_F$,
meaning that convergence is obtained with the faster rate $\mathcal{O}%
(\epsilon^{-1/2})$. On the other hand, there are
no obvious simplifications for the optimization of the trace distance, since
the latter still requires the diagonalization of Eq.~\eqref{tracediag}. For
the trace distance, or its smooth version, only numerical approaches are feasible.

\section{Teleportation processor\label{Sec6}}

One possible (shallow) design for the quantum processor $Q$ is the
teleportation protocol~\cite{bennett1993teleporting} which has to be applied
to a generic program state $\pi$ instead of a maximally entangled state. In
dimension $d$, the program $\pi^{AB}$ is a $d\times d$ state. The
teleportation protocol involves a basis of $d^{2}$ maximally entangled states
$|\Phi_{i}\rangle$ and a basis $\{U_{i}\}$ of teleportation unitary such that
$\mathrm{Tr}(U_{i}^{\dagger}U_{j})=d\delta_{ij}$~\cite{teleREVIEW}. In the
protocol, an input $d$-dimensional state $\rho^{S}$ and the $A$ part of the
program $\pi^{AB}$ are subject to the projector $|\Phi_{i}\rangle\!\left\langle
\Phi_{i}\right\vert $. The classical outcome $i$ is communicated to the $B$
part of $\pi^{AB}$\ where the correction $U_{i}^{-1}$ is applied. In this way,
we implement the following teleportation channel $\mathcal{E}_{\pi}$ from
qudit $S$ to qudit $B$
\begin{equation}
\mathcal{E}_{\pi}(\rho)=\sum_{i}U_{i}^{B}\langle\Phi_{i}^{SA}|\rho^{S}%
\otimes\pi^{AB}|\Phi_{i}^{SA}\rangle U_{i}^{B\dagger}.
\end{equation}

The Choi matrix of the teleportation channel $\mathcal{E}_{\pi}$ can be
written as $\chi_{\pi}=\Lambda_{\text{tele}}(\pi)$, where the map of the
teleportation processor is equal to%
\begin{equation}
\Lambda_{\text{tele}}(\pi)=\frac{1}{d^{2}}\sum_{i}\left(  U_{i}^{\ast}\otimes
U_{i}\right)  \pi\left(  U_{i}^{\ast}\otimes U_{i}\right)  ^{\dagger}.
\label{eqAB}%
\end{equation}
Note that, if the program $\pi$ is teleportation
covariant~\cite{pirandola2017fundamental}, namely if $[\pi,U_{i}^{\ast}\otimes
U_{i}]=0$, then $\pi$ is automatically a fixed point of the map, i.e., we have
$\chi_{\pi}:=\Lambda_{\text{tele}}(\pi)=\pi$. Also note that, the channel in
Eq.~(\ref{eqAB}) is self-dual, i.e., $\Lambda^{\ast}=\Lambda$. As a result,
for any operator $\hat{O}$, we may write%
\begin{equation}
\Lambda_{\text{tele}}^{\ast}(\hat{O})=\frac{1}{d^{2}}\sum_{i}\left(
U_{i}^{\ast}\otimes U_{i}\right)  \hat{O}\left(  U_{i}^{\ast}\otimes
U_{i}\right)  ^{\dagger}. \label{opSS}%
\end{equation}

As an example, assume that the target channel is a unitary $U$, so that its
Choi matrix is $\chi_{U}:=|\chi_{U}\rangle\langle\chi_{U}|$ with $|\chi
_{U}\rangle=\openone\otimes U|\Phi\rangle$ and $|\Phi\rangle$ is maximally
entangled. By using Eq.~(\ref{opSS}) and $U^{\ast}\otimes\openone|\Phi
\rangle=\openone\otimes U^{\dagger}|\Phi\rangle$, we may write the dual
processor map
\begin{align}
&  \Lambda_{\text{tele}}^{\ast}[|\chi_{U}\rangle\langle\chi_{U}|]\nonumber\\
&  =\frac{1}{d^{2}}\sum_{i}\left(  \openone\otimes V_{i}^{U}\right)
|\Phi\rangle\langle\Phi|\left(  \openone\otimes V_{i}^{U}\right)  ^{\dagger
},\label{telelambdadual}%
\end{align}
where $V_{i}^{U}=U_{i}UU_{i}^{\dagger}$. The maximum eigenvector of
$\Lambda_{\text{tele}}^{\ast}[|\chi_{U}\rangle\langle\chi_{U}|]$ represents
the optimal program state $\tilde{\pi}_{F}$ for simulating the unitary
$U$\ via the teleportation processor (according to the fidelity cost
function).
%According to Eq.~(\ref{itera}) this is found in $\mathcal{O}(k^{-2})$ steps.
In some cases, the solution is immediate. For instance, this happens when
$V_{i}^{U}\propto U$ is independent of $i$. This is the case when $U$ is a
teleportation unitary, because it satisfies the Weyl-Heysenberg
algebra~\cite{pirandola2017fundamental}. For a teleportation unitary $U$, we
simply have%
\begin{equation}
\Lambda^{\ast}[|\chi_{U}\rangle\langle\chi_{U}|]=|\chi_{U}\rangle\langle
\chi_{U}|~,
\end{equation}
so that the unique optimal program is $\tilde{\pi}_{F}=|\chi_{U}\rangle
\langle\chi_{U}|$.

\begin{figure}[t]
\centering
\includegraphics[width=0.99\linewidth]{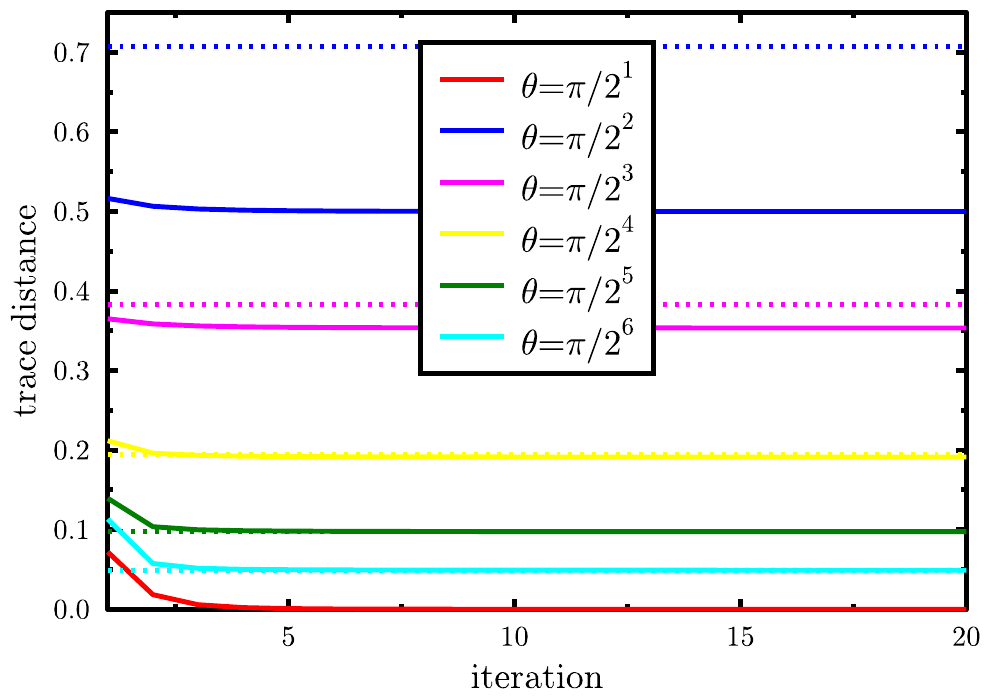}
\caption{Optimization of program states for simulating the rotation
$R(\theta)=e^{i\theta X}$ with a teleportation processor. The optimization is
via the minimization of trace distance $C_{1}$ of Eq.~\eqref{traceD} with the
projected subgradient method in Eq.~\eqref{projsubgrad}. The dashed lines
correspond to the upper bound $\sqrt{1-F[\Lambda(\tilde{\pi}_{F}%
),\chi_{\mathcal{E}}]^{2}}$ of the trace distance, where $\tilde{\pi}_{F}$ is
the optimal program that maximizes the fidelity, namely the eigenvector of
Eq.~\eqref{fidelitygradpure} with maximum eigenvalue. }%
\label{fig:teleunitary}%
\end{figure}

In Fig.~\ref{fig:teleunitary} we show the convergence of the projected
subgradient algorithm using the teleportation processor and target unitaries
$R(\theta)=e^{i\theta X}$, for different values of $\theta$. When $\theta$ is
a multiple of $\pi/2$, then the above unitary is teleportation covariant and
the Frank-Wolfe algorithm converges to zero trace distance. For other values
of $\theta$ perfect simulation is impossible, and we notice that the algorithm
converges to a non zero value of the trace distance \eqref{traceD}. For
comparison, in Fig.~\ref{fig:teleunitary} we also plot the value of the
fidelity upper bound $\sqrt{1-F[\Lambda(\tilde{\pi}_{F}),\chi_{\mathcal{E}%
}]^{2}}$, where $\tilde{\pi}_{F}$ is the optimal program that maximizes the
fidelity of Eq.~\eqref{fidCC}, namely the eigenvector of
Eq.~\eqref{telelambdadual} with maximum eigenvalue.
We note that for $\theta=\pi/2^{\ell}$ the trace
distance decreases for larger $\theta$. The limit case $\ell\rightarrow\infty$
is perfectly simulable as $R(0)$ is teleportation covariant.

\subsection{Pauli channel simulation}

Pauli channels are defined as~\cite{nielsen2000quantum}
\begin{equation}
\mathcal{P}(\rho)=\sum_{i}p_{i}U_{i}\rho U_{i}^{\dagger}~,
\end{equation}
where $U_{i}$ are generalized Pauli operators and $p_{i}$ some probabilities.
For $d=2$ the Pauli operators are the four Pauli matrices $I,X,Y,Z$ and in any
dimension they form the Weyl-Heisenberg group~\cite{nielsen2000quantum}. These
operators are exactly the teleportation unitaries $U_{j}$ defined in the
previous section. The Choi matrix $\chi_{\mathcal{P}}$ of a Pauli channel
$\mathcal{P}$ is diagonal in the Bell basis, i.e., we have
\begin{equation}
\chi_{\mathcal{P}}=\sum_{i}p_{i}|\Phi_{i}\rangle\langle\Phi_{i}|~,
\end{equation}
where $\Phi_{i}=\openone\otimes U_{i}|\Phi\rangle$ and $|\Phi\rangle
=\sum_{j=1}^{d}|jj\rangle/\sqrt{d}$.

We now consider the simulation of a Pauli channel with the teleportation
quantum processor introduced in the previous section. Let
\begin{equation}
\pi=\sum_{ij}\pi_{ij}|\Phi_{i}\rangle\langle\Phi_{j}|~,
\end{equation}
be an arbitrary program state expanded in the Bell basis. For any program
state, the Choi matrix of the teleportation-simulated channel is given by
Eq.~\eqref{eqAB}. Using standard properties of the Pauli matrices we find
\begin{equation}
\chi_{\pi}\equiv\Lambda(\pi)=\sum_{i}\pi_{ii}|\Phi_{i}\rangle\langle\Phi
_{i}|~,
\end{equation}
namely a generic state is transformed into a Bell diagonal state. Therefore,
the cost function
\begin{equation}
C_{1}^{\mathrm{Pauli}}=\Vert\chi_{P}-\chi_{\pi}\Vert_{1}~,
\end{equation}
can be minimized analytically for any Pauli channel by choosing $\pi
_{ij}=p_{i}\delta_{ij}$. With this choice we find $C_{1}^{\mathrm{Pauli}}=0$,
meaning that the simulation is perfect.

From theory~\cite{bowen2001teleportation,cope2017simulation,BDSW} we know that
only Pauli channels can be perfectly simulated in this way. No matter how more
general we can make the states $\pi$, it is
proven~\cite{bowen2001teleportation,cope2017simulation} that these are the
only channels we can perfectly simulate. This is true even if we apply the
Pauli corrections in a probabilistic way, i.e., we assume a classical channel
from the Bell outcomes to the corresponding label of the Pauli correction
operator~\cite{cope2017simulation}.

\section{Port-based teleportation\label{Sec7}}

We now study a design of programmable quantum processor that can potentially
simulate any target quantum channel in the asymptotic limit of an arbitrarily
large program state. This design is
PBT~\cite{ishizaka2008asymptotic,ishizaka2009quantum,ishizaka2015some}, a
generalization of the standard teleportation scheme. For finite-dimensional
programs, a PBT processor cannot achieve a perfect deterministic simulation of
an arbitrary channel~\cite{nielsen1997programmable}. In this realistic
finite-dimensional setting, our study finally establishes the optimal
performance achievable by this type of quantum processor.

\subsection{Basics of PBT}

\begin{figure}[t]
\centering
\includegraphics[width=0.95\linewidth]{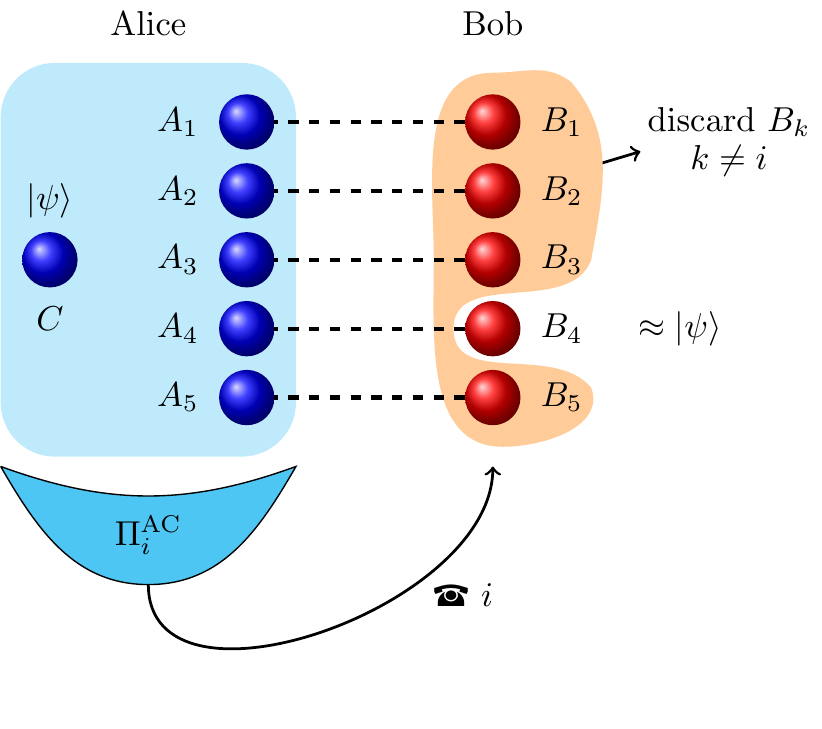} \caption{PBT
scheme. Two distant parties, Alice and Bob, share $N$ maximally entangled
pairs $\{A_{k},B_{k}\}_{k=1}^{N}$. Alice also has another system $C$ in the
state $|\psi\rangle$. To teleport $C$, Alice performs the POVM $\Pi
_{i}^{\mathbf{A}C}$ on all her local systems $\mathbf{A}=\{A_{k}\}_{k=1}^{N}$
and $C$. She then communicates the outcome $i$ to Bob. Bob discards all his
systems $\mathbf{B}=\{B_{k}\}_{k=1}^{N}$ with the exception of $B_{i}$. After
these steps, the state $|\psi\rangle$ is approximately teleported to $B_{i}$.
Similarly, an arbitrary channel $\mathcal{E}$ is simulated with $N$ copies of
the Choi matrix $\chi_{\mathcal{E}}^{A_{k}B_{k}}$. The figure shows an example
with $N=5$, where $i=4$ is selected. }%
\label{fig:pbt}%
\end{figure}

The overall protocol of PBT is illustrated in Fig.~\ref{fig:pbt}.
Unlike standard teleportation protocol, PBT requires that Alice
and Bob share $N$ entangled pairs for the simulation of the identity channel
\cite{ishizaka2008asymptotic}. The protocol is based on a resource state (the
program) given by $\pi_{\mathbf{AB}}=\bigotimes_{k=1}^{N}\Phi_{A_{k}B_{k}}$,
where $|\Phi_{A_{k}B_{k}}\rangle$ are Bell states for Alice's $N$ qudits
$\mathbf{A}=(A_{1},\dots,A_{N})$ and Bob's $N$ qudits $\mathbf{B}=(B_{1}%
,\dots,B_{N})$. After preparing such a state, Alice performs a joint
positive-operator value measure (POVM) $\{\Pi_{i}\}$ on her $\mathbf{A}$-half
of $\pi_{\mathbf{AB}}$ and an input state $|\psi\rangle_{C}$ that she wishes
to teleport. She communicates the outcome $i$ to Bob, who discards all
\textquotedblleft ports\textquotedblright\ $\mathbf{B}$ except $B_{i}%
=B_{\text{\textrm{out}}}$. The resulting PBT channel $\mathcal{P}_{\pi
}:\mathcal{H}_{C}\mapsto\mathcal{H}_{B_{\mathrm{out}}}$ is then
\begin{align}
\mathcal{P}_{\pi}(\rho) &  =\sum_{i=1}^{N}\Tr_{\mathbf{A\bar{B}}_{i}C}\left[
\Pi_{i}(\pi_{\mathbf{AB}}\otimes\rho_{C})\right]  _{B_{i}\rightarrow
B_{\mathrm{out}}}\label{PBTchan}\\
&  =\sum_{i=1}^{N}\Tr_{\mathbf{A\bar{B}}_{i}C}\left[  \sqrt{\Pi_{i}}%
(\pi_{\mathbf{AB}}\otimes\rho_{C})\sqrt{\Pi_{i}}\right]  _{B_{i}\rightarrow
B_{\mathrm{out}}},\nonumber
\end{align}
where $\mathbf{\bar{B}}_{i}=\mathbf{B}\backslash B_{i}=\{B_{k}:k\neq i\}$. In
the limit $N\rightarrow\infty$, PBT approximates an identity channel
$\mathcal{P}_{\pi}(\rho)\approx\rho$.

In the standard PBT protocol~\cite{ishizaka2008asymptotic,ishizaka2009quantum}
the following POVM is used%
\begin{equation}
\Pi_{i}=\tilde{\Pi}_{i}+\frac{1}{N}\left(  \openone-\sum_{k}\tilde{\Pi}%
_{k}\right)  ,\label{POVMchoice}%
\end{equation}
where
\begin{align}
\tilde{\Pi}_{i}  & =\sigma_{\mathbf{A}C}^{-1/2}\Phi_{A_{i}C}{\sigma
}_{\mathbf{A}C}^{-1/2},\\
\sigma_{\mathbf{A}C}  & :=\sum_{i=1}^{N}\Phi_{A_{i}C},
\end{align}
and $\sigma^{-1/2}$ is an operator defined only on the support of $\sigma$.
The PBT\ protocol is formulated for $N\geq2$ ports. However, we also include
here the trivial case for $N=1$, corresponding to the process where Alice's
input is traced out and the output is the reduced state of Bob's port, i.e., a
maximally mixed state.

With the choice of the POVM in Eq.~(\ref{POVMchoice}), the identity channel
$\mathcal{I}$ can be simulated with
fidelity~\cite{ishizaka2008asymptotic,ishizaka2015some}
\begin{equation}
F_{\pi}=1-\mathcal{O}\left(  \frac{1}{N}\right)  ~,\label{e:pbtfid}%
\end{equation}
so perfect simulation is possible only in the limit $N\rightarrow\infty$. More
generally, it has been shown~\cite{pirandola2018fundamental} that simulation
error in diamond norm scales as
\begin{equation}
\Vert\mathcal{I}-\mathcal{P}_{\pi}\Vert_{\diamond}\leq\frac{2d(d-1)}{N}~.
\label{simerr}
\end{equation}

\subsection{Channel simulation via PBT}

Any generic channel $\mathcal{E}$ can be written as a composition
$\mathcal{E}\circ\mathcal{I}$ between $\mathcal{E}$ and the identity channel
$\mathcal{I}$. Channel simulation can be achieved by replacing the identity
channel $\mathcal{I}$ with its PBT simulation $\mathcal{P}_{\pi}$, and then
applying $\mathcal{E}$ to $B_{i}$. However, since Bob does not perform any
post-processing on its systems $\mathbf{B}$, aside from discarding all ports
$B_{k}$ with $k\neq i$, he can also apply \textit{first} the channel
$\mathcal{E}^{\otimes N}$ to all his ports and \textit{then} discard all the
ports $B_{k}$ with $k\neq i$.
%The channel $\mathcal E$ can be included in the protocol since $B$ does not do any post-processing.
In doing so, he changes the program state to
\begin{equation}
\pi_{\mathbf{AB}}=\openone_{A}\otimes\mathcal{E}_{B}^{\otimes N}\left[
\bigotimes_{k=1}^{N}\Phi_{A_{k}B_{k}}\right]  =\bigotimes_{k=1}^{N}%
\chi_{\mathcal{E}}^{A_{k}B_{k}}.\label{e:pbtchoiprogram}%
\end{equation}
In other terms, any channel $\mathcal{E}$ can be PBT-approximated by $N$
copies of its Choi matrix $\chi_{\mathcal{E}}$ as program state. However,
while such a program state is optimal when $N\rightarrow\infty$, for finite
$N$ there may be better alternatives. In general, for any finite $N$, finding
the optimal program state $\pi_{\mathbf{AB}}$ simulating a channel
$\mathcal{E}$ with PBT\ is an open problem, and no explicit solutions or
procedures are known.

We employ our convex optimization procedures to find the optimal program
state. This can be done either exactly by minimizing the diamond distance cost
function $C_{\diamond}$ via SDP, or approximately, by determining the optimal
program state via the minimization of the trace distance cost function $C_{1}$
via the gradient-based ML\ techniques discussed above. For this second
approach, we need to derive the map $\Lambda$ of the PBT processor, between
the program state $\pi$ to output Choi matrix as in Eq.~\eqref{lambdadef}.
From the definition in Eq.~\eqref{PBTchan} we find the following operator sum
decomposition
\begin{align}
\Lambda(\pi) &  =\chi_{\mathcal{P}_{\pi}}=\openone_{D}\otimes\mathcal{P}_{\pi
}[\Phi_{DC}]\nonumber\\
&  =\sum_{i=1}^{N}\Tr_{\mathbf{A\bar{B}}_{i}C}\left[  \sqrt{\Pi_{i}}%
(\pi_{\mathbf{AB}}\otimes\Phi_{DC})\sqrt{\Pi_{i}}\right]  _{B_{i}\rightarrow
B_{\mathrm{out}}}\nonumber\\
&  =\sum_{ik}K_{ik}\pi K_{ik}^{\dagger}~,\label{e:LambdaPBT}%
\end{align}
where the corresponding Kraus operators are
\begin{equation}
K_{ik}^{\mathbf{AB}\rightarrow DB_{\mathrm{out}}}=\langle{e_{k}^{(i)}}%
|\sqrt{\Pi_{i}}\otimes\openone_{\mathbf{B}D}|\Phi_{DC}\rangle~,
\end{equation}
and $|e_{k}^{(i)}\rangle$ span a basis of $\mathbf{A\bar{B}}_{i}C$.

\subsection{Program state compression\label{compressSEC}}

The program state grows exponentially with the number of ports $N$ as $d^{2N}$
where $d$ is the dimension of the Hilbert space. However, as also discussed in
the original proposal~\cite{ishizaka2008asymptotic,ishizaka2009quantum} and more recently in
Ref.~\cite{christandl2018asymptotic}, the resource state of PBT can be chosen
with extra symmetries, so as to reduce the number of free parameters. In
particular, we may consider the set of program states that are symmetric under
the exchange of ports, i.e., such that rearranging any $A$ modes and the
corresponding $B$ modes leaves the program state unchanged.

Let $P_{s}$ be the permutation operator swapping labels 1 to $N$ for the
labels in the sequence $s$, which contains all the numbers 1 to $N$ once each
in some permuted order. Namely $P_{s}$ exchanges all ports according to the
rule $i\mapsto s_{i}$. Since PBT is symmetric under exchange of ports, we may
write%
\begin{equation}
\mathcal{P}_{P_{s}\pi P_{s}^{\dag}}(\rho)=\mathcal{P}_{\pi}\left(
\rho\right)  ~\text{for any }s\text{.}%
\end{equation}
Consider then an arbitrary permutation-symmetric resource state $\pi
_{\mathrm{sym}}$ as
\[
\pi_{\mathrm{sym}}=\frac{1}{N!}\sum_{s}P_{s}\pi P_{s}^{\dag},
\]
where the sum is over all possible sequences $s$ that define independent
permutations and $N!$ is the total number of possible permutations. Clearly
$\mathcal{P}_{\pi_{\mathrm{sym}}}=\mathcal{P}_{\pi}$, so any program state
gives the same PBT channel as some symmetric program state. It therefore
suffices to consider the set of symmetric program states. This is a convex
set: any linear combination of symmetric states is a symmetric state.

To construct a basis of the symmetric space, we note that each element of a
density matrix is the coefficient of a dyadic (of the form $\left\vert
x\right\rangle \left\langle y\right\vert $). If permutation of labels maps one
dyadic to another, the coefficients must be the same. This allows us to
constrain our density matrix using fewer global parameters. For instance, for
$d=2$ we can define the 16 parameters $n_{00,00}$, $n_{00,01}$, $n_{00,10}$,
etc., corresponding to the number of ports in the dyadic of the form
$\left\vert 0_{A}0_{B}\right\rangle \left\langle 0_{A}0_{B}\right\vert $,
$\left\vert 0_{A}0_{B}\right\rangle \left\langle 0_{A}1_{B}\right\vert $,
$\left\vert 0_{A}0_{B}\right\rangle \left\langle 1_{A}0_{B}\right\vert $, etc.
Each element of a symmetric density matrix can then be defined solely in terms
of these parameters, i.e., all elements corresponding to dyadics with the same
values of these parameters have the same value.

For the general qudit case, in which our program state consists of $N$ ports,
each composed of two $d$-dimensional qudits, we can find the number of
independent parameters from the number of independent dyadics. Each port in a
dyadic can be written as $|a_{A},b_{B}\rangle\langle c_{A},d_{B}|$ where the
extra indices $A$ and $B$ describe whether those states are modeling either
qudit $A$ or $B$. There are $d^{4}$ different combinations of $\{a,b,c,d\}$,
so we can place each qudit into one of $d^{4}$ categories based on these
values. If two elements in the density matrix correspond to dyadics with the
same number of ports in each category, they must take the same value. Hence,
the number of independent coefficients is given by the number of ways of
placing $N$ (identical) ports into $d^{4}$ (distinguishable) categories.
%We can consider $N+d^4-1$ positions into which we
%either place on of $N$ ports or one of $d^4-1$ dividers between categories (so
%that any ports to the right of one of the dividers are in that category). Our
%problem is then the number of distinct lists of $d^4-1$ positions (of the
%dividers). Consequently, the number of coefficients is given by a binomial
%coefficient,
This is exactly the binomial coefficient
\begin{equation}
\binom{N+d^{4}-1}{d^{4}-1}=\mathcal{O}(N^{d^{4}-1})~.
\end{equation}
Consequently, exploiting permutation symmetry of the PBT\ protocol, we can
\textit{exponentially} reduce the number of parameters for the optimization
over program states.

The number of parameters can be reduced even further by considering products
of Choi matrices. We may focus indeed on the Choi set
\begin{equation}
\mathcal{C}_{N}=\left\{  \pi:\pi=\sum_{k}p_{k}\chi_{k}^{\otimes N}\right\}
~,\label{Cset}%
\end{equation}
where each $\chi_{k}=\chi_{AB}^{k}$ is a generic Choi matrix, therefore
satisfying $\mathrm{Tr}_{B}\chi_{k}=d^{-1}\openone$, and $p_{k}$ form a
probability distribution. Clearly $\mathcal{C}$ is a convex set. We now show
that this set can be further reduced to just considering $N=1$.

When the program state $\pi=\chi^{\otimes N}$ is directly used in
Eq.~\eqref{e:LambdaPBT} we find
\begin{align}
\Lambda(\pi) &  =\sum_{i=1}^{N}\mathrm{Tr}_{\mathbf{A\bar{B}}_{i}C}\left[
\Pi_{i}\left(  \chi_{AB}^{\otimes N}\otimes\Phi_{DC}\right)  \right]
_{B_{i}\rightarrow B_{\mathrm{out}}}\\
&  =\frac{1}{d^{N-1}}\sum_{i=1}^{N}\mathrm{Tr}_{A_{i}C}\left[  \Pi_{i}\left(
\chi_{A_{i}B_{\mathrm{out}}}\otimes\Phi_{DC}\right)  \right]  \\
&  :=\tilde{\Lambda}(\chi)~,\label{e:LambdaChoi}%
\end{align}
namely that the optimization can be reduced to the $\mathcal{O}(d^{4}$)
dimensional space of Choi matrices $\chi$. Note that, in the above equation,
we used the identity
\begin{equation}
\mathrm{Tr}_{\mathbf{\bar{B}}_{i}}\chi_{AB}^{\otimes N}=\chi_{A_{i}B_{i}%
}\otimes\frac{\openone_{\mathbf{\bar{A}}_{i}}}{d^{N-1}},
\end{equation}
where $\mathbf{\bar{A}}_{i}=\mathbf{A}\backslash A_{i}$.

Now let $\pi$ be a linear combination of tensor products of Choi matrix
states, $\chi_{k}^{\otimes N}$, each with probability $p_{k}$ as in
Eq.~(\ref{Cset}). Then we can write
\begin{align}
\mathrm{Tr}_{\mathbf{\bar{B}}_{i}}\pi_{AB} &  =\mathrm{Tr}_{\mathbf{\bar{B}%
}_{i}}\sum_{k}p_{k}\chi_{k}^{\otimes N}\\
&  =\sum_{k}p_{k}\left(  \chi_{A_{i}B_{i}}^{k}\otimes\frac
{\openone_{\mathbf{\bar{A}}_{i}}}{d^{N-1}}\right)  .
\end{align}
However, this is precisely the partial trace over the tensor product
$\chi^{\prime\otimes N}$ of some other Choi matrix $\chi^{\prime}=\sum
_{k}p_{k}\chi_{k}$. Hence, the program state $\pi=\sum_{k}p_{k}\chi
_{k}^{\otimes N}$ simulates the same channel as the resource state
$\pi^{\prime}=\left(  \sum_{k}p_{k}\chi_{k}\right)  ^{\otimes N}$.

Therefore, the optimization over the convex set $\mathcal{C}_{N}$ can be
reduced to the optimization over products of Choi matrices $\chi^{\otimes N}$.
From Eq.~\eqref{e:LambdaChoi} this can be further reduced to the optimization
of the quantum channel $\tilde{\Lambda}$ over the convex set of single-copy
Choi matrices $\chi$
\begin{equation}
\mathcal{C}_{1}=\left\{  \pi:\pi=\chi_{AB},~\mathrm{Tr}_{B}\chi_{AB}%
=\openone/2\right\}  ,
\end{equation}
which is $\mathcal{O}(d^{4})$. Using $\mathcal{C}_{1}$ drastically reduces the
difficulty of numerical simulations, thus allowing the exploration of
significantly larger values of $N$. Details on how to explicitly construct
$\tilde{\Lambda}$ for $d=2$ are presented in Appendix~\ref{a:pbt}.

\subsection{Numerical examples}

\begin{figure}[t]
\centering
\includegraphics[width=0.95\linewidth]{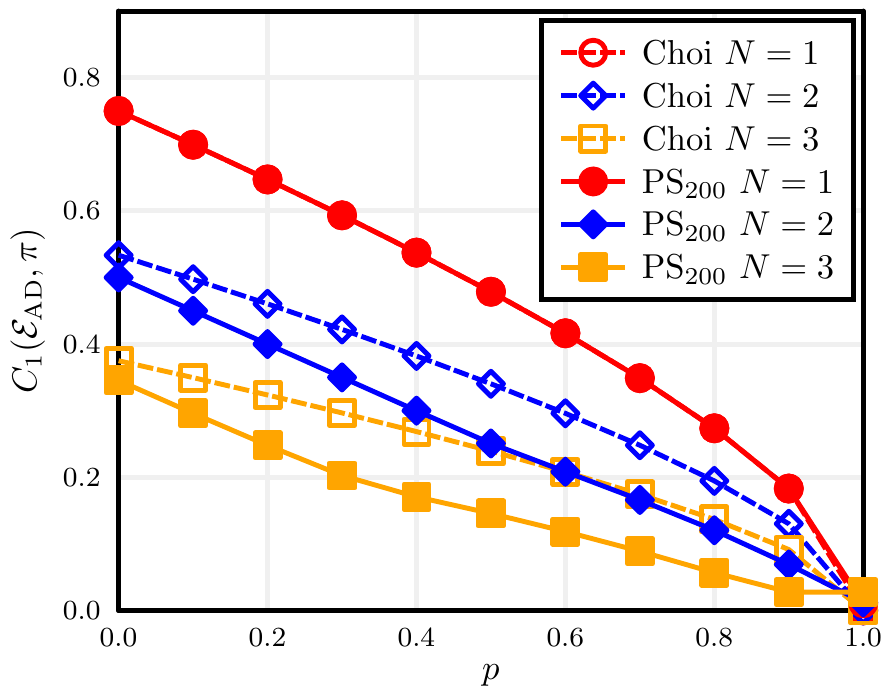} \caption{PBT
Simulation of the amplitude damping channel $\mathcal{E}_{\mathrm{AD}}$ for
various damping rates $p$. Minimization of the trace distance $C_{1}%
(\mathcal{E}_{\mathrm{AD}},\pi)=\Vert\chi_{\mathcal{E}_{\mathrm{AD}}}%
-\chi_{\pi}\Vert_{1}$ between the target channel's Choi matrix and its PBT
simulation with program state $\pi$, for different number of ports $N$. We
consider $N=1,2,3$ and two kinds of programs: copies of the channel's Choi
matrix $\chi_{\mathcal{E}_{\mathrm{AD}}}^{\otimes N}$ and the state
$\tilde{\pi}_{1}$ obtained from the minimization of $C_{1}$ via the projected
subgradient (PS) method after 200 iterations. Note that the simulation error
$C_{1}$ is maximum for the identity channel ($p=0$) and goes to zero for
$p\rightarrow1$.}%
\label{fig:AD_vs_p}%
\end{figure}
%As an example, we
We first consider the simulation of an amplitude damping channel
$\mathcal{E}_{\mathrm{AD}}(\rho)=\sum_{i}K_{i}^{\mathrm{AD}}\rho
K_{i}^{\mathrm{AD}\dagger}$, which is defined by the Kraus operators
\begin{equation}
K_{0}^{\mathrm{AD}}=%
\begin{pmatrix}
1 & 0\\
0 & \sqrt{1-p}%
\end{pmatrix}
,~~K_{1}^{\mathrm{AD}}=%
\begin{pmatrix}
0 & \sqrt{p}\\
0 & 0
\end{pmatrix}
.\label{KrausAD}%
\end{equation}
In Fig.~\ref{fig:AD_vs_p} we study the performance of the PBT simulation of
the amplitude damping channel $\mathcal{E}_{\mathrm{AD}}$ for different
choices of $p$. For $p=0$ the amplitude damping is equal to the identity
channel, while for $p=1$ it is a \textquotedblleft reset\textquotedblright%
\ channel sending all states to $|0\rangle$. We compare the simulation error
with program states $\pi$ either made by products of the channel's Choi matrix
$\chi_{\mathcal{E}_{\mathrm{AD}}}^{\otimes N}$ as in
Eq.~\eqref{e:pbtchoiprogram} or obtained from the minimization of the trace
distance cost function of Eq.~\eqref{traceD} with the projected subgradient
iteration in Eq.~\eqref{projsubgrad}. Alternative methods, like the conjugated
gradient algorithm, perform similarly for this channel. We observe that,
surprisingly, the optimal program $\tilde{\pi}_{1}$ obtained by minimizing the
trace distance $C_{1}$ is always better than the natural choice $\chi
_{\mathcal{E}_{\mathrm{AD}}}^{\otimes N}$.

\begin{figure}[t]
\centering
\includegraphics[width=0.95\linewidth]{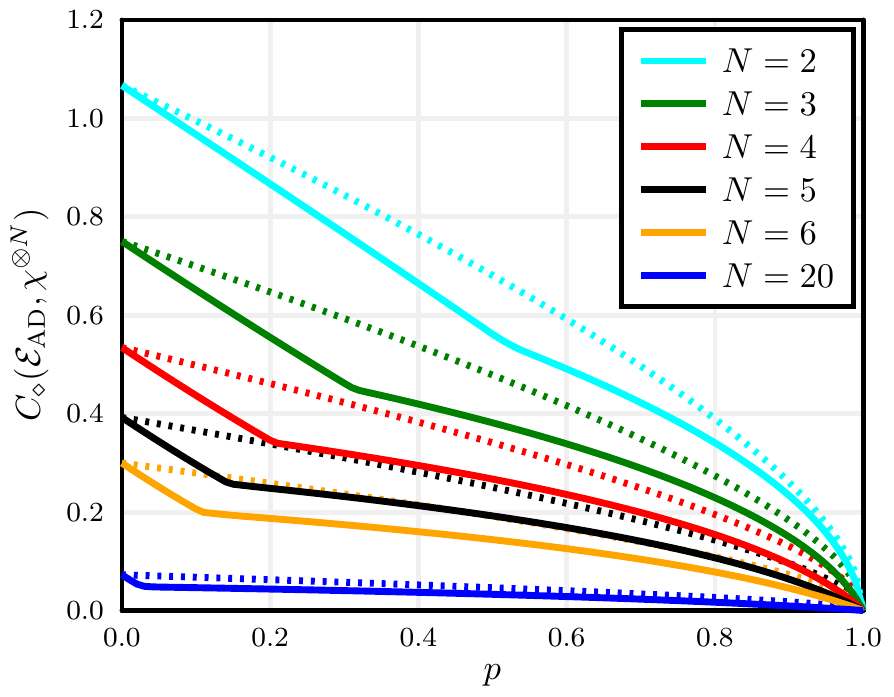} \caption{ PBT
Simulation of the amplitude damping channel $\mathcal{E}_{\mathrm{AD}}$ for
various damping rates $p$. We plot the diamond distance cost function
$C_{\diamond}(\mathcal{E}_{\mathrm{AD}},\pi)=\Vert\mathcal{E}_{\mathrm{AD}%
}-\mathcal{E}_{\mathrm{AD}\text{,}\pi}\Vert_{\diamond}$ between the target
channel $\mathcal{E}_{\mathrm{AD}}$ and its PBT\ simulation $\mathcal{E}%
_{\mathrm{AD}\text{,}\pi}$ with program state $\pi$. In particular, for the
program state we compare the naive choice of the channel's Choi matrix
$\pi=\chi_{\mathcal{E}_{\mathrm{AD}}}^{\otimes N}$ (dotted lines) with the SDP
minimization over the set of generic Choi matrices $\pi=\chi^{\otimes N}$
(solid lines). Different values of $N=2,\dots,6$ and $N=20$ are shown.
}%
\label{fig:ADchoi}%
\end{figure}

In Fig.~\ref{fig:ADchoi} we study the PBT\ simulation of the amplitude damping
channel by considering the subset of program states $\pi=\chi^{\otimes N}$
which is made of tensor products of the $4\times4$ generic Choi matrices
$\chi$ (therefore satisfying $\mathrm{Tr}_{2}\chi=\openone/2$). As discussed
in previous Sec.~\ref{compressSEC}, this is equivalent to optimizing over the
Choi set $\mathcal{C}_{N}$\ and it practically reduces to the convex
optimization of the channel $\tilde{\Lambda}$ over the generic single-copy
Choi matrix $\chi$. Moreover, $\tilde{\Lambda}$ itself can be simplified, as
shown in Appendix \ref{a:pbt}, so the all operations depend polynomially on
the number $N$ of ports. This allows us to numerically explore much larger
values of $N$, even for the minimization of $C_{\diamond}$. In
Fig.~\ref{fig:ADchoi} the dotted lines correspond to the value of
$C_{\diamond}$ when the program $\pi=\chi_{\mathcal{E}_{\mathrm{AD}}}^{\otimes
N}$ is employed, where $\chi_{\mathcal{E}_{\mathrm{AD}}}$ is the channel's
Choi matrix. As Fig.~\ref{fig:ADchoi} shows, the cost $C_{\diamond}$ may be
significantly smaller with an optimal $\chi$, thus showing that the optimal
program may be different from the channel's Choi matrix, especially when $p$
is far from the two boundaries $p=0$ and $p=1$.\begin{figure}[t]
\centering
\includegraphics[width=0.95\linewidth]{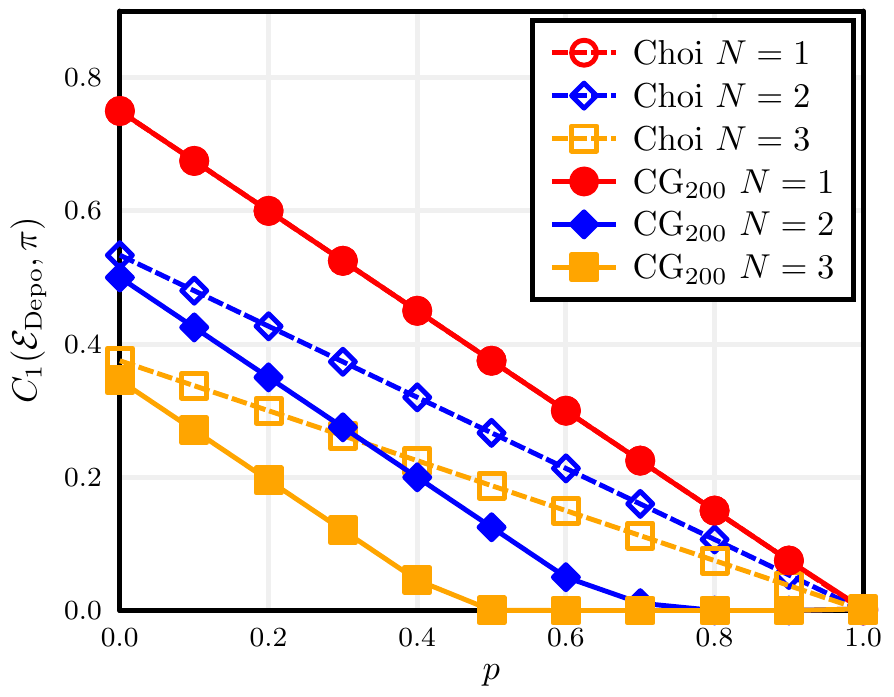} \caption{PBT
Simulation of the qubit depolarizing channel versus probability of
depolarizing $p$. Trace distance $C_{1}(\mathcal{E}_{\mathrm{dep}},\pi
)=\Vert\chi_{\mathcal{E}_{\mathrm{dep}}}-\chi_{\pi}\Vert_{1}$ between the
target channel's Choi matrix and its PBT simulation with program state $\pi$,
for different number of ports $N$. We consider $N=1,2,3$ and two kinds of
programs: copies of the channel's Choi matrix $\pi=\chi_{\mathcal{E}%
_{\mathrm{dep}}}^{\otimes N}$ and the optimal program state $\tilde{\pi}_{1}$
obtained from the minimization of $C_{1}$ via the conjugate gradient (CG)
method after 200 iterations. Note that the simulation error $C_{1}$ is maximum
for the identity channel ($p=0$) and eventually goes to zero for a finite
value of $p$ that decreases for increasing $N$. }%
\label{fig:Depo_vs_p}%
\end{figure}\begin{figure}[t]
\centering
\includegraphics[width=0.95\linewidth]{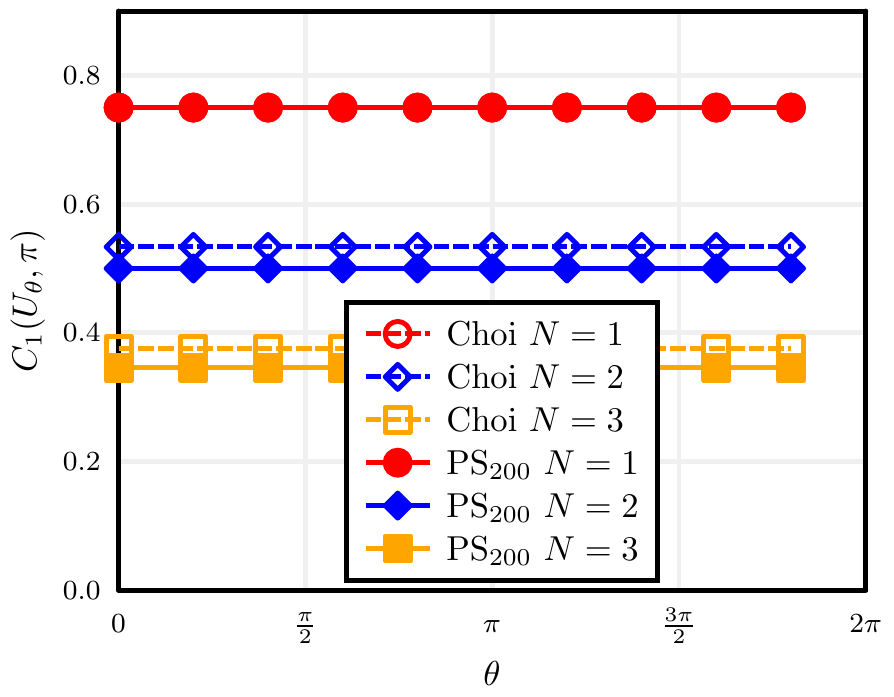} \caption{PBT Simulation
of the unitary gate $U_{\theta}=e^{i\theta X}$ for different angles $\theta$,
where $X$ is the bit-flip Pauli matrix. Trace distance $C_{1}(U_{\theta}%
,\pi)=\Vert\chi_{U_{\theta}}-\chi_{\pi}\Vert_{1}$ between the target Choi
matrix of the unitary and its PBT simulation with program state $\pi$, for
different number of ports $N$. We consider $N=1,2,3$ and two kinds of
programs: copies of the Choi matrix of the unitary $\chi_{U_{\theta}}^{\otimes
N}$ and the program state $\tilde{\pi}_{1}$ obtained from the minimization of
$C_{1}$ via the projected subgradient (PS) method after 200 iterations. }%
\label{fig:UX2}%
\end{figure}

As an another example, we consider the simulation of the depolarizing channel
defined by
\begin{equation}
\mathcal{E}_{\text{\textrm{dep}}}(\rho)=(1-p)\rho+\frac{p}{d}\openone.
\end{equation}
In Fig.~\ref{fig:Depo_vs_p} we study the performance of PBT simulation of the
depolarizing channel in terms of $p$. For $p=0$ the depolarizing channel is
equal to the identity channel, while for $p=1$ it sends all states to the
maximally mixed state. Again we compare the simulation error with program
states either made copies of the channel's Choi matrices $\chi_{\mathcal{E}%
_{\mathrm{dep}}}^{\otimes N}$ or obtained from the minimization of $C_{1}$
with the conjugate gradient method of Eq.~\eqref{FrankWolfe}, which performs
significantly better than the projected subgradient for this channel. Also for
the depolarizing channel we observe that, for any finite $N$, we obtain a
lower error by optimizing over the program states instead of the naive choice
$\chi_{\mathcal{E}_{\mathrm{dep}}}^{\otimes N}$.

Finally, in Fig.~\ref{fig:UX2} we study the PBT simulation of a unitary gate
$U_{\theta}=e^{i\theta X}$ for different values of $\theta$. Unlike the
previous non-unitary channels, in Fig.~\ref{fig:UX2} we observe a flat error
where different unitaries have the same simulation error of the identity
channel $\theta=0$. This is expected because both the trace distance and the
diamond distance are invariant under unitary transformations. In general, we
have the following.

\begin{proposition}
Given a unitary $\mathcal{U}(\rho)=U\rho U^{\dagger}$ and its PBT\ simulation
$\mathcal{U}_{\pi}$ with program $\pi$\ we may write
\begin{equation}
\min_{\pi}||\mathcal{U}-\mathcal{U}_{\pi}||_{\diamond}=\min_{\pi}%
||\mathcal{I}-\mathcal{I}_{\pi}||_{\diamond},
\end{equation}
where $\mathcal{I}_{\pi}$ is the PBT\ simulation of the identity channel.
\end{proposition}

\noindent\textit{Proof}.~~In fact, we simultaneously prove
\begin{equation}
\min_{\pi}||\mathcal{I}-\mathcal{I}_{\pi}||_{\diamond}\overset{(1)}{\leq}%
\min_{\pi}||\mathcal{U}-\mathcal{U}_{\pi}||_{\diamond}\overset{(2)}{\leq}%
\min_{\pi}||\mathcal{I}-\mathcal{I}_{\pi}||_{\diamond},
\end{equation}
where (1) comes from the fact that $||\mathcal{U}-\mathcal{U}_{\pi
}||_{\diamond}=||\mathcal{U}^{-1}\mathcal{U}-\mathcal{U}^{-1}\mathcal{U}_{\pi
}||_{\diamond}=||\mathcal{I}-\mathcal{U}^{-1}\mathcal{U}_{\pi}||_{\diamond}$
and $\mathcal{U}^{-1}\mathcal{U}_{\pi}$ is a possible PBT\ simulation of the
identity $\mathcal{I}$ with program state $\mathcal{I}\otimes(\mathcal{U}%
^{-1})^{\otimes N}(\pi)$ once $\mathcal{U}^{-1}$ is swapped with the filtering
of the ports; then (2) comes from the fact that the composition $\mathcal{U}%
\circ\mathcal{I}_{\pi}$ is a possible simulation of the unitary $\mathcal{U}$
with program state $\mathcal{I}\otimes\mathcal{U}^{\otimes N}(\pi)$ and we
have the inequality$\ ||\mathcal{U}\circ\mathcal{I}-\mathcal{U}\circ
\mathcal{I}_{\pi}||_{\diamond}\leq||\mathcal{I}-\mathcal{I}_{\pi}||_{\diamond
}$.~$\blacksquare$

%\bigskip

%\textbf{SP\ comment}: Simulation of the identity for increasing N to be
%compared with previous results.

\begin{figure}[t]
    \centering
    \includegraphics[width=0.95\linewidth]{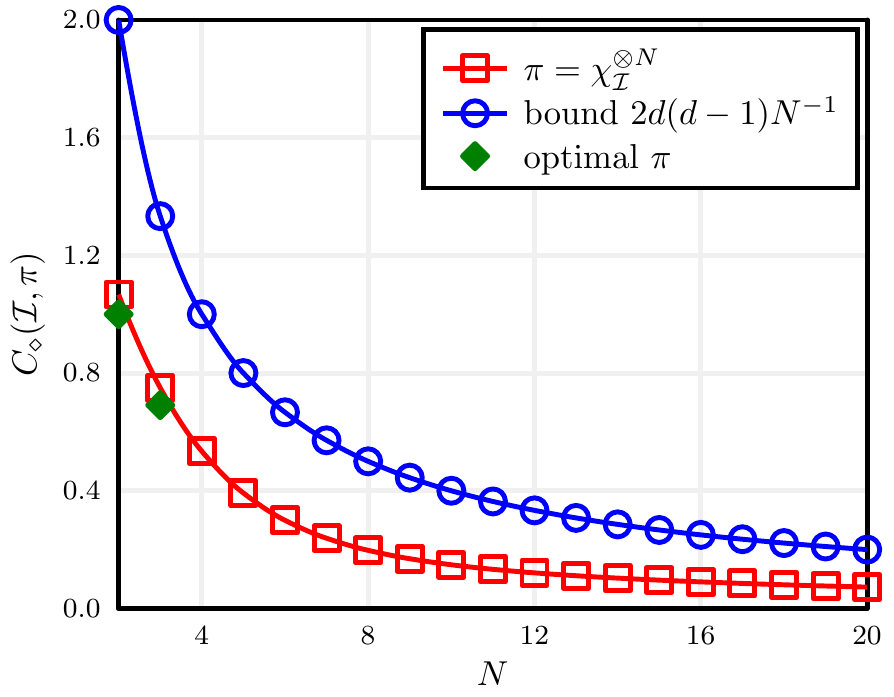}
    \caption{ PBT Simulation of the identity channel for different number of ports $N$.
        For the identity channel the optimal Choi matrix coincides with the channel's Choi
        matrix $\chi_{\mathcal I}$. The optimal $\pi$ has been obtained by minimising
        $C_\diamond$ via SDP. The upper bound corresponds to Eq.~\eqref{simerr}.
    }
    \label{fig:id}
\end{figure}

The scaling of $||\mathcal{I}-\mathcal{I}_{\pi}||_{\diamond}$ for different values of
$N$ is plotted in Fig.~\ref{fig:id} where numerical values are obtained from SDP, while
the upper bound is given by Eq.~\eqref{simerr}.

\section{Parametric quantum circuits\label{Sec8}}

We now study another design of universal quantum processor that can simulate
any target quantum channel in the asymptotic limit of an arbitrarily large
program state. This is based on a suitable reformulation of the PQCs, which
are known to simulate any quantum computation with a limited set of quantum
gates~\cite{lloyd1995almost,lloyd1996universal}.

\subsection{Basic idea}

A PQC is composed of a sequence of unitary matrices $U_{j}(\theta_{j})$, each
depending on a classical parameter $\theta$. The resulting unitary operation
is then
\begin{equation}
U(\theta)=U_{N}(\theta_{N})\dots U_{2}(\theta_{2})U_{1}(\theta_{1}%
).\label{Utheta}%
\end{equation}
A convenient choice is via $U_{j}(\theta_{j})=\exp(i\theta_{j}H_{j})$, where
each elementary gate corresponds to a Schr\"{o}dinger evolution with
Hamiltonian $H_{j}$ for a certain time interval $\theta_{j}$. For certain
choices of $H_{j}$ and suitably large $N$ the above circuit is
universal~\cite{lloyd1996universal}, namely any unitary can be obtained with
$U(\theta)$ and a suitable choice of $\theta_{j}$. The optimal parameters can
be found with numerical algorithms~\cite{khaneja2005optimal}, e.g. by minimizing
the cost function $C(\theta)=|\Tr[U_{\rm target}^\dagger U(\theta)]|$.
However, the above cost function is not convex, so the numerical algorithms
are not guaranteed to converge to the global optimum.

As a first step, we show that the task of learning the optimal parameters in a
PQC can be transformed into a convex optimization problem by using a quantum
program. This allows us to use SDP and gradient-based ML methods for finding
the global optimum solution.

\subsection{Convex reformulation}

Consider a program state $|\pi\rangle=|\theta_{1},\dots,\theta_{N}\rangle$
composed by $N$ registers $R_{j}$, each in a separable state $|\theta
_{j}\rangle$. We can transform the classical parameters in Eq.~\eqref{Utheta}
into quantum parameters via the conditional gates
\begin{equation}
\hat{U}_{j}=\exp\left(  iH_{j}\otimes\sum_{\theta_{j}}\theta_{j}|\theta
_{j}\rangle\langle\theta_{j}|\right)  ,\label{condUni}%
\end{equation}
that acts non-trivially on system and register $R_{j}$. If the parameters
$\theta_{j}$ are continuous, then we can replace the sum with an integral.
\begin{figure}[t]
\centering
\includegraphics[width=0.9\linewidth]{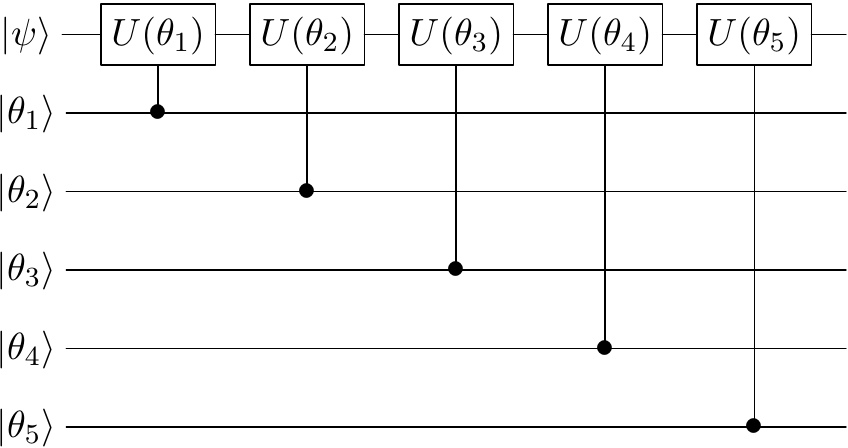} \caption{Convex
reformulation of a PQC as a coherent programmable quantum processor that
applies a sequence of conditional gates as in Eq.~\eqref{condUni} depending on
the program state $|\pi\rangle=|\theta_{1},\dots,\theta_{N}\rangle$. The
program state is not destroyed and can be reused. }%
\label{fig:PQC}%
\end{figure}With the above gates we define the parametric quantum channel
\begin{equation}
Q_{\pi}(\rho)=\mathrm{Tr}_{R}\left[  \prod_{j=1}^{N}\hat{U_{j}}\left(
\rho\otimes\pi\right)  \prod_{j=1}^{N}\hat{U_{j}}^{\dagger}\right]
,\label{PQCchan}%
\end{equation}
whose action on a generic state $|\psi\rangle$ is shown in Fig.~\ref{fig:PQC}.
For a pure separable program $|\pi\rangle=|\theta_{1},\dots,\theta_{N}\rangle
$, we obtain the standard result, i.e.,
\begin{equation}
Q_{|\theta_{1},\dots,\theta_{N}\rangle}(\rho)=U(\theta)\rho U(\theta
)^{\dagger},\label{Qchan}%
\end{equation}
where $U(\theta)$ is defined in Eq.~\eqref{Utheta}. The parametric quantum
processor $Q_{\pi}$ in Eq.~(\ref{PQCchan}) is capable of simulating any
parametric quantum channels, but it is more general, as it allows entangled
quantum parameters and also parameters in quantum superposition.

An equivalent measurement-based protocol is obtained by performing the trace
in Eq.~\eqref{Qchan} over the basis $|\theta_{1},\dots,\theta_{N}\rangle$, so
that
\begin{equation}
Q_{\pi}(\theta)=\sum_{\{\theta_{j}\}}U(\theta)\rho U(\theta)^{\dagger}%
\langle\theta_{1},\dots,\theta_{N}|\pi|\theta_{1},\dots,\theta_{N}%
\rangle,\label{Mchan}%
\end{equation}
where $U(\theta)$ is defined in Eq.~\eqref{Utheta}. In this alternative, yet
equivalent formulation, at a certain iteration $j$, the processor measures the
qubit register $R_{j}$. Depending on the measurement outcome $\theta_{j}$, the
processor then applies a different unitary $U(\theta_{j})$ on the system.
However, in this formulation the program state $|\pi\rangle$ is destroyed
after each channel use. From Eq.~\eqref{Mchan} we note that $Q_{\pi}$ depends
on $\pi$ only via the probability distribution $\langle\theta_{1},\dots
,\theta_{N}|\pi|\theta_{1},\dots,\theta_{N}\rangle$. As such any advantage in
using quantum states can only come from the capability of quantum systems to
model computationally hard probability distributions
\cite{boixo2018characterizing}.

\subsection{Universal channel simulation via PQCs}

The universality of PQCs can be employed for universal channel simulation.
Indeed, thanks to Stinespring's dilation theorem, any channel can be written
as a unitary evolution on a bigger space, where the system is paired to an
extra register $R_{0}$
\begin{equation}
\mathcal{E}(\rho_{A})=\mathrm{Tr}_{R_{0}}[U(\rho_{A}\otimes\theta
_{0})U^{\dagger}],\label{stinespring}%
\end{equation}
where $\theta_{0}$ belongs to $R_{0}$, and $U$ acts on system $A$ and register
$R_{0}$. In Ref.~\cite{lloyd1995almost} it was shown that two quantum gates
are universal for quantum computation. Specifically, given $U_{0}%
=e^{it_{0}H_{0}}$ and $U_{B}=e^{it_{1}H_{1}}$ for fixed times $t_{i}$ and
Hamiltonians $H_{j}$, it is possible to write any unitary as
\begin{equation}
U\approx\cdots U_{1}^{m_{4}}U_{0}^{m_{3}}U_{1}^{m_{2}}U_{0}^{m_{1}%
},\label{qcdecomposition}%
\end{equation}
for some integers $m_{j}$. Under suitable conditions, it was shown that with
$M=\sum_{j}m_{j}=\mathcal{O}(d^{2}\epsilon^{-d})$ it is possible to
approximate any unitary $U$ with a precision $\epsilon$. More precisely, the
conditions are the following

\begin{enumerate}
\item[i)] The Hamiltonians $H_{0}$ and $H_{1}$ are generators of the full Lie
algebra, namely $H_{0}$, $H_{1}$ and their repeated commutators generate all
the elements of su(d).

\item[ii)] The eigenvalues of $U_{0}$ and $U_{1}$ have phases that are
irrationally related to $\pi$.
\end{enumerate}

The decomposition in Eq.~\eqref{qcdecomposition} is a particular case of
Eq.~\eqref{Utheta} where $\theta_{j}$ can only take binary values $\theta
_{j}=0,1$. As such we can write the conditional gates of Eq.~\eqref{condUni}
as
\begin{equation}
\hat{U}_{j}=\exp\left(  it_{0}H_{0}\otimes|0\rangle_{j}{}_{j}\langle
0|+it_{1}H_{1}\otimes|1\rangle_{j}{}_{j}\langle1|\right)  ,\label{condAB}%
\end{equation}
for some times $t_{j}$. Channel simulation is then obtained by replacing the
unitary evolution $U$ of Eq.~\eqref{stinespring} with the approximate form in
Eq.~\eqref{qcdecomposition} and its simulation in Eq.~\eqref{Qchan}.
\begin{figure}[t]
\centering
\includegraphics[width=0.7\linewidth]{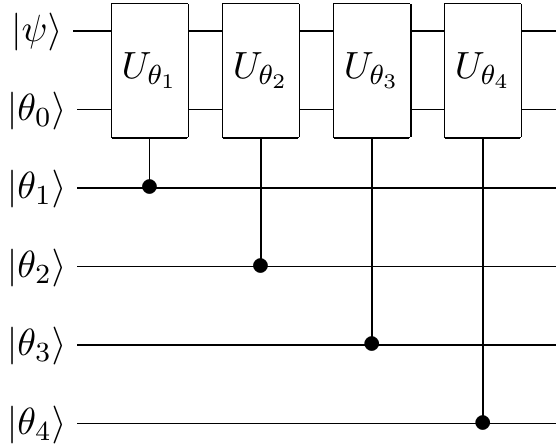} \caption{Simulation of
a quantum channel via Stinespring decomposition together with unitary
simulation as in Fig.~\ref{fig:PQC}. }%
\label{fig:PQC3}%
\end{figure}The result is illustrated in Fig.~\ref{fig:PQC3} and described by
the following channel
\begin{equation}
Q_{\pi}(\rho)=\mathrm{Tr}_{\mathbf{R}}\left[  \prod_{j=1}^{N}\hat{U_{j}%
}_{A,R_{0},R_{j}}\left(  \rho_{A}\otimes\pi\right)  \prod_{j=1}^{N}\hat{U_{j}%
}_{A,R_{0},R_{j}}^{\dagger}\right]  ,
\label{Qchan}
\end{equation}
where the program state $\pi$ is defined over $\mathbf{R}=(R_{0},\dots,R_{N})$
and each $\hat{H}_{j}$ acts on the input system $A$ and two ancillary qubits
$R_{0}$ and $R_{j}$. The decomposition of Eq.~\eqref{qcdecomposition} assures
that, with the program
\begin{equation}
|\pi\rangle=|\theta_{0}\rangle\otimes\cdots\otimes|1\rangle^{\otimes m_{2}%
}\otimes|0\rangle^{\otimes m_{1}},
\end{equation}
the product of unitaries approximates $U$ in Eq.~\eqref{stinespring} with
precision $\epsilon$. This is possible in general, provided that the program
state has dimension $\mathcal{O}(d^{2}\epsilon^{-d})$.
However, the channel \eqref{Qchan} is more general, as it allows both quantum
superposition and entanglement.

The processor map $\Lambda$ is then simply obtained as
\begin{equation}
\Lambda(\pi)=\mathrm{Tr}_{\mathbf{R}}\left[  \hat{U}_{A\mathbf{R}}\left(
\Phi_{BA}\otimes\pi_{\mathbf{R}}\right)  \hat{U}_{A\mathbf{R}}^{\dagger
}\right]  ,
\end{equation}
where
\begin{equation}
\hat{U}_{A\mathbf{R}}=\openone_{B}\otimes\prod_{j=1}^{N}\hat{U_{j}}%
_{A,R_{0},R_{j}},
\end{equation}
while the (non-trace-preserving) dual channel may be written as
\begin{equation}
\Lambda^{\ast}(X)=\langle\Phi_{BA}|\hat{U}_{A\mathbf{R}}^{\dagger}\left(
X_{BA}\otimes\openone_{\mathbf{R}}\right)  \hat{U}_{A\mathbf{R}}|\Phi
_{BA}\rangle.
\end{equation}
This channel requires $2N$ quantum gates at each iteration and can be employed
for the calculation of gradients, following Theorem~\ref{t:gradients}. When we
are interested in simulating a unitary channel $U$ via the quantum fidelity,
then following the results of Section~\ref{Sec5}, the corresponding optimal
program $\tilde{\pi}_{F}$ is simply the eigenvector $\Lambda^{\ast}[|\chi
_{U}\rangle\langle\chi_{U}|]$ with maximum eigenvalue, where $|\chi_{U}%
\rangle=\openone\otimes U|\Phi\rangle$. Note also that $\Lambda^{\ast}%
[|\chi_{U}\rangle\langle\chi_{U}|]=Z^{\dagger}Z$ where
\begin{equation}
Z=\left(  \langle\chi_{U}|_{BA}\otimes\openone_{\mathbf{R}}\right)  \hat
{U}_{A\mathbf{R}}\left(  |\Phi_{BA}\rangle\otimes\openone_{\mathbf{R}}\right)
,
\end{equation}
so the optimal program $\tilde{\pi}_{F}$ is the principal component of $Z$.
Since there are quantum algorithms for principal component analysis
\cite{lloyd2014quantum}, the optimization may be efficiently performed on a
quantum computer.

\subsection{Numerical examples}

As an example we study the simulation of an amplitude damping channel, with
Kraus operators in Eq.~\eqref{KrausAD}. A possible Stinespring dilation for
this channel is obtained with $|\theta_{0}\rangle=|0\rangle$ and
\begin{equation}
U=%
\begin{pmatrix}
1 & 0 & 0 & 0\\
0 & \sqrt{1-p} & \sqrt{p} & 0\\
0 & -\sqrt{p} & \sqrt{1-p} & 0\\
0 & 0 & 0 & 1
\end{pmatrix}
=e^{iH_{\mathrm{AD}}},
\end{equation}
where the Hamiltonian is given by
\begin{equation}
H_{\mathrm{AD}}=\frac{\arcsin(\sqrt{p})}{2}(Y\otimes X-X\otimes Y),
\end{equation}
with $X$ and $Y$ being Pauli operators. We may construct a PQC simulation by
taking
\begin{equation}
U_{0}=e^{i\alpha(Y\otimes X-X\otimes Y)},
\end{equation}
for some $\alpha$ and taking $U_{1}$ to be a different unitary that makes the
pair $U_{0}$, $U_{1}$ universal. Here we may choose $\alpha=\sqrt{2}$ and
$U_{1}=e^{iH_{1}}$ with
\begin{equation}
H_{1}=(\sqrt{2}Z+\sqrt{3}Y+\sqrt{5}X)\otimes(Y+\sqrt{2}Z).
\end{equation}
\begin{figure}[t]
\centering
\includegraphics[width=0.9\linewidth]{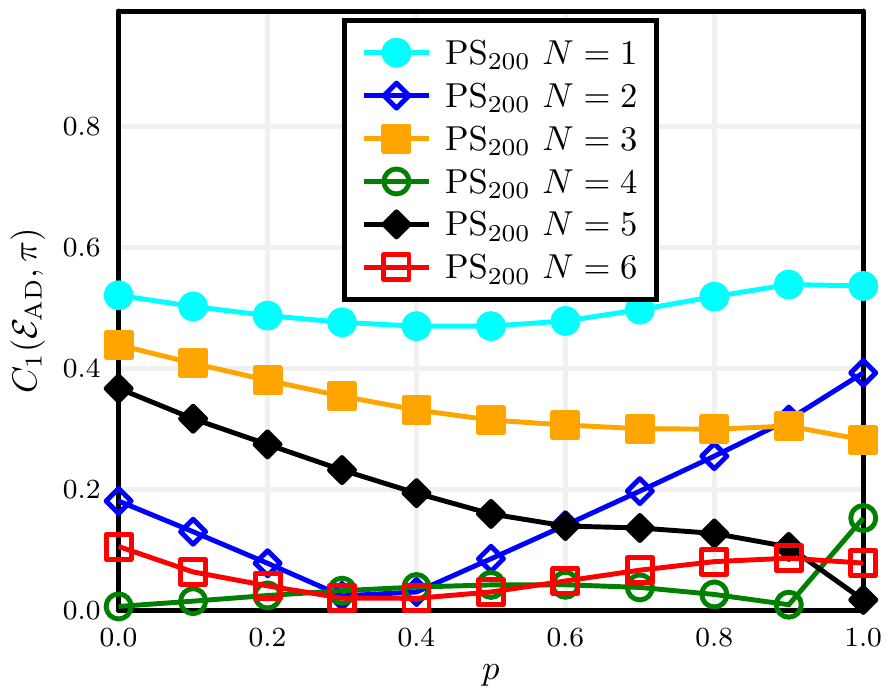} \caption{ PQC
simulation of the amplitude damping channel. Trace distance $C_{1}%
(\mathcal{E}_{\mathrm{AD}},\pi)=\Vert\chi_{\mathcal{E}_{\mathrm{AD}}}%
-\chi_{\pi}\Vert_{1}$ between the target channel's Choi matrix and its PQC
simulation with program state $\pi$, for different numbers of register qubits
$N$. The optimal program is obtained from the minimization of $C_{1}$ via the
projected subgradient (PS) method after 200 iterations. }%
\label{fig:UQC_AD_vs_p}%
\end{figure}

Results are shown in Fig.~\ref{fig:UQC_AD_vs_p}. Compared with the similar PBT
simulation of Fig.~\ref{fig:AD_vs_p}, we observe that PQC simulation displays
a non-monotonic behavior as a function of $N$. PBT with $N$ pairs requires a
register of $2N$ qubits, while PQC requires $N+1$ qubits, namely $N$ qubits
from the conditional gates and an extra one coming from Stinespring
decomposition (see Fig.~\ref{fig:PQC3}). We observe that, with a comparable
yet finite register size, PQC can outperform PBT in simulating the amplitude
damping channel. \begin{figure}[t]
\centering
\includegraphics[width=0.9\linewidth]{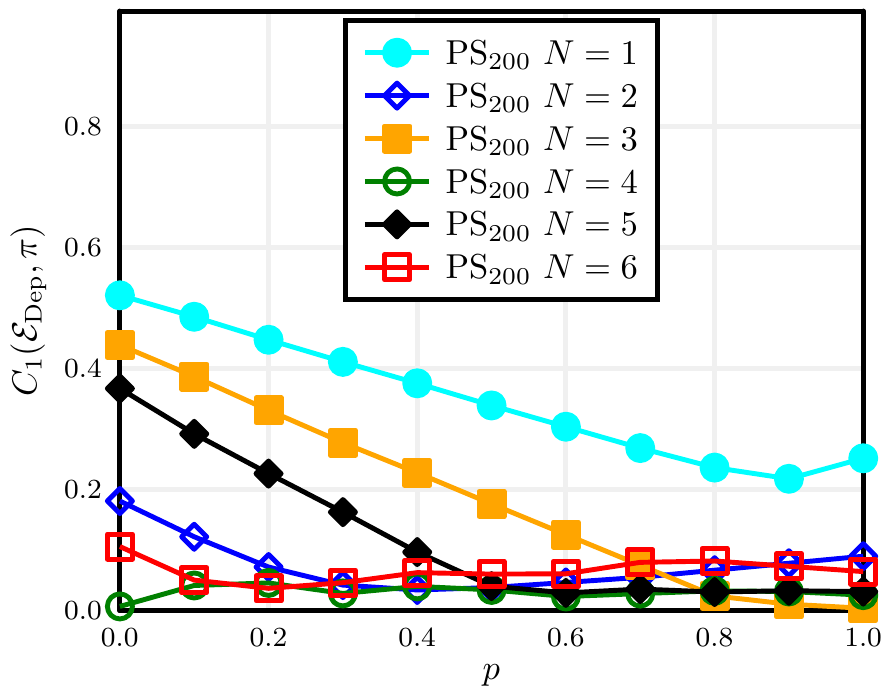} \caption{ PQC
simulation of the depolarizing channel. Trace distance $C_{1}(\mathcal{E}%
_{\mathrm{Dep}},\pi)=\Vert\chi_{\mathcal{E}_{\mathrm{Dep}}}-\chi_{\pi}%
\Vert_{1}$ between the target channel's Choi matrix and its UPQC simulation
with program state $\pi$, for different numbers of register qubits $N$. The
optimal program is obtained from the minimization of $C_{1}$ via the projected
subgradient (PS) method after 200 iterations. }%
\label{fig:UQC_Dep_vs_p}%
\end{figure}In Fig.~\ref{fig:UQC_Dep_vs_p} we also study the PQC simulation of
the depolarizing channel for different values of $p$. Although the gates
$U_{0}$ and $U_{1}$ were chosen with inspiration from the Stinespring
decomposition of the amplitude damping channel, those gates are universal and
capable of simulating other channels. Indeed, we observe in
Fig.~\ref{fig:UQC_Dep_vs_p} that a depolarizing channel is already well
simulated with $N=4$ for all values of $p$.

\section{Conclusions\label{Sec9}}

In this work we have considered a general, finite-dimensional, model of
programmable quantum processor, which is a fundamental scheme for quantum
computing and also a primitive tool for other areas of quantum information. By
introducing suitable cost functions, based on the diamond distance, trace
distance and quantum fidelity, we have shown how to characterize the optimal
performance of this processor in the simulation of an arbitrary quantum gate
or channel. In fact, we have shown that the minimization of these cost
functions is a convex optimization problem that can always be solved.

In particular, by minimizing the diamond distance via SDP, we can always
determine the optimal program state for the simulation of an arbitrary
channel. Alternatively, we may minimize the simpler but larger cost functions
in terms of trace distance and quantum fidelity via gradient-based ML methods,
so as to provide a very good approximation of the optimal program state. This
other approach can also provide closed analytical solutions, as is the case
for the simulation of arbitrary unitaries, for which the minimization of the
fidelity cost function corresponds to compute an eigenvector.

We have then applied our results to various designs of programmable quantum
processor, from a shallow teleportation-based scheme to deeper and
asymptotically-universal designs that are based on PBT and PQCs. We have
explicitly benchmarked the performances of these quantum processors by
considering the simulation of unitary gates, depolarizing and amplitude
damping channels, showing that the optimal program states may differ from the
naive choice based on the Choi matrix of the target channel.

An immediate application of our work may be the development of a model of
\textquotedblleft programmable\textquotedblright\ blind quantum computation,
where a client has an input state to be processed by a quantum server which is
equipped with a programmable quantum processor. The client classically informs
the server about what type of computation it needs (e.g., some specified
quantum algorithm) and the server generates an optimal program state which
closely approximates the overall quantum channel to be applied to the input.
The server then accepts the input from the client, processes it, and returns
the output together with the value of a cost function quantifying how close
the computation was with respect to the client's request.

Our results may also be useful in areas beyond quantum computing, wherever
channel simulation is a basic problem. For instance this is the case of
quantum communication, for the derivation of quantum and private communication
capacities, and quantum metrology and hypothesis testing, for the
simplification of adaptive protocols and the analysis of the ultimate
discrimination and estimation performance with quantum channels.

\bigskip

\noindent\textbf{Acknowledgements.~} L.B. acknowledges support by
the program ``Rita Levi Montalcini'' for your young researchers.
S.P. acknowledges support by the EPSRC via the `UK Quantum
Communications Hub' (EP/M013472/1) and the European Union via the
project `Continuous Variable Quantum Communications' (CiViQ, no
820466). S.P. would like to thank George Zweig, Jacques Carolan,
John Watrous, and Dirk Englund for discussions and feedback.

\appendix

\section{Matrix calculus\label{a:calculus}}

\subsection{Matrix differentiation\label{matrixDIFF}}

For a general overview of these techniques, the reader may consult
Ref.~\cite{stickel1987frechet}. Thanks to Cauchy's theorem, a matrix function
can be written as
\begin{equation}
f(A)=\frac{1}{2\pi i}\int_{\Gamma}d\lambda\,f(\lambda)(\lambda\openone-A)^{-1}%
~.\label{e.f}%
\end{equation}
For the same reason
\begin{equation}
f^{\prime}(A)=\frac{1}{2\pi i}\int_{\Gamma}d\lambda\,f(\lambda)(\lambda
\openone-A)^{-2}~.\label{e.fp}%
\end{equation}
Applying a basic rule of matrix differentiation, $d(A^{-1})=-A^{-1}(dA)A^{-1}$
we obtain
\begin{equation}
df(A)=\frac{1}{2\pi i}\int_{\Gamma}d\lambda\,f(\lambda)(\lambda
\openone-A)^{-1}dA(\lambda\openone-A)^{-1}~.\label{e.df}%
\end{equation}
Clearly, $df(A)=f^{\prime}(A)dA$ only when $[A,dA]=0$. In general $df(A)$ is a
superoperator that depends on $A$ and is applied to $dA$. The explicit form is
easily computed using the eigenvalue decomposition or other techniques
\cite{stickel1987frechet}. Note that in some cases the expressions are simple.
Indeed, using the cyclic invariance of the trace, we have
\begin{equation}
d\text{\textrm{Tr}}[f(A)]=\text{\textrm{Tr}}[f^{\prime}(A)dA],\label{e:nablag}%
\end{equation}
while in general $d$\textrm{Tr}$[Bf(A)]\neq$\textrm{Tr}$[Bf^{\prime}(A)dA]$.

\subsection{Differential of the quantum fidelity\label{fidelityDIFF}}

The quantum fidelity can be expanded as
\begin{align}
F(X,Y) &  =\text{\textrm{Tr}}\sqrt{\sqrt{X}Y\sqrt{X}}\\
&  =\frac{1}{2\pi i}\int_{\Gamma}d\lambda\,\sqrt{\lambda}\text{\textrm{Tr}%
}[(\lambda\openone-\sqrt{X}Y\sqrt{X})^{-1}]~,\nonumber
\end{align}
where in the second line we have applied Eq.~\eqref{e.f}. Taking the
differential with respect to $Y$ and using the cyclic property of the trace we
get%
\begin{align}
d_{Y}F &  :=F(X,Y+dY)-F(X,Y)\nonumber\\
&  \overset{(1)}{=}\frac{1}{2\pi i}\int_{\Gamma}d\lambda\,\sqrt{\lambda
}\text{\textrm{Tr}}[(\lambda\openone-\sqrt{X}Y\sqrt{X})^{-2}\sqrt{X}dY\sqrt
{X}]\nonumber\\
&  \overset{(2)}{=}\frac{1}{2}\text{\textrm{Tr}}[(\sqrt{X}Y\sqrt{X}%
)^{-\frac{1}{2}}\sqrt{X}dY\sqrt{X}]\nonumber\\
&  \overset{(3)}{=}\frac{1}{2}\text{\textrm{Tr}}[\sqrt{X}(\sqrt{X}Y\sqrt
{X})^{-\frac{1}{2}}\sqrt{X}\;dY]~,\label{dFdY}%
\end{align}
where in (1) we use Eq.~\eqref{e.df} and the cyclic property of the trace; in
(2) we use Eq.~\eqref{e.fp} with $f(\lambda)=\sqrt{\lambda}$, so $f^{\prime
}(\lambda)=\frac{1}{2}\lambda^{-1/2}$; and in (3) we use the cyclic property
of the trace. See also Lemma 11 in \cite{coutts2018certifying}.

\subsection{Differential of the trace distance\label{traceDIFF}}

The trace norm for a Hermitian operator $X$ is defined as%
\begin{align}
t(X)  &  =\Vert X\Vert_{1}:=\text{\textrm{Tr}}\sqrt{X^{\dagger}X}%
=\text{\textrm{Tr}}[\sqrt{X^{2}}]\nonumber\\
&  =\frac{1}{2\pi i}\int_{\Gamma}d\lambda\,\sqrt{\lambda}\text{\textrm{Tr}%
}[(\lambda\openone-XX)^{-1}]~,
\end{align}
where in the second line we applied Eq.~\eqref{e.f}. From the spectral
decomposition $X=U\lambda U^{\dagger}$ we find $t(X)=\sum_{j} |\lambda_{j}|$,
so the trace distance reduces to the absolute value function for
one-dimensional Hilbert spaces. The absolute value function $|\lambda|$ is
differentiable at every points, except $\lambda=0$. Therefore, for any
$\lambda\neq0$ the subgradient of the absolute value function is made by its
gradient, namely
\begin{equation}
\partial|\lambda| = \{ \mathrm{sign}(\lambda) \} ~~~ \mathrm{for} ~~
\lambda\neq0~. \label{dabsvaln}%
\end{equation}
For $\lambda=0$ we can use the definition \eqref{subgradient} to write
\begin{equation}
\partial|\lambda|_{\lambda=0} = \{z: |\sigma|\ge z \sigma\mathrm{~~ for~
all~}\sigma\}~,
\end{equation}
which is true iff $-1\le z\le1$. Therefore,
\begin{equation}
\partial|\lambda|_{\lambda=0} = [-1,1]~.
\end{equation}
The sign function in \eqref{dabsvaln} can be extended to $\lambda=0$ in
multiple ways (common choices are $\mathrm{sign}(0)=-1,0,1$). From the above
equation, it appears that for any extension of the sign function, provided
that $\mathrm{sign}(0)\in[-1,1]$ we may write the general form
\begin{equation}
\mathrm{sign}(\lambda) \in\partial|\lambda|~, \label{signinabs}%
\end{equation}
which is true for any value of $\lambda$.

With the same spirit we extend the above argument to any matrix dimension,
starting from the case where $X$ is an invertible operator (no zero
eigenvalues). Taking the differential with respect to $X$ we find%
\begin{align}
dt(X) &  :=t(X+dX)-t(X)=\nonumber\\
&  \overset{(1)}{=}\frac{1}{2\pi i}\int_{\Gamma}d\lambda\,\sqrt{\lambda
}\text{\textrm{Tr}}[(\lambda\openone-X^{2})^{-2}(X(dX)+(dX)X)]\nonumber\\
&  \overset{(2)}{=}\frac{1}{2}\text{\textrm{Tr}}[(X^{2})^{-\frac{1}{2}%
}(X(dX)+(dX)X)]\nonumber\\
&  \overset{(3)}{=}\text{\textrm{Tr}}[(X^{2})^{-\frac{1}{2}}X\;(dX)]
\end{align}
where in (1) we use Eq.~\eqref{e.df}, the cyclic property of the trace and the
identity $dX^{2}=X(dX)+(dX)X$; in (2) we use Eq.~\eqref{e.fp} with
$f(\lambda)=\sqrt{\lambda}$, so $f^{\prime}(\lambda)=\frac{1}{2}\lambda
^{-1/2}$; and in (3) we use the cyclic property of the trace and the
commutation of $X$ and $\sqrt{X^{2}}$. Let
\begin{equation}
X=\sum_{k}\lambda_{k}P_{k}~,
\end{equation}
be the eigenvalue decomposition of $X$ with eigenvalues $\lambda_{k}$ and
eigenprojectors $P_{k}$. For non-zero eigenvalues we may write
\begin{equation}
(X^{2})^{-\frac{1}{2}}X=\sum_{k}\mathrm{sign}(\lambda_{k})P_{k}=:\mathrm{sign}%
(X)~,
\end{equation}
and accordingly
\begin{align}
dt(X) &  :=\Vert X+dX\Vert_{1}-\Vert X\Vert_{1}\nonumber\label{dtdX}\\
&  =\sum_{k}\mathrm{sign}(\lambda_{k})\text{\textrm{Tr}}[P_{k}\;dX]~.
\end{align}
Therefore, for invertible operators we may write
\[
\partial t(X)=\{\nabla t(X)\}~,\nabla t(X)=\mathrm{sign}(X)~.
\]
We now consider the general case where some eigenvalues of $X$ may be zero. We
do this by generalizing Eq.~\eqref{signinabs}, namely we show that even if
$\partial t(X)$ may contain multiple elements, it is always true that $\nabla
t\in\partial t$, provided that $-\openone\leq\mathrm{sign}(X)\leq\openone$.
Following \eqref{subgradient} we may write, for fixed $X$ and arbitrary $Y$,%
\begin{gather}
t(Y)-t(X)-\text{\textrm{Tr}}[\nabla t(X)(Y-X)]\nonumber\\
\overset{(1)}{=}t(Y)-t(X)-\text{\textrm{Tr}}[\nabla t(X)Y]+t(X)\nonumber\\
\overset{(2)}{\geq}t(Y)-\text{\textrm{Tr}}[Y]=\sum_{j}(|\lambda_{j}%
|-\lambda_{j})\geq0~,\label{tsubdiffineq}%
\end{gather}
where in (1) we use the property $\Vert X\Vert_{1}=\Tr[{\rm sign}(X)X]$ and in
(2) we use the assumption $-\openone\leq\mathrm{sign}(X)\leq\openone$.
%$z\le|z|$;
%in (3) we use the definition \cite{coutts2018certifying}
%\begin{equation}
%t(Y) = \|Y\|_1=\sup_X \{|\Tr[XY]| : \|X\|_\infty=1\} \ge |\Tr[XY]|~,
%\end{equation}
%for any $X$ with $\|X\|_{\infty}=1$
%$\|{\rm sign}(X)\|_2\le1$,
From the definition of the subgradient \eqref{subgradient}, the above equation
shows that $\mathrm{sign}(X)\in\partial t(X)$, so we may always use $\nabla
t(X)=\mathrm{sign}(X)$ in the projected subgradient algorithm
\eqref{projsubgrad}.

\section{Smoothing techniques\label{SmoothAPP}}

\subsection{Stochastic smoothing\label{a:stocsmooth} }

The conjugate gradient algorithm converges after $\mathcal{O}(c/\epsilon)$
steps \cite{jaggi2011convex,jaggi2013revisiting}, where $\epsilon$ is the
desired precision and $c$ is a curvature constant that depends on the
function. However, it is known that $c$ could diverge for non-smooth
functions. This is the case for the trace norm, as shown in Example 0.1 in
\cite{ravi2017deterministic}.

A general solution, valid for arbitrary functions, is via stochastic smoothing
\cite{yousefian2012stochastic}. In this approach the non-smooth function
$C(\pi)$ is replaced by the average
\begin{equation}
C_{\eta}(\pi)=\mathbb{E}_{\sigma}[C(\pi+\eta\sigma)]~.
\end{equation}
where $\sigma$ is such that $\Vert\sigma\Vert_\infty\leq1$. If $|C(x)-C(y)|\leq
M\Vert x-y\Vert_{\infty}$, then
\begin{equation}
C(\pi)\leq C_{\eta}(\pi)\leq C(\pi)+M\eta~,
\end{equation}
so that $C_{\eta}(\pi)$ provides a good approximation for $C(\pi)$. Moreover,
$C_{\eta}$ is differentiable at any point, so we may apply the conjugate
gradient algorithm. A modified conjugate gradient algorithm with adaptive
stochastic approximation was presented in Ref.~\cite{lan2013complexity}, At
each iteration $k$ the algorithm reads
\[%
\begin{array}
[c]{l}%
1)~\text{\textrm{Sample some operators }}\sigma_{1},\dots,\sigma_{k},\\
2)~\text{\textrm{Evaluate }}\bar{g}_{k}=\frac{1}{k}\sum_{j=1}^{k}g(\pi
_{k}+\eta_{k}\sigma_{j})\text{\textrm{ for }}\eta_{k}\propto k^{-1/2},\\
3)~\text{\textrm{Find the smallest eigenvalue }}\left\vert \sigma
_{k}\right\rangle \text{\textrm{ of~}}\bar{g}_{k},\\
4)~\pi_{k+1}=\frac{k}{k+2}\pi_{k}+\frac{2}{k+2}\left\vert \sigma
_{k}\right\rangle \left\langle \sigma_{k}\right\vert .
\end{array}
\]
where $g$ denotes any element of the subgradient $\partial C$. The above
algorithm converges after $\mathcal{O}(\epsilon^{2})$ iterations. Since
Eqs.~\eqref{fidelitygrad} and \eqref{traceDproof} provide an element of the
subgradient, the above algorithm can be applied to both fidelity and trace
distance. However, this algorithm requires $k$ evaluation of the subgradient
to perform the averages, so it may be impractical when the number of
iterations get larger. In the following we study an alternative that does not
require any average.

\subsection{Nesterov's smoothing\label{a:smoothtrace}}

An alternative smoothing scheme is based on Nesterov's dual
formulation~\cite{nesterov2005smooth}. Suppose that the non-smooth objective
function $f$ admits a dual representation as follows
\begin{equation}
f(x)=\sup_{y}[\langle x,y\rangle-g(y)],
\end{equation}
for some inner product $\langle\cdot,\cdot\rangle$. Nesterov's approximation
consists in adding a strongly convex function $d$ to the dual
\begin{equation}
f_{\mu}(x)=\sum_{y}[\langle x,y\rangle-g(y)-\mu d(y)].
\end{equation}
The resulting $\mu$-approximation is smooth and satisfies
\begin{equation}
f_{\mu}(x)\leq f(x)\leq f_{\mu}(x)+\mu\sup_{y}d(y).\label{fmuineq}%
\end{equation}

The trace norm admits the dual representation~\cite{watrous2004advanced}
\begin{equation}
t(X)=\Vert X\Vert_{1}=\sup_{\Vert Y\Vert_\infty\leq1}\langle Y,X\rangle,
\end{equation}
where $\langle Y,X\rangle$ is the Hilbert Schmidt product. This can be
regularized with any strongly convex function $d$. A convenient choice
\cite{liu2013tensor} that enables an analytic solution is via $d(X)=\frac
{1}{2}\Vert X\Vert_{2}^{2}:=\frac{1}{2}\langle X,X\rangle$ so
\begin{equation}
t_{\mu}(X)=\max_{\Vert Y\Vert_\infty\leq1}\left[  \langle Y,X\rangle-\frac{\mu}%
{2}\Vert Y\Vert_{2}^{2}\right]  .\label{tmu}%
\end{equation}
This function is smooth and its gradient is given by \cite{liu2013tensor}%
\begin{align*}
\nabla t_{\mu}(X) &  =\argmax_{\Vert Y\Vert_\infty\leq1}\left[  \langle
Y,X\rangle-\frac{\mu}{2}\Vert Y\Vert_{2}^{2}\right]  \\
&  =\argmin_{\Vert Y\Vert_\infty\leq1}\Vert\mu Y-X\Vert_{2}^{2}=U\Sigma_{\mu
}V^{\dagger},
\end{align*}
where $X=U\Sigma V^{\dagger}$ is the singular value decomposition of $X$ and
$\Sigma_{\mu}$ is a diagonal matrix with diagonal entries $(\Sigma_{\mu}%
)_{i}=\min\{\Sigma_{i}/\mu,1\}$. Plugging this into Eq.~\eqref{tmu} we get
\begin{equation}
t_{\mu}(X)=\Tr\left[  \Sigma_{\mu}\left(  \Sigma-\frac{\mu}{2}\Sigma_{\mu
}\right)  \right]  .\label{e.tmu}%
\end{equation}
For a diagonalizable matrix $X$ with spectral decomposition $X=U\lambda
U^{\dagger}$, the singular value decomposition is obtained with $\Sigma
=|\lambda|$ and $V=U\mathrm{sign}(\lambda)$. Inserting these expressions in
\eqref{e.tmu} we find
\begin{equation}
t_{\mu}(X)=\sum_{j}h_{\mu}(\lambda_{j})=\Tr[h_\mu(X)],
\end{equation}
where $h_{\mu}$ is the so called Huber penalty function
\begin{equation}
h_{\mu}(x)=%
\begin{cases}
\frac{x^{2}}{2\mu} & \mathrm{~if~}|x|<\mu,\\
|x|-\frac{\mu}{2} & \mathrm{~if~}|x|\geq\mu.
\end{cases}
\end{equation}
The gradient $\nabla t_{\mu}$ is then $h_{\mu}^{\prime}(X)\equiv Uh^{\prime
}(\lambda)U^{\dagger}$, where
\begin{equation}
h_{\mu}^{\prime}(x)=%
\begin{cases}
\frac{x}{\mu} & \mathrm{~if~}|x|<\mu,\\
\mathrm{sign}(x) & \mathrm{~if~}|x|\geq\mu.
\end{cases}
\end{equation}

We find then that
via the smooth trace norm $t_{\mu}$ we can define the smooth trace distance of
Eq.~\eqref{Dmu} that is differentiable at every point
\begin{equation}
C_{\mu}(\pi)=\Tr\left[  h_{\mu}\left(  \chi_{\pi}-\chi_{\mathcal{E}}\right)
\right]  .
\end{equation}
Thanks to the inequalities in \eqref{fmuineq}, the smooth trace distance
bounds the cost $C_{1}$ as
\begin{equation}
C_{\mu}(\pi)\leq C_{1}(\pi)\leq C_{\mu}(\pi)+\frac{\mu d}{2},
\end{equation}
where we employed the identity $\sup_{\Vert Y\Vert_\infty\leq1}\Vert Y\Vert^2_{2}\le d$ to
get the upper bound. Moreover, we find the following

\begin{lemma}
The smooth trace distance, defined in Eq.~\eqref{Dmu}, is a convex function of
$\pi$.
\end{lemma}

\noindent\textit{Proof.}~~From the definition and Eq.~\eqref{tmu} we find%
\begin{align}
C_{\mu}(\pi)  & =t_{\mu}\left[  \Lambda(\pi)-\chi_{\mathcal{E}}\right]
\nonumber\\
& =\max_{\Vert Y\Vert_\infty\leq1}\left[  \langle Y,\Lambda(\pi)-\chi_{\mathcal{E}%
}\rangle-\frac{\mu}{2}\Vert Y\Vert_{2}^{2}\right]  .
\end{align}
Now for $\bar{\pi}=p\pi_{1}+(1-p)\pi_{2}$ linearity implies $f(\bar{\pi
}):=\langle Y,\Lambda(\bar{\pi})-\chi_{\mathcal{E}}\rangle=pf(\pi
_{1})+(1-p)f(\pi_{2})$. Therefore%
\begin{align}
C_{\mu}(\bar{\pi})  & =\max_{\Vert Y\Vert_\infty\leq1}\left[  pf(\pi_{1}%
)+(1-p)f(\pi_{2})-\frac{\mu}{2}\Vert Y\Vert_{2}^{2}\right]  \nonumber\\
& \leq p\max_{\Vert Y\Vert_\infty\leq1}\left[  \langle Y,\Lambda(\pi_{1}%
)-\chi_{\mathcal{E}}\rangle-\frac{\mu}{2}\Vert Y\Vert_{2}^{2}\right]
\nonumber\\
& +(1-p)\max_{\Vert Z\Vert_\infty\leq1}\left[  \langle Z,\Lambda(\pi_{2}%
)-\chi_{\mathcal{E}}\rangle-\frac{\mu}{2}\Vert Z\Vert_{2}^{2}\right]
\nonumber\\
& =pC_{\mu}(\pi_{1})+(1-p)C_{\mu}(\pi_{2}),
\end{align}
showing the convexity.~$\blacksquare$

Then, using the definitions from \cite{nesterov2005smooth}, the following
theorem bounds on the growth of the gradient

\begin{theorem}
The gradient of the smooth trace norm is Lipschitz continuous
%, namely
%\begin{equation}
%\Vert\nabla C_{\mu}(\pi)-\nabla C_{\mu}(\pi^{\prime})\Vert_{1}\leq L\Vert
%\pi-\pi^{\prime}\Vert,
%\end{equation}
with Lipschitz constant
\begin{equation}
L=\frac{d}{\mu}.
\end{equation}
\end{theorem}

In particular, being the gradient Lipschitz continuous, the smooth trace norm
satisfies the following inequality for any state $\pi,\sigma$
\begin{equation}
C_{\mu}(\sigma)\leq C_{\mu}(\pi)+\langle\nabla C_{\mu}(\pi),\sigma-\pi
\rangle+\frac{L}{2}\Vert\sigma-\pi\Vert_2^{2}.
\end{equation}
\textit{Proof.}~~Given the linearity of the quantum channel $\Lambda$, we can
apply theorem 1 from \cite{nesterov2005smooth} to find
\begin{equation}
L=\frac{1}{\mu}\sup_{\Vert x\Vert_2=1,\Vert y\Vert_2=1}\langle y,\Lambda
(x)\rangle.
\end{equation}
Since all eigenvalues of $y$ are smaller or equal to 1, we can write $y\leq1$
and as such
\begin{equation}
L\leq\frac{1}{\mu}\sup_{\Vert x\Vert_2=1}\Tr[\Lambda(x)]=\frac{1}{\mu}%
\sup_{\Vert x\Vert_2=1}\Tr[x]\leq\frac{d}{\mu}.~
\end{equation}
\hfill$\blacksquare$

\section{PBT reduced channel\label{a:pbt}}

Here we provide an explicit expression for the reduced map $\tilde{\Lambda}%
$\ of Eq.~(\ref{e:LambdaChoi}) in the case of qubits. For $d=2$ we can rewrite
PBT in a language that can be more easily formulated from representations of
SU(2). For simplicity of notation, here we do not use bold letters for
vectorial quantities.

Let us modify the POVM in Eq.~\eqref{PBTchan} as
\begin{align}
\tilde{\Pi}_{i} &  =\sigma_{AC}^{-1/2}\Psi_{A_{i}C}^{-}{\sigma}_{AC}^{-1/2},\\
\sigma_{AC} &  =\sum_{i=1}^{N}\Psi_{A_{i}C}^{-},\\
\Pi_{i} &  =\tilde{\Pi}_{i}+\Delta,\\
\Delta &  =\frac{1}{N}\left(  \openone-\sum_{j}\tilde{\Pi}_{j}\right)
,\label{DeltaOp}%
\end{align}
where $|\Psi^{-}\rangle=(|01\rangle-|10\rangle)/\sqrt{2}$ is a singlet state.
%We can separate the projectors in
%\eqref{PBTchan} to write
%\begin{equation}
%   \mathcal P_\pi(\rho) = \sum_{ki} E_{ki} (\rho\otimes\pi) E_{ki}^\dagger +
%   \sum_{ki} D_{ki} (\rho\otimes\pi) D_{ki}^\dagger~,
%\end{equation}
%where
%\begin{align}
%   E_{ik} &= \langle{e^{(i)}_k}| \sqrt{\tilde{\Pi}_i}\otimes \openone_{B}~,
%                &
%   D_{ik} &= \langle{e^{(i)}_k}| \sqrt{\Delta}\otimes \openone_{B}~.
%\end{align}
%Alternatively, we may write
%\begin{align}
%   E_{ik\ell} &= \bra{\Psi^-_{A_i,C},a_k^{(i)},b_\ell^{(i)}} \sigma_{AC}^{-1/2}~,
%   \label{eop}\\
%   D_{iz\ell} &= N^{-1/2}\bra{\Phi^0_z,b_\ell^{(i)}}~,
%   \label{dop}
%\end{align}
%where $\ket{a_k^{(i)}}$ and $\ket{b_\ell^{(i)}}$ and form a basis of $\bar A_i$ and
%$\bar B_i$ and $\ket{\Phi^0_z}$ are the eigenvectors of $\sigma_{AC}$ with zero
%eigenvalues.
For $\pi=\chi^{\otimes n}$ the quantum channel is simplified. In fact, since
$\Tr_{B}\chi=\openone/2$, we may write
\begin{align}
\mathcal{P}_{\pi} &  =\sum_{i=1}^{N}\frac{1}{2^{N-1}}\mathrm{Tr}_{AC}\left[
\sqrt{\Pi_{i}}\left(  \rho_{C}\otimes\chi_{A_{i}B}\otimes\openone_{\bar{A}%
_{i}}\right)  \sqrt{\Pi_{i}}\right]  \nonumber\\
&  =\sum_{\ell}K_{\ell}^{0}(\rho_{C}\otimes\chi)K_{\ell}^{0}{}^{\dagger}%
+\sum_{\ell^{\prime}}K_{\ell}^{1}(\rho_{C}\otimes\chi)K_{\ell}^{1}{}^{\dagger
},\label{PBTasd}%
\end{align}
where $\ell$ and $\ell^{\prime}$ are multi-indices and, in defining the Kraus
operators, we have separated the contributions from $\tilde{\Pi}_{i}$ and
$\Delta$ (see below).

In order to express these operators, we write
\begin{equation}
|\psi_{CA_{i}}^{-}\rangle\!\langle\psi_{CA_{i}}^{-}|=\frac{\openone-\vec
{\sigma}_{C}\cdot\vec{\sigma}_{A_{i}}}{4},
\end{equation}
so that%
\begin{align}
\sigma_{AC}  & =\sum_{i=1}^{N}|\psi_{CA_{i}}^{-}\rangle\!\langle\psi_{CA_{i}%
}^{-}|=\frac{N}{4}-\vec{S}_{C}\cdot\vec{S}_{A}\nonumber\\
& =\frac{N}{4}-\frac{\vec{S}_{\text{\textrm{tot}}}^{2}-\vec{S}_{C}^{2}-\vec
{S}_{A}^{2}}{2},
\end{align}
where $\vec{S}=\vec{\sigma}/2$ is a vector of spin operators, $\vec{S}%
_{A}=\sum_{j}\vec{S}_{A_{j}}$ and $\vec{S}_{\mathrm{tot}}=\vec{S}_{C}+\vec
{S}_{A}$. The eigenvalues of $\sigma_{AC}$ are then obtained from the
eigenvalues of the three commuting Casimir operators
\begin{equation}
\lambda(s_{A})=\frac{N}{4}-\frac{S_{\mathrm{tot}}(S_{\mathrm{tot}}%
+1)-s_{A}(s_{A}+1)-3/4}{2}~,
\end{equation}
where $S_{\mathrm{tot}}=s_{A}\pm1/2$.

Substituting the definition of $S_{\mathrm{tot}}$, we find two classes of
eigenvalues
\begin{equation}
\lambda^{+}(s_{A})=\frac{N-2s_{A}}{4}~,\lambda^{-}(s_{A})=\frac{N+2s_{A}+2}%
{4}~,\label{eigv2}%
\end{equation}
with corresponding eigenvectors
\begin{equation}
|\pm,s_{A},M,\alpha\rangle=\sum_{k,m}\Gamma_{s_{A}\pm\frac{1}{2},s_{A}%
}^{M,m,k}|k\rangle_{C}|s_{A},m,\alpha\rangle_{A}~,\label{esigma}%
\end{equation}
where $-\frac{N+1}{2}\leq M\leq\frac{N+1}{2}$, $\alpha=1,\dots,g^{[N]}(s)$
describes the degeneracy, $g^{[N]}(s)$ is the size of the degenerate subspace,
and
\begin{equation}
\Gamma_{S,s}^{M,m,k}=\langle S,M;s,1/2|1/2,1/2-k;s,m\rangle\label{CG}%
\end{equation}
are Clebsch-Gordan coefficients.

Note that the Clebsch-Gordan coefficients define a unitary transformation
between the two bases $|s_{1},m_{1};s_{2};m_{2}\rangle$ and $|S,M;s_{1}%
,s_{2}\rangle$. From the orthogonality relations of these coefficients we find
the equalities%
\begin{align}
\sum_{S,M}\Gamma_{S,s}^{M,m,i}\Gamma_{S,s}^{M,m^{\prime},i^{\prime}}  &
=\delta_{i,i^{\prime}}\delta_{m,m^{\prime}},\\
\sum_{m,i}\Gamma_{S,s}^{M,m,i}\Gamma_{S^{\prime},s}^{M^{\prime},m,i}  &
=\delta_{M,M^{\prime}}\delta(S,S^{\prime},s),
\end{align}
where $\delta(S,S^{\prime},s)=1$ iff $S=S^{\prime}$ and $|s-1/2|\leq S\leq
s+1/2$. The eigenvalues in Eq.~\eqref{eigv2} are zero iff $S_{\mathrm{tot}%
}=S_{A}+1/2$ and $S_{A}=N/2$. These eigenvalues have degeneracy
$2S_{\mathrm{tot}}+1=N+2$ and the corresponding eigenvectors are
\begin{equation}
|\perp,M,\alpha\rangle=|+,N/2,M,\alpha\rangle~.
\end{equation}
Thus, the operator $\Delta$ from Eq.~\eqref{DeltaOp} may be written as
\begin{equation}
\Delta=\frac{1}{N}\sum_{M=-\frac{N+1}{2}}^{\frac{N+1}{2}}\sum_{\alpha}%
|\perp,M,\alpha\rangle\!\langle\perp,M,\alpha|~.
\end{equation}

%and from the above we can write \eqref{dop} as
%\begin{equation}
%   D_{i,M_A,\alpha,s_{\bar B_i}, m_{\bar B_i},\beta_i} =
%   \frac1{\sqrt N}\bra{\perp M,\alpha} \otimes \bra{s_{\bar B_i},m_{\bar B_i},\beta_i}~,
%   \label{Dexp}
%\end{equation}
%where $k=(M_A,\alpha,s_{\bar B_i}, m_{\bar B_i},\beta_i)$.
%Similarly we may write \eqref{eop} as
%\begin{equation}
%   \sqrt{\tilde{\Pi}_i} = \ket{\psi^-_{A_iC}}\!\!\bra{\psi^-_{A_iC}}\otimes
%   \sum_{s_{\bar A_i},m_{\bar A_i},\alpha_i}
%   \frac{c(s_{\bar A_i})}{2^{\frac{N-1}2}}
%   \ket{s_{\bar A_i},m_{\bar A_i},\alpha_i}\!\!\bra{s_{\bar A_i},m_{\bar A_i},\alpha_i}~,
%\end{equation}
%so
%\begin{align}
%   E_{i,s_{\bar A_i},m_{\bar A_i},\alpha_i,s_{\bar B_i},m_{\bar B_i},\beta_i} =
%%  \frac{c(s_{\bar A_i})}{2^{\frac{N-1}2}}
%%  \bra{\psi^-_{A_iC}}\otimes
%%  \bra{s_{\bar A_i},m_{\bar A_i},\alpha_i}\otimes
%%  \bra{s_{\bar B_i},m_{\bar B_i},\beta_i}~,
%   \bra{\psi^-_{A_iC}}\otimes
%   \bra{s_{\bar A_i},m_{\bar A_i},\alpha_i}\otimes
%   \bra{s_{\bar B_i},m_{\bar B_i},\beta_i} \; \sigma_{AC}^{-1/2}
%   ~,
%   \label{ei}
%\end{align}
To finish the calculation we need to perform the partial trace over all spins
except those in port $i$.
%To study this we focus on the operator
%\begin{equation}
%   \tilde Q(s_{\bar i},m_{\bar i}) =
%   \bra{s_{\bar i},m_{\bar i},\alpha_i} \sum_{sm\alpha} \frac1{\sqrt{g^{[N]}(s)}}
%   \ket{s,m,\alpha}\!\bra{s,m}
%\end{equation}
We use $s_{\bar{A}_{i}}$, $m_{\bar{A}_{i}}$ and $\alpha_{i}$ to model the
state of the total spin in ports ${A_{j}}$ with $j\neq i$. These refer to the
value of total spin and the projection along the $z$ axis, as well as the
degeneracy. Moreover, since $S_{\bar{A}_{i}}$ commutes with both $S_{A}^{2}$
and $S_{A}^{z}$, we may select a basis for the degeneracy that explicitly
contains $s_{\bar{A}_{i}}$. We may write then $\alpha=(s_{\bar{A}_{i}}%
,\tilde{\alpha}_{i})$ where $\tilde{\alpha}_{i}$ represents some other degrees
of freedom.

With the above definitions, when we insert several resolutions of the identity
in Eq.~\eqref{PBTasd}, we may write the Kraus operators as \begin{widetext}
\begin{align}
    K^0_{i,s_{\bar{A}_i},m_{\bar{A}_i},\alpha_i,s'_{\bar{A}_i},m'_{\bar{A}_i},\alpha'_i} &=
    2^{-\frac{N-1}2}
    \bra{s_{\bar{A}_i},m_{\bar{A}_i},\alpha_i}\otimes \bra{\psi^-_{A_iC}}\sigma_{AC}^{-1/2}
    \ket{s'_{\bar{A}_i},m'_{\bar{A}_i},\alpha'_i}
    \cr \nonumber &=
    2^{-\frac{N-1}2} \sum_{\pm,s_A,M,\alpha} \lambda_\pm(s_A)^{-1/2}
\bra{\psi^-_{A_iC}}
    \bra{s_{\bar{A}_i},m_{\bar{A}_i},\alpha_i}  {\pm,s_A,M,\alpha}\rangle \!
    \bra{\pm,s_A,M,\alpha} {s'_{\bar{A}_i},m'_{\bar{A}_i},\alpha'_i}\rangle,
\\
    K^1_{i,M,\alpha,s'_{\bar{A}_i},m'_{\bar{A}_i},\alpha'_i} &=
    2^{-\frac{N-1}2} N^{-1/2}
    \bra{+,N/2,M,\alpha} {s'_{\bar{A}_i},m'_{\bar{A}_i},\alpha'_i}\rangle,
\end{align}
\end{widetext}where each set of states $|s_{\bar{A}_{i}},m_{\bar{A}_{i}%
},\alpha_{i}\rangle$ represent a basis of the space corresponding to all ports
$j$ with $j\neq i$. To simplify the Kraus operators we study the overlap%
\begin{align}
& \langle s_{\bar{\imath}},m_{\bar{\imath}},\alpha_{i}|{\pm,S,M,\alpha}%
\rangle\nonumber\\
& =\sum_{k,m}|k\rangle_{C}\langle s_{\bar{\imath}},m_{\bar{\imath}},\alpha
_{i}|\Gamma_{S\pm\frac{1}{2},S}^{M,m,k}|S,m,\alpha\rangle_{A}\nonumber\\
& =\sum_{k,m}|k\rangle_{C}\langle s_{\bar{\imath}},m_{\bar{\imath}},\alpha
_{i}|\Gamma_{S\pm\frac{1}{2},S}^{M,m,k}\sum_{\ell}|\ell\rangle_{i}%
|s_{\bar{\imath}}^{\prime},m_{\bar{\imath}}^{\prime},\alpha_{i}^{\prime
}\rangle_{\bar{\imath}}\Gamma_{S,s_{\bar{\imath}}^{\prime}}^{m,m_{\bar{\imath
}}^{\prime},k}\nonumber\\
& =\sum_{k,\ell,m}|k\rangle_{C}|\ell\rangle_{A_{i}}\Gamma_{S\pm\frac{1}{2}%
,S}^{M,m,k}\Gamma_{S,s_{\bar{\imath}}}^{m,m_{\bar{\imath}},\ell}\equiv\hat
{Q}_{\pm,s,M}^{s_{\bar{\imath}},m_{\bar{\imath}}}.
\end{align}
In the last line we find that the overlap is independent on $\alpha$ and
$\alpha_{i}$, though with constraints $\alpha=(s_{\bar{\imath}},\alpha_{i})$,
which requires $\alpha_{i}=\alpha_{i}^{\prime}$. Therefore, different Kraus
operators provide exactly the same operation and, accordingly, we can sum over
these equivalent Kraus operators to reduce the number of indices. After this
process we get%
\begin{align}
K_{\ell}^{0}  & \equiv K_{s_{\bar{\imath}},m_{\bar{\imath}},m_{\bar{\imath}%
}^{\prime}}^{0}\nonumber\\
& =2^{-\frac{N-1}{2}}\sqrt{N}\sum_{\pm,s_{A},M}\lambda_{\pm}(s_{A}%
)^{-1/2}\sqrt{g^{[N-1]}(s_{\bar{\imath}})}\times\nonumber\\
& \times\left(  \langle\psi_{AC}^{-}|\hat{Q}_{\pm,s_{A},M}^{s_{\bar{\imath}%
},m_{\bar{\imath}}}\hat{Q}_{\pm,s_{A},M}^{s_{\bar{\imath}},m_{\bar{\imath}%
}^{\prime}}{}^{\dagger}\right)  \otimes\openone_{B},\\
K_{\ell}^{1}  & \equiv K_{M,s_{\bar{\imath}},m_{\bar{\imath}}}^{1}\nonumber\\
& =\sqrt{\frac{g^{[N-1]}(s_{\bar{\imath}})}{2^{N-1}}}\;\hat{Q}_{+,N/2,M}%
^{s_{\bar{\imath}},m_{\bar{\imath}}^{\prime}}{}^{\dagger}\otimes\openone_{B}.
\end{align}
The Kraus operators of the reduced channel $\tilde{\Lambda}$ are obtained as
$(K_{\ell}^{u}\otimes\openone_{D})(|\Psi_{CD}^{-}\rangle\otimes\openone_{AB}%
)$. It is simple to check that the above operators define a CPTP-map.

\bibliography{biblio}

\begin{thebibliography}{57}%
\makeatletter
\providecommand \@ifxundefined [1]{%
 \@ifx{#1\undefined}
}%
\providecommand \@ifnum [1]{%
 \ifnum #1\expandafter \@firstoftwo
 \else \expandafter \@secondoftwo
 \fi
}%
\providecommand \@ifx [1]{%
 \ifx #1\expandafter \@firstoftwo
 \else \expandafter \@secondoftwo
 \fi
}%
\providecommand \natexlab [1]{#1}%
\providecommand \enquote  [1]{``#1''}%
\providecommand \bibnamefont  [1]{#1}%
\providecommand \bibfnamefont [1]{#1}%
\providecommand \citenamefont [1]{#1}%
\providecommand \href@noop [0]{\@secondoftwo}%
\providecommand \href [0]{\begingroup \@sanitize@url \@href}%
\providecommand \@href[1]{\@@startlink{#1}\@@href}%
\providecommand \@@href[1]{\endgroup#1\@@endlink}%
\providecommand \@sanitize@url [0]{\catcode `\\12\catcode `\$12\catcode
  `\&12\catcode `\#12\catcode `\^12\catcode `\_12\catcode `\%12\relax}%
\providecommand \@@startlink[1]{}%
\providecommand \@@endlink[0]{}%
\providecommand \url  [0]{\begingroup\@sanitize@url \@url }%
\providecommand \@url [1]{\endgroup\@href {#1}{\urlprefix }}%
\providecommand \urlprefix  [0]{URL }%
\providecommand \Eprint [0]{\href }%
\providecommand \doibase [0]{http://dx.doi.org/}%
\providecommand \selectlanguage [0]{\@gobble}%
\providecommand \bibinfo  [0]{\@secondoftwo}%
\providecommand \bibfield  [0]{\@secondoftwo}%
\providecommand \translation [1]{[#1]}%
\providecommand \BibitemOpen [0]{}%
\providecommand \bibitemStop [0]{}%
\providecommand \bibitemNoStop [0]{.\EOS\space}%
\providecommand \EOS [0]{\spacefactor3000\relax}%
\providecommand \BibitemShut  [1]{\csname bibitem#1\endcsname}%
\let\auto@bib@innerbib\@empty
%</preamble>
\bibitem [{\citenamefont {Nielsen}\ and\ \citenamefont
  {Chuang}(2000)}]{nielsen2000quantum}%
  \BibitemOpen
  \bibfield  {author} {\bibinfo {author} {\bibfnamefont {M.~A.}\ \bibnamefont
  {Nielsen}}\ and\ \bibinfo {author} {\bibfnamefont {I.~L.}\ \bibnamefont
  {Chuang}},\ }\href@noop {} {\emph {\bibinfo {title} {Quantum Computation and
  Quantum Information}}}\ (\bibinfo  {publisher} {Cambridge University Press,
  Cambridge},\ \bibinfo {year} {2000})\BibitemShut {NoStop}%
\bibitem [{\citenamefont {Bishop}(2006)}]{MLbook}%
  \BibitemOpen
  \bibfield  {author} {\bibinfo {author} {\bibfnamefont {C.~M.}\ \bibnamefont
  {Bishop}},\ }\href@noop {} {\emph {\bibinfo {title} {Pattern Recognition and
  Machine Learning}}}\ (\bibinfo  {publisher} {Springer},\ \bibinfo {year}
  {2006})\BibitemShut {NoStop}%
\bibitem [{\citenamefont {Wittek}(2014)}]{QML1}%
  \BibitemOpen
  \bibfield  {author} {\bibinfo {author} {\bibfnamefont {P.}~\bibnamefont
  {Wittek}},\ }\href@noop {} {\emph {\bibinfo {title} {Quantum Machine
  Learning: What Quantum Computing Means to Data Mining}}}\ (\bibinfo
  {publisher} {Academic Press, Elsevier},\ \bibinfo {year} {2014})\BibitemShut
  {NoStop}%
\bibitem [{\citenamefont {Biamonte}\ \emph {et~al.}(2017)\citenamefont
  {Biamonte}, \citenamefont {Wittek}, \citenamefont {Pancotti}, \citenamefont
  {Rebentrost}, \citenamefont {Wiebe},\ and\ \citenamefont {Lloyd}}]{QML2}%
  \BibitemOpen
  \bibfield  {author} {\bibinfo {author} {\bibfnamefont {J}~\bibnamefont
  {Biamonte}}, \bibinfo {author} {\bibfnamefont {P.}~\bibnamefont {Wittek}},
  \bibinfo {author} {\bibfnamefont {N.}~\bibnamefont {Pancotti}}, \bibinfo
  {author} {\bibfnamefont {P.}~\bibnamefont {Rebentrost}}, \bibinfo {author}
  {\bibfnamefont {N.}~\bibnamefont {Wiebe}}, \ and\ \bibinfo {author}
  {\bibfnamefont {S.}~\bibnamefont {Lloyd}},\ }\bibfield  {title} {\enquote
  {\bibinfo {title} {Quantum machine learning},}\ }\href@noop {} {\bibfield
  {journal} {\bibinfo  {journal} {Nature (London)}\ }\textbf {\bibinfo {volume}
  {549}},\ \bibinfo {pages} {195} (\bibinfo {year} {2017})}\BibitemShut
  {NoStop}%
\bibitem [{\citenamefont {Dunjko}\ and\ \citenamefont
  {Briegel}(2018)}]{dunjko2018machine}%
  \BibitemOpen
  \bibfield  {author} {\bibinfo {author} {\bibfnamefont {V.}~\bibnamefont
  {Dunjko}}\ and\ \bibinfo {author} {\bibfnamefont {H.~J.}\ \bibnamefont
  {Briegel}},\ }\bibfield  {title} {\enquote {\bibinfo {title} {Machine
  learning \& artificial intelligence in the quantum domain: a review of recent
  progress},}\ }\href@noop {} {\bibfield  {journal} {\bibinfo  {journal} {Rep.
  Prog. Phys.}\ }\textbf {\bibinfo {volume} {81}},\ \bibinfo {pages} {074001}
  (\bibinfo {year} {2018})}\BibitemShut {NoStop}%
\bibitem [{\citenamefont {Schuld}\ \emph {et~al.}(2015)\citenamefont {Schuld},
  \citenamefont {Sinayskiy},\ and\ \citenamefont
  {Petruccione}}]{schuld2015introduction}%
  \BibitemOpen
  \bibfield  {author} {\bibinfo {author} {\bibfnamefont {M.}~\bibnamefont
  {Schuld}}, \bibinfo {author} {\bibfnamefont {I.}~\bibnamefont {Sinayskiy}}, \
  and\ \bibinfo {author} {\bibfnamefont {F.}~\bibnamefont {Petruccione}},\
  }\bibfield  {title} {\enquote {\bibinfo {title} {An introduction to quantum
  machine learning},}\ }\href@noop {} {\bibfield  {journal} {\bibinfo
  {journal} {Contemporary Physics}\ }\textbf {\bibinfo {volume} {56}},\
  \bibinfo {pages} {172--185} (\bibinfo {year} {2015})}\BibitemShut {NoStop}%
\bibitem [{\citenamefont {Ciliberto}\ \emph {et~al.}(2018)\citenamefont
  {Ciliberto}, \citenamefont {Herbster}, \citenamefont {Ialongo}, \citenamefont
  {Pontil}, \citenamefont {Rocchetto}, \citenamefont {Severini},\ and\
  \citenamefont {Wossnig}}]{ciliberto2018quantum}%
  \BibitemOpen
  \bibfield  {author} {\bibinfo {author} {\bibfnamefont {C.}~\bibnamefont
  {Ciliberto}}, \bibinfo {author} {\bibfnamefont {M.}~\bibnamefont {Herbster}},
  \bibinfo {author} {\bibfnamefont {Alessandro~D.}\ \bibnamefont {Ialongo}},
  \bibinfo {author} {\bibfnamefont {M.}~\bibnamefont {Pontil}}, \bibinfo
  {author} {\bibfnamefont {A.}~\bibnamefont {Rocchetto}}, \bibinfo {author}
  {\bibfnamefont {S.}~\bibnamefont {Severini}}, \ and\ \bibinfo {author}
  {\bibfnamefont {L.}~\bibnamefont {Wossnig}},\ }\bibfield  {title} {\enquote
  {\bibinfo {title} {Quantum machine learning: a classical perspective},}\
  }\href@noop {} {\bibfield  {journal} {\bibinfo  {journal} {Proceedings of the
  Royal Society A: Mathematical, Physical and Engineering Sciences}\ }\textbf
  {\bibinfo {volume} {474}},\ \bibinfo {pages} {20170551} (\bibinfo {year}
  {2018})}\BibitemShut {NoStop}%
\bibitem [{\citenamefont {Tang}(2018{\natexlab{a}})}]{QIML1}%
  \BibitemOpen
  \bibfield  {author} {\bibinfo {author} {\bibfnamefont {E}~\bibnamefont
  {Tang}},\ }\bibfield  {title} {\enquote {\bibinfo {title} {A quantum-inspired
  classical algorithm for recommendation systems},}\ }\href@noop {} {\bibfield
  {journal} {\bibinfo  {journal} {preprint arXiv:1807.04271}\ } (\bibinfo
  {year} {2018}{\natexlab{a}})}\BibitemShut {NoStop}%
\bibitem [{\citenamefont {Tang}(2018{\natexlab{b}})}]{QIML2}%
  \BibitemOpen
  \bibfield  {author} {\bibinfo {author} {\bibfnamefont {E}~\bibnamefont
  {Tang}},\ }\bibfield  {title} {\enquote {\bibinfo {title} {Quantum-inspired
  classical algorithms for principal component analysis and supervised
  clustering},}\ }\href@noop {} {\bibfield  {journal} {\bibinfo  {journal}
  {preprint arXiv:1811.00414}\ } (\bibinfo {year}
  {2018}{\natexlab{b}})}\BibitemShut {NoStop}%
\bibitem [{\citenamefont {Nielsen}\ and\ \citenamefont
  {Chuang}(1997)}]{nielsen1997programmable}%
  \BibitemOpen
  \bibfield  {author} {\bibinfo {author} {\bibfnamefont {M.~A.}\ \bibnamefont
  {Nielsen}}\ and\ \bibinfo {author} {\bibfnamefont {I.~L.}\ \bibnamefont
  {Chuang}},\ }\bibfield  {title} {\enquote {\bibinfo {title} {Programmable
  quantum gate arrays},}\ }\href@noop {} {\bibfield  {journal} {\bibinfo
  {journal} {Phys. Rev. Lett.}\ }\textbf {\bibinfo {volume} {79}},\ \bibinfo
  {pages} {321} (\bibinfo {year} {1997})}\BibitemShut {NoStop}%
\bibitem [{\citenamefont {Ishizaka}\ and\ \citenamefont
  {Hiroshima}(2008)}]{ishizaka2008asymptotic}%
  \BibitemOpen
  \bibfield  {author} {\bibinfo {author} {\bibfnamefont {S.}~\bibnamefont
  {Ishizaka}}\ and\ \bibinfo {author} {\bibfnamefont {T.}~\bibnamefont
  {Hiroshima}},\ }\bibfield  {title} {\enquote {\bibinfo {title} {Asymptotic
  teleportation scheme as a universal programmable quantum processor},}\
  }\href@noop {} {\bibfield  {journal} {\bibinfo  {journal} {Phys. Rev. Lett.}\
  }\textbf {\bibinfo {volume} {101}},\ \bibinfo {pages} {240501} (\bibinfo
  {year} {2008})}\BibitemShut {NoStop}%
\bibitem [{\citenamefont {Ishizaka}\ and\ \citenamefont
  {Hiroshima}(2009)}]{ishizaka2009quantum}%
  \BibitemOpen
  \bibfield  {author} {\bibinfo {author} {\bibfnamefont {S.}~\bibnamefont
  {Ishizaka}}\ and\ \bibinfo {author} {\bibfnamefont {T.}~\bibnamefont
  {Hiroshima}},\ }\bibfield  {title} {\enquote {\bibinfo {title} {Quantum
  teleportation scheme by selecting one of multiple output ports},}\
  }\href@noop {} {\bibfield  {journal} {\bibinfo  {journal} {Phys. Rev. A}\
  }\textbf {\bibinfo {volume} {79}},\ \bibinfo {pages} {042306} (\bibinfo
  {year} {2009})}\BibitemShut {NoStop}%
\bibitem [{\citenamefont {Ishizaka}(2015)}]{ishizaka2015some}%
  \BibitemOpen
  \bibfield  {author} {\bibinfo {author} {\bibfnamefont {S.}~\bibnamefont
  {Ishizaka}},\ }\bibfield  {title} {\enquote {\bibinfo {title} {Some remarks
  on port-based teleportation},}\ }\href@noop {} {\bibfield  {journal}
  {\bibinfo  {journal} {preprint arXiv:1506.01555}\ } (\bibinfo {year}
  {2015})}\BibitemShut {NoStop}%
\bibitem [{\citenamefont {Pirandola}\ \emph
  {et~al.}(2018{\natexlab{a}})\citenamefont {Pirandola}, \citenamefont
  {Laurenza},\ and\ \citenamefont {Lupo}}]{pirandola2018fundamental}%
  \BibitemOpen
  \bibfield  {author} {\bibinfo {author} {\bibfnamefont {S.}~\bibnamefont
  {Pirandola}}, \bibinfo {author} {\bibfnamefont {R.}~\bibnamefont {Laurenza}},
  \ and\ \bibinfo {author} {\bibfnamefont {C.}~\bibnamefont {Lupo}},\
  }\bibfield  {title} {\enquote {\bibinfo {title} {Fundamental limits to
  quantum channel discrimination},}\ }\href@noop {} {\bibfield  {journal}
  {\bibinfo  {journal} {preprint arXiv:1803.02834}\ } (\bibinfo {year}
  {2018}{\natexlab{a}})}\BibitemShut {NoStop}%
\bibitem [{\citenamefont {Boyd}\ \emph {et~al.}(2003)\citenamefont {Boyd},
  \citenamefont {Xiao},\ and\ \citenamefont {Mutapcic}}]{boyd2003subgradient}%
  \BibitemOpen
  \bibfield  {author} {\bibinfo {author} {\bibfnamefont {S.}~\bibnamefont
  {Boyd}}, \bibinfo {author} {\bibfnamefont {L.}~\bibnamefont {Xiao}}, \ and\
  \bibinfo {author} {\bibfnamefont {A.}~\bibnamefont {Mutapcic}},\ }\href@noop
  {} {\emph {\bibinfo {title} {Subgradient methods}}}\ (\bibinfo {year}
  {2003})\BibitemShut {NoStop}%
\bibitem [{\citenamefont {Jaggi}(2011)}]{jaggi2011convex}%
  \BibitemOpen
  \bibfield  {author} {\bibinfo {author} {\bibfnamefont {M.}~\bibnamefont
  {Jaggi}},\ }\bibfield  {title} {\enquote {\bibinfo {title} {Convex
  optimization without projection steps},}\ }\href@noop {} {\bibfield
  {journal} {\bibinfo  {journal} {preprint arXiv:1108.1170}\ } (\bibinfo {year}
  {2011})}\BibitemShut {NoStop}%
\bibitem [{\citenamefont {Jaggi}(2013)}]{jaggi2013revisiting}%
  \BibitemOpen
  \bibfield  {author} {\bibinfo {author} {\bibfnamefont {M.}~\bibnamefont
  {Jaggi}},\ }\bibfield  {title} {\enquote {\bibinfo {title} {Revisiting
  frank-wolfe: projection-free sparse convex optimization},}\ }in\ \href@noop
  {} {\emph {\bibinfo {booktitle} {Proceedings of the 30th International
  Conference on International Conference on Machine Learning-Volume 28}}}\
  (\bibinfo {organization} {JMLR. org},\ \bibinfo {year} {2013})\ pp.\ \bibinfo
  {pages} {1--427}\BibitemShut {NoStop}%
\bibitem [{\citenamefont {Duchi}\ \emph {et~al.}(2008)\citenamefont {Duchi},
  \citenamefont {Shalev-Shwartz}, \citenamefont {Singer},\ and\ \citenamefont
  {Chandra}}]{duchi2008efficient}%
  \BibitemOpen
  \bibfield  {author} {\bibinfo {author} {\bibfnamefont {J.}~\bibnamefont
  {Duchi}}, \bibinfo {author} {\bibfnamefont {S.}~\bibnamefont
  {Shalev-Shwartz}}, \bibinfo {author} {\bibfnamefont {Y.}~\bibnamefont
  {Singer}}, \ and\ \bibinfo {author} {\bibfnamefont {T.}~\bibnamefont
  {Chandra}},\ }\bibfield  {title} {\enquote {\bibinfo {title} {Efficient
  projections onto the l 1-ball for learning in high dimensions},}\ }in\
  \href@noop {} {\emph {\bibinfo {booktitle} {Proceedings of the 25th
  international conference on Machine learning}}}\ (\bibinfo {organization}
  {ACM},\ \bibinfo {year} {2008})\ pp.\ \bibinfo {pages} {272--279}\BibitemShut
  {NoStop}%
\bibitem [{\citenamefont {Liu}\ \emph {et~al.}(2013)\citenamefont {Liu},
  \citenamefont {Musialski}, \citenamefont {Wonka},\ and\ \citenamefont
  {Ye}}]{liu2013tensor}%
  \BibitemOpen
  \bibfield  {author} {\bibinfo {author} {\bibfnamefont {J.}~\bibnamefont
  {Liu}}, \bibinfo {author} {\bibfnamefont {P.}~\bibnamefont {Musialski}},
  \bibinfo {author} {\bibfnamefont {P.}~\bibnamefont {Wonka}}, \ and\ \bibinfo
  {author} {\bibfnamefont {J.}~\bibnamefont {Ye}},\ }\bibfield  {title}
  {\enquote {\bibinfo {title} {Tensor completion for estimating missing values
  in visual data},}\ }\href@noop {} {\bibfield  {journal} {\bibinfo  {journal}
  {IEEE transactions on pattern analysis and machine intelligence}\ }\textbf
  {\bibinfo {volume} {35}},\ \bibinfo {pages} {208--220} (\bibinfo {year}
  {2013})}\BibitemShut {NoStop}%
\bibitem [{\citenamefont {Lloyd}(1996)}]{lloyd1996universal}%
  \BibitemOpen
  \bibfield  {author} {\bibinfo {author} {\bibfnamefont {S}~\bibnamefont
  {Lloyd}},\ }\bibfield  {title} {\enquote {\bibinfo {title} {Universal quantum
  simulators},}\ }\href@noop {} {\bibfield  {journal} {\bibinfo  {journal}
  {Science}\ }\textbf {\bibinfo {volume} {273}},\ \bibinfo {pages} {1073--1078}
  (\bibinfo {year} {1996})}\BibitemShut {NoStop}%
\bibitem [{\citenamefont {Pirandola}\ \emph {et~al.}(2017)\citenamefont
  {Pirandola}, \citenamefont {Laurenza}, \citenamefont {Ottaviani},\ and\
  \citenamefont {Banchi}}]{pirandola2017fundamental}%
  \BibitemOpen
  \bibfield  {author} {\bibinfo {author} {\bibfnamefont {S.}~\bibnamefont
  {Pirandola}}, \bibinfo {author} {\bibfnamefont {R.}~\bibnamefont {Laurenza}},
  \bibinfo {author} {\bibfnamefont {C.}~\bibnamefont {Ottaviani}}, \ and\
  \bibinfo {author} {\bibfnamefont {L.}~\bibnamefont {Banchi}},\ }\bibfield
  {title} {\enquote {\bibinfo {title} {Fundamental limits of repeaterless
  quantum communications},}\ }\href@noop {} {\bibfield  {journal} {\bibinfo
  {journal} {Nat. Commun.}\ }\textbf {\bibinfo {volume} {8}},\ \bibinfo {pages}
  {15043} (\bibinfo {year} {2017})}\BibitemShut {NoStop}%
\bibitem [{\citenamefont {Pirandola}\ \emph
  {et~al.}(2018{\natexlab{b}})\citenamefont {Pirandola}, \citenamefont
  {Braunstein}, \citenamefont {Laurenza}, \citenamefont {Ottaviani},
  \citenamefont {Cope}, \citenamefont {Spedalieri},\ and\ \citenamefont
  {Banchi}}]{commREVIEW}%
  \BibitemOpen
  \bibfield  {author} {\bibinfo {author} {\bibfnamefont {S.}~\bibnamefont
  {Pirandola}}, \bibinfo {author} {\bibfnamefont {S.~L.}\ \bibnamefont
  {Braunstein}}, \bibinfo {author} {\bibfnamefont {R.}~\bibnamefont
  {Laurenza}}, \bibinfo {author} {\bibfnamefont {C.}~\bibnamefont {Ottaviani}},
  \bibinfo {author} {\bibfnamefont {T.~P.~W.}\ \bibnamefont {Cope}}, \bibinfo
  {author} {\bibfnamefont {G.}~\bibnamefont {Spedalieri}}, \ and\ \bibinfo
  {author} {\bibfnamefont {L.}~\bibnamefont {Banchi}},\ }\bibfield  {title}
  {\enquote {\bibinfo {title} {Theory of channel simulation and bounds for
  private communication},}\ }\href@noop {} {\bibfield  {journal} {\bibinfo
  {journal} {Quant. Sci. Tech.}\ }\textbf {\bibinfo {volume} {3}},\ \bibinfo
  {pages} {035009} (\bibinfo {year} {2018}{\natexlab{b}})}\BibitemShut
  {NoStop}%
\bibitem [{\citenamefont {Watrous}(2018)}]{watrous2018theory}%
  \BibitemOpen
  \bibfield  {author} {\bibinfo {author} {\bibfnamefont {J.}~\bibnamefont
  {Watrous}},\ }\href@noop {} {\emph {\bibinfo {title} {The theory of quantum
  information}}}\ (\bibinfo  {publisher} {Cambridge Univ. Press},\ \bibinfo
  {year} {2018})\ \bibinfo {note} {freely available at
  \url{https://cs.uwaterloo.ca/~watrous/TQI/}}\BibitemShut {NoStop}%
\bibitem [{\citenamefont {Kitaev}\ \emph {et~al.}(2002)\citenamefont {Kitaev},
  \citenamefont {Shen},\ and\ \citenamefont {Vyalyi}}]{kitaev2002classical}%
  \BibitemOpen
  \bibfield  {author} {\bibinfo {author} {\bibfnamefont {A.~Y.}\ \bibnamefont
  {Kitaev}}, \bibinfo {author} {\bibfnamefont {A.}~\bibnamefont {Shen}}, \ and\
  \bibinfo {author} {\bibfnamefont {M.~N.}\ \bibnamefont {Vyalyi}},\
  }\href@noop {} {\emph {\bibinfo {title} {Classical and quantum
  computation}}},\ \bibinfo {number} {47}\ (\bibinfo  {publisher} {American
  Mathematical Society, Providence, Rhode Island},\ \bibinfo {year} {2002})\
  \bibinfo {note} {sec. 11}\BibitemShut {NoStop}%
\bibitem [{\citenamefont {Watrous}(2004)}]{watrous2004advanced}%
  \BibitemOpen
  \bibfield  {author} {\bibinfo {author} {\bibfnamefont {J.}~\bibnamefont
  {Watrous}},\ }\href@noop {} {\emph {\bibinfo {title} {Advanced Topics in
  Quantum Information Processing}}}\ (\bibinfo  {publisher} {Lecture notes},\
  \bibinfo {year} {2004})\BibitemShut {NoStop}%
\bibitem [{\citenamefont {Knill}\ \emph {et~al.}(2001)\citenamefont {Knill},
  \citenamefont {Laflamme},\ and\ \citenamefont {Milburn}}]{knill2001scheme}%
  \BibitemOpen
  \bibfield  {author} {\bibinfo {author} {\bibfnamefont {E.}~\bibnamefont
  {Knill}}, \bibinfo {author} {\bibfnamefont {R.}~\bibnamefont {Laflamme}}, \
  and\ \bibinfo {author} {\bibfnamefont {G.~J}\ \bibnamefont {Milburn}},\
  }\bibfield  {title} {\enquote {\bibinfo {title} {A scheme for efficient
  quantum computation with linear optics},}\ }\href@noop {} {\bibfield
  {journal} {\bibinfo  {journal} {Nature}\ }\textbf {\bibinfo {volume} {409}},\
  \bibinfo {pages} {46} (\bibinfo {year} {2001})}\BibitemShut {NoStop}%
\bibitem [{\citenamefont {Watrous}(2013)}]{Watrous}%
  \BibitemOpen
  \bibfield  {author} {\bibinfo {author} {\bibfnamefont {J.}~\bibnamefont
  {Watrous}},\ }\bibfield  {title} {\enquote {\bibinfo {title} {Simpler
  semidefinite programs for completely bounded norms},}\ }\href@noop {}
  {\bibfield  {journal} {\bibinfo  {journal} {Chicago Journal of Theoretical
  Computer Science}\ }\textbf {\bibinfo {volume} {8}},\ \bibinfo {pages}
  {1--19} (\bibinfo {year} {2013})}\BibitemShut {NoStop}%
\bibitem [{\citenamefont {Fuchs}\ and\ \citenamefont {van~de
  Graaf}(1999)}]{fuchs1999cryptographic}%
  \BibitemOpen
  \bibfield  {author} {\bibinfo {author} {\bibfnamefont {C.~A.}\ \bibnamefont
  {Fuchs}}\ and\ \bibinfo {author} {\bibfnamefont {J.}~\bibnamefont {van~de
  Graaf}},\ }\bibfield  {title} {\enquote {\bibinfo {title} {Cryptographic
  distinguishability measures for quantum-mechanical states},}\ }\href@noop {}
  {\bibfield  {journal} {\bibinfo  {journal} {IEEE Trans. Info. Theory}\
  }\textbf {\bibinfo {volume} {45}},\ \bibinfo {pages} {1216--1227} (\bibinfo
  {year} {1999})}\BibitemShut {NoStop}%
\bibitem [{\citenamefont {Pinsker}(1964)}]{pinsker1964information}%
  \BibitemOpen
  \bibfield  {author} {\bibinfo {author} {\bibfnamefont {M.~S.}\ \bibnamefont
  {Pinsker}},\ }\href@noop {} {\emph {\bibinfo {title} {Information and
  information stability of random variables and processes}}}\ (\bibinfo
  {publisher} {Holden-Day, San Francisco},\ \bibinfo {year} {1964})\BibitemShut
  {NoStop}%
\bibitem [{\citenamefont {Carlen}\ and\ \citenamefont
  {Lieb}(2012)}]{carlen2012bounds}%
  \BibitemOpen
  \bibfield  {author} {\bibinfo {author} {\bibfnamefont {E.~A.}\ \bibnamefont
  {Carlen}}\ and\ \bibinfo {author} {\bibfnamefont {E.~H.}\ \bibnamefont
  {Lieb}},\ }\bibfield  {title} {\enquote {\bibinfo {title} {Bounds for
  entanglement via an extension of strong subadditivity of entropy},}\
  }\href@noop {} {\bibfield  {journal} {\bibinfo  {journal} {Lett. Math.
  Phys.}\ }\textbf {\bibinfo {volume} {101}},\ \bibinfo {pages} {1--11}
  (\bibinfo {year} {2012})}\BibitemShut {NoStop}%
\bibitem [{\citenamefont {Uhlmann}(1976)}]{uhlmann1976transition}%
  \BibitemOpen
  \bibfield  {author} {\bibinfo {author} {\bibfnamefont {A.}~\bibnamefont
  {Uhlmann}},\ }\bibfield  {title} {\enquote {\bibinfo {title} {The transition
  probability...}}\ }\href@noop {} {\bibfield  {journal} {\bibinfo  {journal}
  {Rep. Math. Phys.}\ }\textbf {\bibinfo {volume} {9}},\ \bibinfo {pages}
  {273--279} (\bibinfo {year} {1976})}\BibitemShut {NoStop}%
\bibitem [{\citenamefont {Watrous}(2009)}]{watrous2009semidefinite}%
  \BibitemOpen
  \bibfield  {author} {\bibinfo {author} {\bibfnamefont {John}\ \bibnamefont
  {Watrous}},\ }\bibfield  {title} {\enquote {\bibinfo {title} {Semidefinite
  programs for completely bounded norms},}\ }\href@noop {} {\bibfield
  {journal} {\bibinfo  {journal} {Theory of Computing}\ }\textbf {\bibinfo
  {volume} {5}},\ \bibinfo {pages} {217--238} (\bibinfo {year}
  {2009})}\BibitemShut {NoStop}%
\bibitem [{\citenamefont {Nechita}\ \emph {et~al.}(2018)\citenamefont
  {Nechita}, \citenamefont {Pucha{\l}a}, \citenamefont {Pawela},\ and\
  \citenamefont {\.{Z}yczkowski}}]{Karol}%
  \BibitemOpen
  \bibfield  {author} {\bibinfo {author} {\bibfnamefont {I.}~\bibnamefont
  {Nechita}}, \bibinfo {author} {\bibfnamefont {Z.}~\bibnamefont {Pucha{\l}a}},
  \bibinfo {author} {\bibfnamefont {{\L}.}~\bibnamefont {Pawela}}, \ and\
  \bibinfo {author} {\bibfnamefont {K.}~\bibnamefont {\.{Z}yczkowski}},\
  }\bibfield  {title} {\enquote {\bibinfo {title} {Almost all quantum channels
  are equidistant},}\ }\href@noop {} {\bibfield  {journal} {\bibinfo  {journal}
  {J. Math. Phys.}\ }\textbf {\bibinfo {volume} {59}},\ \bibinfo {pages}
  {052201} (\bibinfo {year} {2018})}\BibitemShut {NoStop}%
\bibitem [{\citenamefont {Nesterov}(2013)}]{nesterov2013introductory}%
  \BibitemOpen
  \bibfield  {author} {\bibinfo {author} {\bibfnamefont {Y.}~\bibnamefont
  {Nesterov}},\ }\href@noop {} {\emph {\bibinfo {title} {Introductory lectures
  on convex optimization: A basic course}}},\ Vol.~\bibinfo {volume} {87}\
  (\bibinfo  {publisher} {Springer Science \& Business Media, New York},\
  \bibinfo {year} {2013})\BibitemShut {NoStop}%
\bibitem [{\citenamefont {Coutts}\ \emph {et~al.}(2018)\citenamefont {Coutts},
  \citenamefont {Girard},\ and\ \citenamefont
  {Watrous}}]{coutts2018certifying}%
  \BibitemOpen
  \bibfield  {author} {\bibinfo {author} {\bibfnamefont {B.}~\bibnamefont
  {Coutts}}, \bibinfo {author} {\bibfnamefont {M.}~\bibnamefont {Girard}}, \
  and\ \bibinfo {author} {\bibfnamefont {J.}~\bibnamefont {Watrous}},\
  }\bibfield  {title} {\enquote {\bibinfo {title} {Certifying optimality for
  convex quantum channel optimization problems},}\ }\href@noop {} {\bibfield
  {journal} {\bibinfo  {journal} {preprint arXiv:1810.13295}\ } (\bibinfo
  {year} {2018})}\BibitemShut {NoStop}%
\bibitem [{\citenamefont {Duchi}\ \emph {et~al.}(2011)\citenamefont {Duchi},
  \citenamefont {Hazan},\ and\ \citenamefont {Singer}}]{duchi2011adaptive}%
  \BibitemOpen
  \bibfield  {author} {\bibinfo {author} {\bibfnamefont {J.}~\bibnamefont
  {Duchi}}, \bibinfo {author} {\bibfnamefont {E.}~\bibnamefont {Hazan}}, \ and\
  \bibinfo {author} {\bibfnamefont {Y.}~\bibnamefont {Singer}},\ }\bibfield
  {title} {\enquote {\bibinfo {title} {Adaptive subgradient methods for online
  learning and stochastic optimization},}\ }\href@noop {} {\bibfield  {journal}
  {\bibinfo  {journal} {Journal of Machine Learning Research}\ }\textbf
  {\bibinfo {volume} {12}},\ \bibinfo {pages} {2121--2159} (\bibinfo {year}
  {2011})}\BibitemShut {NoStop}%
\bibitem [{\citenamefont {Garber}\ and\ \citenamefont
  {Hazan}(2015)}]{garber2015faster}%
  \BibitemOpen
  \bibfield  {author} {\bibinfo {author} {\bibfnamefont {D.}~\bibnamefont
  {Garber}}\ and\ \bibinfo {author} {\bibfnamefont {E.}~\bibnamefont {Hazan}},\
  }\bibfield  {title} {\enquote {\bibinfo {title} {Faster rates for the
  frank-wolfe method over strongly-convex sets},}\ }in\ \href@noop {} {\emph
  {\bibinfo {booktitle} {Proceedings of the 32nd International Conference on
  International Conference on Machine Learning-Volume 37}}}\ (\bibinfo
  {organization} {JMLR. org},\ \bibinfo {year} {2015})\ pp.\ \bibinfo {pages}
  {541--549}\BibitemShut {NoStop}%
\bibitem [{\citenamefont {Nesterov}(2005)}]{nesterov2005smooth}%
  \BibitemOpen
  \bibfield  {author} {\bibinfo {author} {\bibfnamefont {Y.}~\bibnamefont
  {Nesterov}},\ }\bibfield  {title} {\enquote {\bibinfo {title} {Smooth
  minimization of non-smooth functions},}\ }\href@noop {} {\bibfield  {journal}
  {\bibinfo  {journal} {Mathematical programming}\ }\textbf {\bibinfo {volume}
  {103}},\ \bibinfo {pages} {127--152} (\bibinfo {year} {2005})}\BibitemShut
  {NoStop}%
\bibitem [{\citenamefont {Bhatia}(2013)}]{bhatia2013matrix}%
  \BibitemOpen
  \bibfield  {author} {\bibinfo {author} {\bibfnamefont {R.}~\bibnamefont
  {Bhatia}},\ }\href@noop {} {\emph {\bibinfo {title} {Matrix analysis}}},\
  Vol.\ \bibinfo {volume} {169}\ (\bibinfo  {publisher} {Springer Science \&
  Business Media, New York},\ \bibinfo {year} {2013})\BibitemShut {NoStop}%
\bibitem [{\citenamefont {Banchi}\ \emph {et~al.}(2016)\citenamefont {Banchi},
  \citenamefont {Pancotti},\ and\ \citenamefont {Bose}}]{banchi2016quantum}%
  \BibitemOpen
  \bibfield  {author} {\bibinfo {author} {\bibfnamefont {L.}~\bibnamefont
  {Banchi}}, \bibinfo {author} {\bibfnamefont {N.}~\bibnamefont {Pancotti}}, \
  and\ \bibinfo {author} {\bibfnamefont {S.}~\bibnamefont {Bose}},\ }\bibfield
  {title} {\enquote {\bibinfo {title} {Quantum gate learning in qubit networks:
  Toffoli gate without time-dependent control},}\ }\href@noop {} {\bibfield
  {journal} {\bibinfo  {journal} {npj Quantum Inf.}\ }\textbf {\bibinfo
  {volume} {2}},\ \bibinfo {pages} {16019} (\bibinfo {year}
  {2016})}\BibitemShut {NoStop}%
\bibitem [{\citenamefont {Innocenti}\ \emph {et~al.}(2018)\citenamefont
  {Innocenti}, \citenamefont {Banchi}, \citenamefont {Ferraro}, \citenamefont
  {Bose},\ and\ \citenamefont {Paternostro}}]{innocenti2018supervised}%
  \BibitemOpen
  \bibfield  {author} {\bibinfo {author} {\bibfnamefont {L.}~\bibnamefont
  {Innocenti}}, \bibinfo {author} {\bibfnamefont {L.}~\bibnamefont {Banchi}},
  \bibinfo {author} {\bibfnamefont {A.}~\bibnamefont {Ferraro}}, \bibinfo
  {author} {\bibfnamefont {S.}~\bibnamefont {Bose}}, \ and\ \bibinfo {author}
  {\bibfnamefont {M.}~\bibnamefont {Paternostro}},\ }\bibfield  {title}
  {\enquote {\bibinfo {title} {Supervised learning of time-independent
  hamiltonians for gate design},}\ }\href@noop {} {\bibfield  {journal}
  {\bibinfo  {journal} {preprint arXiv:1803.07119}\ } (\bibinfo {year}
  {2018})}\BibitemShut {NoStop}%
\bibitem [{\citenamefont {Mitarai}\ \emph {et~al.}(2018)\citenamefont
  {Mitarai}, \citenamefont {Negoro}, \citenamefont {Kitagawa},\ and\
  \citenamefont {Fujii}}]{mitarai2018quantum}%
  \BibitemOpen
  \bibfield  {author} {\bibinfo {author} {\bibfnamefont {K.}~\bibnamefont
  {Mitarai}}, \bibinfo {author} {\bibfnamefont {M.}~\bibnamefont {Negoro}},
  \bibinfo {author} {\bibfnamefont {M.}~\bibnamefont {Kitagawa}}, \ and\
  \bibinfo {author} {\bibfnamefont {K.}~\bibnamefont {Fujii}},\ }\bibfield
  {title} {\enquote {\bibinfo {title} {Quantum circuit learning},}\ }\href@noop
  {} {\bibfield  {journal} {\bibinfo  {journal} {Phys. Rev. A}\ }\textbf
  {\bibinfo {volume} {98}},\ \bibinfo {pages} {032309} (\bibinfo {year}
  {2018})}\BibitemShut {NoStop}%
\bibitem [{\citenamefont {Arrazola}\ \emph {et~al.}(2019)\citenamefont
  {Arrazola}, \citenamefont {Bromley}, \citenamefont {Izaac}, \citenamefont
  {Myers}, \citenamefont {Br{\'a}dler},\ and\ \citenamefont
  {Killoran}}]{arrazola2018machine}%
  \BibitemOpen
  \bibfield  {author} {\bibinfo {author} {\bibfnamefont {J.~M.}\ \bibnamefont
  {Arrazola}}, \bibinfo {author} {\bibfnamefont {T.~R.}\ \bibnamefont
  {Bromley}}, \bibinfo {author} {\bibfnamefont {J.}~\bibnamefont {Izaac}},
  \bibinfo {author} {\bibfnamefont {C.~R.}\ \bibnamefont {Myers}}, \bibinfo
  {author} {\bibfnamefont {K.}~\bibnamefont {Br{\'a}dler}}, \ and\ \bibinfo
  {author} {\bibfnamefont {N.}~\bibnamefont {Killoran}},\ }\bibfield  {title}
  {\enquote {\bibinfo {title} {Machine learning method for state preparation
  and gate synthesis on photonic quantum computers},}\ }\href@noop {}
  {\bibfield  {journal} {\bibinfo  {journal} {Quantum Sci. Technol.}\ }\textbf
  {\bibinfo {volume} {4}},\ \bibinfo {pages} {024004} (\bibinfo {year}
  {2019})}\BibitemShut {NoStop}%
\bibitem [{\citenamefont {Bennett}\ \emph {et~al.}(1993)\citenamefont
  {Bennett}, \citenamefont {Brassard}, \citenamefont {Cr{\'e}peau},
  \citenamefont {Jozsa}, \citenamefont {Peres},\ and\ \citenamefont
  {Wootters}}]{bennett1993teleporting}%
  \BibitemOpen
  \bibfield  {author} {\bibinfo {author} {\bibfnamefont {C.~H.}\ \bibnamefont
  {Bennett}}, \bibinfo {author} {\bibfnamefont {G.}~\bibnamefont {Brassard}},
  \bibinfo {author} {\bibfnamefont {C.}~\bibnamefont {Cr{\'e}peau}}, \bibinfo
  {author} {\bibfnamefont {R.}~\bibnamefont {Jozsa}}, \bibinfo {author}
  {\bibfnamefont {A.}~\bibnamefont {Peres}}, \ and\ \bibinfo {author}
  {\bibfnamefont {W.~K.}\ \bibnamefont {Wootters}},\ }\bibfield  {title}
  {\enquote {\bibinfo {title} {Teleporting an unknown quantum state via dual
  classical and einstein-podolsky-rosen channels},}\ }\href@noop {} {\bibfield
  {journal} {\bibinfo  {journal} {Phys. Rev. Lett.}\ }\textbf {\bibinfo
  {volume} {70}},\ \bibinfo {pages} {1895} (\bibinfo {year}
  {1993})}\BibitemShut {NoStop}%
\bibitem [{\citenamefont {Pirandola}\ \emph {et~al.}(2015)\citenamefont
  {Pirandola}, \citenamefont {Eisert}, \citenamefont {Weedbrook}, \citenamefont
  {Furusawa},\ and\ \citenamefont {Braunstein}}]{teleREVIEW}%
  \BibitemOpen
  \bibfield  {author} {\bibinfo {author} {\bibfnamefont {S.}~\bibnamefont
  {Pirandola}}, \bibinfo {author} {\bibfnamefont {J.}~\bibnamefont {Eisert}},
  \bibinfo {author} {\bibfnamefont {C.}~\bibnamefont {Weedbrook}}, \bibinfo
  {author} {\bibfnamefont {A.}~\bibnamefont {Furusawa}}, \ and\ \bibinfo
  {author} {\bibfnamefont {S.~L.}\ \bibnamefont {Braunstein}},\ }\bibfield
  {title} {\enquote {\bibinfo {title} {Advances in quantum teleportation},}\
  }\href@noop {} {\bibfield  {journal} {\bibinfo  {journal} {Nat. Photon.}\
  }\textbf {\bibinfo {volume} {9}},\ \bibinfo {pages} {641--652} (\bibinfo
  {year} {2015})}\BibitemShut {NoStop}%
\bibitem [{\citenamefont {Bowen}\ and\ \citenamefont
  {Bose}(2001)}]{bowen2001teleportation}%
  \BibitemOpen
  \bibfield  {author} {\bibinfo {author} {\bibfnamefont {G.}~\bibnamefont
  {Bowen}}\ and\ \bibinfo {author} {\bibfnamefont {S.}~\bibnamefont {Bose}},\
  }\bibfield  {title} {\enquote {\bibinfo {title} {Teleportation as a
  depolarizing quantum channel, relative entropy, and classical capacity},}\
  }\href@noop {} {\bibfield  {journal} {\bibinfo  {journal} {Phys. Rev. Lett.}\
  }\textbf {\bibinfo {volume} {87}},\ \bibinfo {pages} {267901} (\bibinfo
  {year} {2001})}\BibitemShut {NoStop}%
\bibitem [{\citenamefont {Cope}\ \emph {et~al.}(2017)\citenamefont {Cope},
  \citenamefont {Hetzel}, \citenamefont {Banchi},\ and\ \citenamefont
  {Pirandola}}]{cope2017simulation}%
  \BibitemOpen
  \bibfield  {author} {\bibinfo {author} {\bibfnamefont {T.~P.~W.}\
  \bibnamefont {Cope}}, \bibinfo {author} {\bibfnamefont {L.}~\bibnamefont
  {Hetzel}}, \bibinfo {author} {\bibfnamefont {L.}~\bibnamefont {Banchi}}, \
  and\ \bibinfo {author} {\bibfnamefont {S.}~\bibnamefont {Pirandola}},\
  }\bibfield  {title} {\enquote {\bibinfo {title} {Simulation of non-pauli
  channels},}\ }\href@noop {} {\bibfield  {journal} {\bibinfo  {journal} {Phys.
  Rev. A}\ }\textbf {\bibinfo {volume} {96}},\ \bibinfo {pages} {022323}
  (\bibinfo {year} {2017})}\BibitemShut {NoStop}%
\bibitem [{\citenamefont {Bennett}\ \emph {et~al.}(1996)\citenamefont
  {Bennett}, \citenamefont {DiVincenzo}, \citenamefont {Smolin},\ and\
  \citenamefont {Wootters}}]{BDSW}%
  \BibitemOpen
  \bibfield  {author} {\bibinfo {author} {\bibfnamefont {C.~H.}\ \bibnamefont
  {Bennett}}, \bibinfo {author} {\bibfnamefont {D.~P.}\ \bibnamefont
  {DiVincenzo}}, \bibinfo {author} {\bibfnamefont {J.~A.}\ \bibnamefont
  {Smolin}}, \ and\ \bibinfo {author} {\bibfnamefont {W.~K.}\ \bibnamefont
  {Wootters}},\ }\bibfield  {title} {\enquote {\bibinfo {title} {Mixed-state
  entanglement and quantum error correction},}\ }\href@noop {} {\bibfield
  {journal} {\bibinfo  {journal} {Phys. Rev. A}\ }\textbf {\bibinfo {volume}
  {58}},\ \bibinfo {pages} {3824} (\bibinfo {year} {1996})}\BibitemShut
  {NoStop}%
\bibitem [{\citenamefont {Christandl}\ \emph {et~al.}(2018)\citenamefont
  {Christandl}, \citenamefont {Leditzky}, \citenamefont {Majenz}, \citenamefont
  {Smith}, \citenamefont {Speelman},\ and\ \citenamefont
  {Walter}}]{christandl2018asymptotic}%
  \BibitemOpen
  \bibfield  {author} {\bibinfo {author} {\bibfnamefont {M.}~\bibnamefont
  {Christandl}}, \bibinfo {author} {\bibfnamefont {F.}~\bibnamefont
  {Leditzky}}, \bibinfo {author} {\bibfnamefont {C.}~\bibnamefont {Majenz}},
  \bibinfo {author} {\bibfnamefont {G.}~\bibnamefont {Smith}}, \bibinfo
  {author} {\bibfnamefont {F.}~\bibnamefont {Speelman}}, \ and\ \bibinfo
  {author} {\bibfnamefont {M.}~\bibnamefont {Walter}},\ }\bibfield  {title}
  {\enquote {\bibinfo {title} {Asymptotic performance of port-based
  teleportation},}\ }\href@noop {} {\bibfield  {journal} {\bibinfo  {journal}
  {preprint arXiv:1809.10751}\ } (\bibinfo {year} {2018})}\BibitemShut
  {NoStop}%
\bibitem [{\citenamefont {Lloyd}(1995)}]{lloyd1995almost}%
  \BibitemOpen
  \bibfield  {author} {\bibinfo {author} {\bibfnamefont {S.}~\bibnamefont
  {Lloyd}},\ }\bibfield  {title} {\enquote {\bibinfo {title} {Almost any
  quantum logic gate is universal},}\ }\href@noop {} {\bibfield  {journal}
  {\bibinfo  {journal} {Phys. Rev. Lett.}\ }\textbf {\bibinfo {volume} {75}},\
  \bibinfo {pages} {346} (\bibinfo {year} {1995})}\BibitemShut {NoStop}%
\bibitem [{\citenamefont {Khaneja}\ \emph {et~al.}(2005)\citenamefont
  {Khaneja}, \citenamefont {Reiss}, \citenamefont {Kehlet}, \citenamefont
  {Schulte-Herbr{\"u}ggen},\ and\ \citenamefont {Glaser}}]{khaneja2005optimal}%
  \BibitemOpen
  \bibfield  {author} {\bibinfo {author} {\bibfnamefont {N.}~\bibnamefont
  {Khaneja}}, \bibinfo {author} {\bibfnamefont {T.}~\bibnamefont {Reiss}},
  \bibinfo {author} {\bibfnamefont {C.}~\bibnamefont {Kehlet}}, \bibinfo
  {author} {\bibfnamefont {T.}~\bibnamefont {Schulte-Herbr{\"u}ggen}}, \ and\
  \bibinfo {author} {\bibfnamefont {S.~J.}\ \bibnamefont {Glaser}},\ }\bibfield
   {title} {\enquote {\bibinfo {title} {Optimal control of coupled spin
  dynamics: design of nmr pulse sequences by gradient ascent algorithms},}\
  }\href@noop {} {\bibfield  {journal} {\bibinfo  {journal} {J. Magn. Reson.}\
  }\textbf {\bibinfo {volume} {172}},\ \bibinfo {pages} {296--305} (\bibinfo
  {year} {2005})}\BibitemShut {NoStop}%
\bibitem [{\citenamefont {Boixo}\ \emph {et~al.}(2018)\citenamefont {Boixo},
  \citenamefont {Isakov}, \citenamefont {Smelyanskiy}, \citenamefont {Babbush},
  \citenamefont {Ding}, \citenamefont {Jiang}, \citenamefont {Bremner},
  \citenamefont {Martinis},\ and\ \citenamefont
  {Neven}}]{boixo2018characterizing}%
  \BibitemOpen
  \bibfield  {author} {\bibinfo {author} {\bibfnamefont {S.}~\bibnamefont
  {Boixo}}, \bibinfo {author} {\bibfnamefont {S.~V.}\ \bibnamefont {Isakov}},
  \bibinfo {author} {\bibfnamefont {V.~N.}\ \bibnamefont {Smelyanskiy}},
  \bibinfo {author} {\bibfnamefont {R.}~\bibnamefont {Babbush}}, \bibinfo
  {author} {\bibfnamefont {N.}~\bibnamefont {Ding}}, \bibinfo {author}
  {\bibfnamefont {Z.}~\bibnamefont {Jiang}}, \bibinfo {author} {\bibfnamefont
  {M.~J.}\ \bibnamefont {Bremner}}, \bibinfo {author} {\bibfnamefont {J.~M.}\
  \bibnamefont {Martinis}}, \ and\ \bibinfo {author} {\bibfnamefont
  {H.}~\bibnamefont {Neven}},\ }\bibfield  {title} {\enquote {\bibinfo {title}
  {Characterizing quantum supremacy in near-term devices},}\ }\href@noop {}
  {\bibfield  {journal} {\bibinfo  {journal} {Nat. Phys.}\ }\textbf {\bibinfo
  {volume} {14}},\ \bibinfo {pages} {595} (\bibinfo {year} {2018})}\BibitemShut
  {NoStop}%
\bibitem [{\citenamefont {Lloyd}\ \emph {et~al.}(2014)\citenamefont {Lloyd},
  \citenamefont {Mohseni},\ and\ \citenamefont
  {Rebentrost}}]{lloyd2014quantum}%
  \BibitemOpen
  \bibfield  {author} {\bibinfo {author} {\bibfnamefont {Seth}\ \bibnamefont
  {Lloyd}}, \bibinfo {author} {\bibfnamefont {Masoud}\ \bibnamefont {Mohseni}},
  \ and\ \bibinfo {author} {\bibfnamefont {Patrick}\ \bibnamefont
  {Rebentrost}},\ }\bibfield  {title} {\enquote {\bibinfo {title} {Quantum
  principal component analysis},}\ }\href@noop {} {\bibfield  {journal}
  {\bibinfo  {journal} {Nat. Phys.}\ }\textbf {\bibinfo {volume} {10}},\
  \bibinfo {pages} {631} (\bibinfo {year} {2014})}\BibitemShut {NoStop}%
\bibitem [{\citenamefont {Stickel}(1987)}]{stickel1987frechet}%
  \BibitemOpen
  \bibfield  {author} {\bibinfo {author} {\bibfnamefont {E.}~\bibnamefont
  {Stickel}},\ }\bibfield  {title} {\enquote {\bibinfo {title} {On the
  fr{\'e}chet derivative of matrix functions},}\ }\href@noop {} {\bibfield
  {journal} {\bibinfo  {journal} {Linear Algebra and its Applications}\
  }\textbf {\bibinfo {volume} {91}},\ \bibinfo {pages} {83--88} (\bibinfo
  {year} {1987})}\BibitemShut {NoStop}%
\bibitem [{\citenamefont {Ravi}\ \emph {et~al.}(2017)\citenamefont {Ravi},
  \citenamefont {Collins},\ and\ \citenamefont
  {Singh}}]{ravi2017deterministic}%
  \BibitemOpen
  \bibfield  {author} {\bibinfo {author} {\bibfnamefont {S.~N.}\ \bibnamefont
  {Ravi}}, \bibinfo {author} {\bibfnamefont {M.~D.}\ \bibnamefont {Collins}}, \
  and\ \bibinfo {author} {\bibfnamefont {V.}~\bibnamefont {Singh}},\ }\bibfield
   {title} {\enquote {\bibinfo {title} {A deterministic nonsmooth frank wolfe
  algorithm with coreset guarantees},}\ }\href@noop {} {\bibfield  {journal}
  {\bibinfo  {journal} {preprint arXiv:1708.06714}\ } (\bibinfo {year}
  {2017})}\BibitemShut {NoStop}%
\bibitem [{\citenamefont {Yousefian}\ \emph {et~al.}(2012)\citenamefont
  {Yousefian}, \citenamefont {Nedi{\'c}},\ and\ \citenamefont
  {Shanbhag}}]{yousefian2012stochastic}%
  \BibitemOpen
  \bibfield  {author} {\bibinfo {author} {\bibfnamefont {F.}~\bibnamefont
  {Yousefian}}, \bibinfo {author} {\bibfnamefont {A.}~\bibnamefont
  {Nedi{\'c}}}, \ and\ \bibinfo {author} {\bibfnamefont {U.~V.}\ \bibnamefont
  {Shanbhag}},\ }\bibfield  {title} {\enquote {\bibinfo {title} {On stochastic
  gradient and subgradient methods with adaptive steplength sequences},}\
  }\href@noop {} {\bibfield  {journal} {\bibinfo  {journal} {Automatica}\
  }\textbf {\bibinfo {volume} {48}},\ \bibinfo {pages} {56--67} (\bibinfo
  {year} {2012})}\BibitemShut {NoStop}%
\bibitem [{\citenamefont {Lan}(2013)}]{lan2013complexity}%
  \BibitemOpen
  \bibfield  {author} {\bibinfo {author} {\bibfnamefont {G.}~\bibnamefont
  {Lan}},\ }\bibfield  {title} {\enquote {\bibinfo {title} {The complexity of
  large-scale convex programming under a linear optimization oracle},}\
  }\href@noop {} {\bibfield  {journal} {\bibinfo  {journal} {preprint
  arXiv:1309.5550}\ } (\bibinfo {year} {2013})}\BibitemShut {NoStop}%
\end{thebibliography}%

\end{document}